\documentclass[
 reprint, 
 superscriptaddress,
 amsmath,
 amssymb,
 aps
]{revtex4-2}

\usepackage{graphicx}
\usepackage{dcolumn}
\usepackage{bm}
\usepackage{url}
\usepackage{braket}

\usepackage{color}
\usepackage{esvect}
\usepackage{graphicx}
\usepackage{dcolumn}
\usepackage[utf8]{inputenc}
\usepackage{graphicx}
\usepackage{epstopdf}
\usepackage{amsthm}
\usepackage{amsmath}
\usepackage{empheq}
\usepackage{amssymb}
\usepackage{tabularx}
\usepackage[usenames,dvipsnames]{xcolor}
\usepackage[colorlinks=true,citecolor=Cerulean,linkcolor=RubineRed,urlcolor=Cerulean]{hyperref}

\usepackage[noend]{algpseudocode}
\usepackage{algorithm}

\DeclareMathOperator*{\argmin}{arg\,min}

\begin{document}

\preprint{APS/123-QED}

\title{Discovering hydrodynamic equations of many-body quantum systems
}
\author{Yaroslav Kharkov}
\affiliation{Joint Quantum Institute and Joint Center for Quantum Information and Computer Science,
NIST/University of Maryland, College Park, Maryland 20742, USA}
\author{Oles Shtanko}
\thanks{Currently at IBM Quantum, MIT-IBM Watson AI Lab, Cambridge MA, 02139 US}
\affiliation{Joint Quantum Institute and Joint Center for Quantum Information and Computer Science,
NIST/University of Maryland, College Park, Maryland 20742, USA}
\author{Alireza Seif}
\affiliation{Pritzker School of Molecular Engineering, University of Chicago, Chicago, IL 60637}
\author{Przemyslaw Bienias}
\affiliation{Joint Quantum Institute and Joint Center for Quantum Information and Computer Science,
NIST/University of Maryland, College Park, Maryland 20742, USA}
\author{Mathias Van Regemortel}
\affiliation{Joint Quantum Institute and Joint Center for Quantum Information and Computer Science,
NIST/University of Maryland, College Park, Maryland 20742, USA}
\author{Mohammad Hafezi}
\affiliation{Joint Quantum Institute and Joint Center for Quantum Information and Computer Science,
NIST/University of Maryland, College Park, Maryland 20742, USA}
\author{Alexey V. Gorshkov}
\affiliation{Joint Quantum Institute and Joint Center for Quantum Information and Computer Science,
NIST/University of Maryland, College Park, Maryland 20742, USA}


\begin{abstract}
\textbf{
Simulating and predicting dynamics of quantum many-body systems is extremely challenging, even for state-of-the-art computational methods, due to the spread of entanglement across the system.
However, in the long-wavelength limit, quantum systems often admit a simplified description, which involves a small set of physical observables and requires only a few  parameters such as sound velocity or viscosity. Unveiling the  relationship between these hydrodynamic equations and the underlying microscopic theory usually requires a great effort by condensed matter theorists. In the present paper, we develop a new machine-learning framework for automated discovery of effective equations from a limited set of available data, thus bypassing complicated analytical derivations. The data can be generated from  numerical simulations or come from experimental quantum simulator platforms. Using integrable models, where direct comparisons can be made, we reproduce  previously known hydrodynamic equations,  strikingly discover novel equations and provide their derivation whenever possible. 
We discover new hydrodynamic equations describing dynamics of interacting systems, for which the derivation remains an outstanding challenge. 
Our approach provides a new interpretable method to study properties of quantum materials and quantum simulators in non-perturbative regimes. 
}
\end{abstract}

\maketitle

The discovery of analytical formulations of physical laws requires profound research intuition combined with domain expertise and ingenuity, hence, ultimately relying on human talent and insight.
Finding new ways to automate scientists' thinking process by leveraging machine learning methods could significantly accelerate research progress.
Machine learning algorithms are already achieving superhuman performance across a broad range of industrial applications and becoming widely utilized in various domains of science~\cite{carleo2019machine}. However, the potential of machine learning as a tool for automated derivation of previously unknown mathematical models or equations from numerical simulations or experimental data remains almost untapped.

In the context of classical physics, machine learning algorithms have been applied to extracting equations of classical mechanics from experimental data~\cite{schmidt2009distilling}, rediscovering physical concepts and conservation laws~\cite{iten2020discovering,seif2021machine,bondesan2019learning, wetzel2020snn,udrescu2020ai}, 
and finding ordinary or partial differential equations describing dynamics of complex classical systems/fluids \cite{brunton2016discovering, rudy2017data, Kaheman2020Sindy, schaeffer2017learning}. 
The focus of these works was to demonstrate capabilities of learning algorithms rather than discovering previously unknown equations.

Meanwhile, the power of symbolic-level discovery algorithms has not been explored in the quantum setting, namely in the context of quantum many-body transport phenomena, where, as we show below, they can be most fruitful and lead to new nontrivial analytical results. 
Understanding dynamics of many-particle quantum systems represents a long-standing challenge, since analytical techniques remain scarce and numerical methods have a limited evolution time horizon. While for generic quantum evolution there is no way around the curse of dimensionality of the Hilbert space, in many physically relevant cases,  the long-wavelength dynamics of local observables can  be described  by a small set of partial differential equations, of hydrodynamic nature, respecting the fundamental conservation laws. Such solutions are of immense importance, with applications ranging from quantum critical matter in solids~\cite{custers2003break,fermi2007lohneysen} and ultracold atoms~\cite{many2008bloch,greiner2002quantum} to the quark-gluon plasma~\cite{kovtun2005viscosity}.
Prominent successful examples of hydrodynamic models of many-body quantum systems include electron transport in graphene, where electron-electron interactions result in a viscous electron flow~\cite{levitov2016electron, bandurin2016negative, narozhny2015hydrodynamics},  and a generalized quantum hydrodynamics description of integrable systems~\cite{bulchandani2017solvable, castro2016emergent, bettelheim2008quantum, ruggiero2020quantum, borsi2020prx}.
Analytical derivation from first principles of such hydrodynamic equations is a formidable task, especially in non-perturbative regimes.

\begin{figure*}[htbp]
    \centering
    \includegraphics[width=0.945\linewidth]{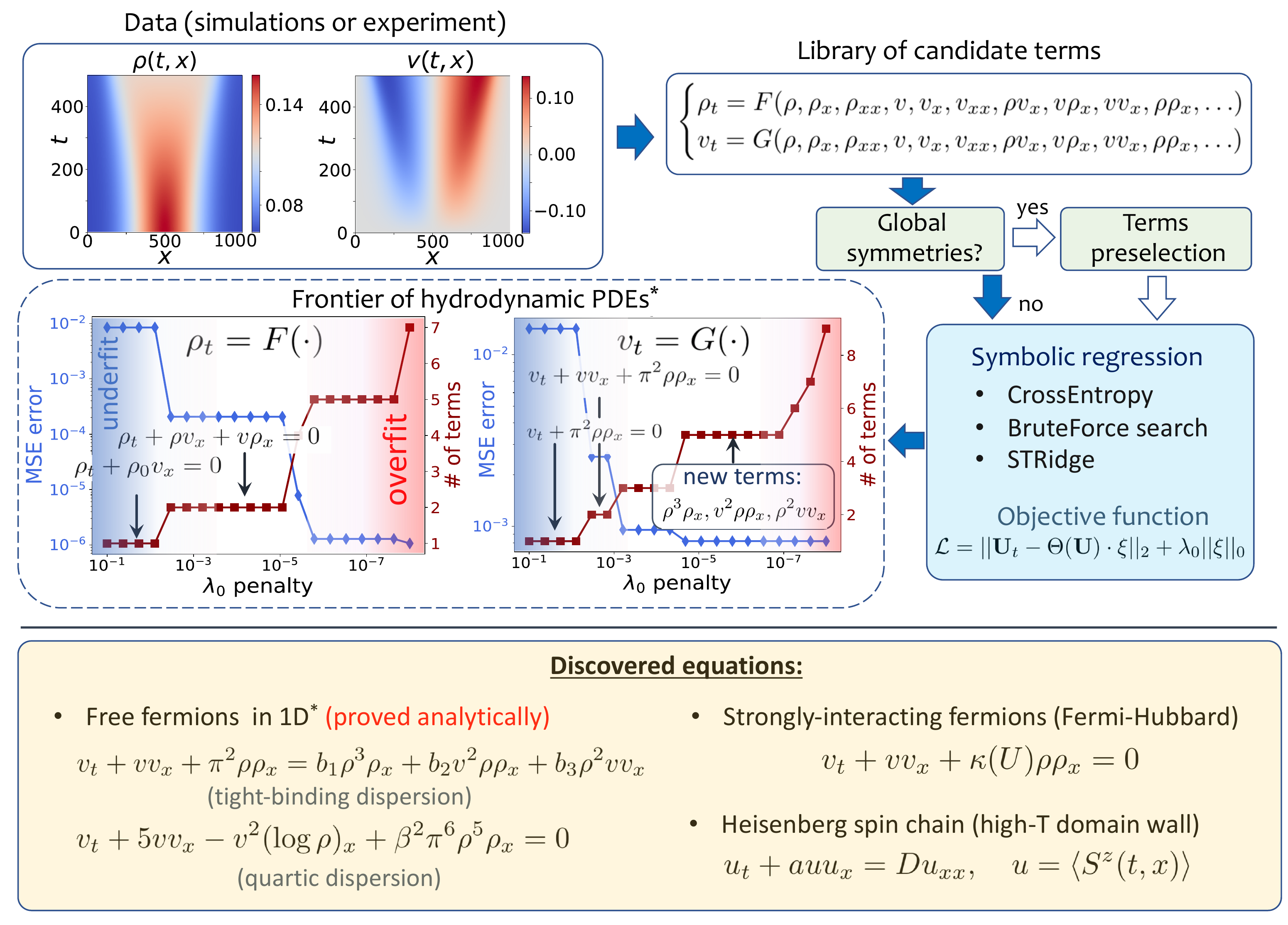}
    \caption{Framework for data-driven discovery of PDEs in many-body quantum systems from real-time dynamics. Data for the evolution of physical observables (e.g.~particle density $\rho$ and velocity $v$) is obtained either from numerical simulations or directly from experimental measurements. A general ansatz for an unknown PDE is constructed by performing long-wavelength expansion and considering leading order non-linear terms. 
    Compact PDEs are recovered by utilizing a combinations of symbolic regression algorithms.
    If a priori knowledge of underlying symmetries is available, the candidate terms are preselected on the basis of global symmetries, e.g.~parity ($P$) or time-reversal ($T$) symmetries. While scanning across values of the penalty parameter $\lambda_0$, we discover a frontier of PDEs with an increasing number of terms and a higher accuracy until we enter the overfit regime. 
    The list of previously unknown PDEs discovered with our algorithm is shown in the yellow box.
    Example of the frontier of semiclassical hydrodynamic PDEs for a free fermion gas and newly discovered correction terms are presented in the dashed blue box.
    }
    \label{fig:collage}
\end{figure*}

In this work, we employ symbolic regression methods to deduce partial differential equations (PDEs) governing the evolution of physical observables in many-body quantum systems. 
In addition to reproducing well-established equations and finding semiclassical approximations of exact equations, our algorithm discovered several new PDEs, where the most striking results correspond to hydrodynamics in fermionic systems and in the Heisenberg spin chain.
The  list of newly discovered equations, as well as the schematic workflow of hydrodynamic PDE-learning, are displayed in Fig.~\ref{fig:collage}.
The spatiotemporal data for the evolution of physical observables is either generated from numerical simulations or directly taken from experiment.
Next, we form a library of candidate terms from observables and their spatial derivatives. 
In cases when a priori knowledge about global symmetries of the system is available, we can significantly reduce the size of the search space by considering only symmetry-preserving terms.
Finally, a symbolic sparse regression problem is solved by searching through a large number of combinations of terms in the PDE and reconstructing the Pareto frontier---a sequence of the best matching hydrodynamic equations with progressively increasing complexity.

Until now the applications of machine learning to quantum problems were mostly restricted to black-box-type neural-network models. The list of use cases includes, among many others, identification of phases of matter~\cite{rodriguez2019identifying,carrasquilla2017machine,van2017learning},
neural-network wave function ansatze~\cite{carleo2017solving,sharir2020deep,choo2019two, hibat2020recurrent,levine_2019_quantum, carrasquilla2020machine,carleo_2019_machine,dunjko2018machine,mohseni2021deep,Mazza2021}, forecasting the dynamics of physical observables~\cite{luchnikov2020machine, banchi2018modelling},
experimental data processing and design~\cite{bohrdt2019classifying,rem2019identifying,zhang2019machine,kalantre2019machine,melnikov2018active,krenn2016automated}, and
quantum computing and quantum tomography~\cite{torlai2018neural,agresti2019pattern, torlai2017neural,Bienias2021}.
Nevertheless, these examples were either limited to problems with a
known solution, or the solution provided by machine learning tools lacked interpretability and, hence, provided only limited analytical insight into the underlying physical system.
Symbolic regression-based methods  offer a solution to these limitations as they (1) are interpretable by construction, (2)  provide an additional insight to the physics of the phenomena by recovering exact or approximate evolution equations, (3) are robust to noise, (4) are data-efficient with a single trajectory being sufficient in most cases, and 
(5) have a low computational cost for training and solving the evolution forward in time.

\textit{Reconstruction of PDEs via sparse regression}.---Fundamental equations in physics usually have a compact form and contain only a small number of relevant terms, thus following Occam's razor principle. 
Suppose that we want to infer the PDE in the functional form
\begin{equation}
\label{eq:ut_gen_form}
    u_t = F(u, u_x, u_{xx}, \ldots, x, t),
\end{equation}
where $u(t,x)=\langle O(t,x) \rangle$ is a scalar observable corresponding to a quantum operator $O$, and $F$ is an unknown function that we are trying to learn.
A PDE in the form of Eq.~(\ref{eq:ut_gen_form}) is a natural ansatz in the long-wavelength limit for the dynamics of many-body quantum systems with smooth initial conditions.
It is convenient to represent the function $F$ as a linear combination of individual terms, such that the terms could be nonlinear in $u$ and its spatial derivatives, e.g., $u u_x$, $u^2 u_x$, $(u_x)^2$, $u_x u_{xx}$ etc., which span the library of candidate terms.
We convert the PDE-learning problem to a sparse regression problem following the general method proposed in Refs.~\cite{rudy2017data, brunton2016discovering}. 
The dataset is given as a matrix of observables on the discretized spatiotemporal grid $u(t_i, x_j)$,  which we further vectorize to form a single vector: $\mathbf{U}=\textrm{vec}[u(t_i, x_j)]$. 
The library matrix $\Theta(\mathbf{U})$ is created by combining $M$ columns corresponding to candidate terms from $F$:
\begin{equation}
    \Theta(\mathbf{U}) = [1 \quad \mathbf{U} \quad \mathbf{U}_x \quad \mathbf{U}_{xx} \quad \mathbf{U U}_x  \ldots].
\end{equation}
Thus the task of PDE learning is reduced to the identification of active terms in the library and extraction of the values of the corresponding coefficients. 

The PDE identification problem could be rewritten as a minimization problem for the following objective function:
\begin{eqnarray}
\label{eq:L0_regr}
    &&\mathcal{L}(\mathbf{\xi}) = ||\mathbf{U}_t - \Theta(\mathbf{U}) \cdot \xi||_2 + \lambda_0 ||\xi||_0,\\
    &&\xi_{best} =\mathrm{argmin}_{\xi}\mathcal{L}(\xi), 
    \end{eqnarray}
where $\xi\in \mathbb{C}^M$ is the vector of regression coefficients. The objective function is given by the sum of an error term and an $L_0$ penalty term (proportional to the number of non-zero terms), which promotes sparse solutions, i.e.~parsimonious solutions with a small number of nonzero terms. 
The presence of the $L_0$ penalty  in Eq.~(\ref{eq:L0_regr}) results in a non-convex optimization landscape, 
and the optimization problem (\ref{eq:L0_regr}) is NP-hard in the general case due to an exponential growth with $M$ of the total number of  linear combinations of terms~\cite{natarajan1995sparse,nphardproof}.
However, the hydrodynamic nature of PDE ansatze, in
conjunction with symmetry-based analysis, leads to a rather limited library, and the search remains within reach for most of the problems considered.
While the convex relaxation of the optimization problem (\ref{eq:L0_regr}) with an $L_1$ penalty instead of the $L_0$ penalty  makes the problem tractable,
such an approach results in poor PDE reconstruction quality of nonlinear equations when there are strong correlations 
between columns of the matrix $\Theta$~\cite{rudy2017data}. The average-case hardness of the optimization problem (\ref{eq:L0_regr}) over the set of physically relevant PDEs remains an open question and a subject for future research.  

Naively, setting aside the sparsity requirement, the regression problem could be simply solved via the least-squares method.  The  least squares regression solution will generally have no vanishing coefficients suggesting a PDE containing all the terms presented in the library, thus violating  the assumption on model sparsity. In addition, the least-squares problem is usually poorly conditioned in the presence of nearly-collinear terms~\cite{rudy2017data}.
Most importantly, inactive nonlinear terms may induce a significant bias in the values of coefficients corresponding to the true terms.
Thus naive least-squares regression, when applied to nonlinear problems, will fail. 

Development of new symbolic regression algorithms is an active area of research, see e.g.~\cite{dubvcakova2011eureqa, rudy2017data, petersen2019deep, udrescu2020ai, messenger2020weak, almomani2020entropic}. 
Sequential Thresholding Ridge regression (STRidge)~\cite{rudy2017data} is among the state-of-the-art methods for symbolic PDE-learning and is closely related to the SINDy (Sparse Identification of Nonlinear
Dynamics) algorithm~\cite{brunton2016discovering}.
While STRidge~\cite{rudy2017data} has demonstrated  strong performance for certain PDE-learning tasks, we found that it is prone to getting stuck in local optima when applied to challenging nonlinear problems.
To overcome this difficulty, we employ a rather straightforward BruteForce search algorithm, that combines exhaustive combinatorial search for relevant terms with linear regression. 
As a scalable alternative to BruteForce, we propose a novel CrossEntropy algorithm.  It is based, like BruteForce, on the minimization of the objective function in Eq.~(\ref{eq:L0_regr}) and relies on the sampling-based cross-entropy method for combinatorial optimization~\cite{rubinstein1999cross, de2005tutorial}.
We find that CrossEntropy and BruteForce produce the most reliable results in our tests and use them as primary  PDE-reconstruction tools.  

\textit{Interacting spins}.---One of the paradigmatic examples of many-body dynamics is the one-dimensional XXZ spin model, which is of great interest for realizing models of quantum magnetism using quantum simulators, such as ultracold atoms in optical lattices~\cite{gross2017quantum}, Rydberg atoms~\cite{bernien2017probing}, cold polar molecules~\cite{hazzard2013far}, superconducting qubits~\cite{aleiner2020accurately} and trapped ions~\cite{monroe2021programmable}. 
We focus on the ferromagnetic XXZ chain described by the Hamiltonian
\begin{equation}
\label{eq:xxz_ham_main}
    H_{\rm S} =  - \sum_{i}\Bigl[S^x_i S^x_{i+1} + S^y_i S^y_{i+1} + \Delta S^z_i S^z_{i+1}\Bigl] + \sum_i B_i S_i^z,
\end{equation}
where $\Delta$ is the anisotropy parameter, $B_i$ are local magnetic fields, ${S}_i^{\mu=x,y,z}=\mathbf{\sigma}_i^\mu/2$ is the spin operator, and $\sigma^\mu_i$ are Pauli matrices associated with spin $i$.
The ground state phase diagram of the XXZ model without a magnetic field is controlled by the value of the anisotropy parameter:  $\Delta<1$ corresponds to a $U(1)$-symmetric gapless phase, whereas, for $\Delta>1$, the system is in a gapped Ising phase.
The spin dynamics in the general case can be described by a system of interacting magnons, such that the total magnetization is conserved.
In spite of the fact that the XXZ model (\ref{eq:xxz_ham_main}) is integrable by Bethe ansatz,  the derivation of closed-form equations describing dynamics of observables is a notoriously difficult task in many physically interesting cases~\cite{giamarchi2003quantum}.
Although our PDE-learning methodology is applicable beyond integrable quantum models,  integrability helps benchmarking symbolic regression algorithms against known analytical solutions.

In order to illustrate the methodology of PDE reconstruction, we start with the quench dynamics of a wave packet in the single-magnon excitation sector of the XXZ model. The initial state $|\psi_0\rangle$ is prepared by deforming a ferromagnetic product state: 
$|\psi_0\rangle  =  \sum_{n} f(n) \ket{\theta_n, \phi_n}_n \prod_{j \neq n} \ket{\downarrow}_{j}$, where  $| \theta_i, \phi_i \rangle$ is a rotated spin state at site $i$, parametrized by two Bloch angles $\theta_i$, $\phi_i$, and $f(n)$ corresponds to an envelope function, such that the Bloch angles and the wave packet envelope are smoothly varying across the spin chain.
We choose the physical observable of interest to be $u(t,x_i) =  \frac{1}{2}\left(\langle \sigma_i^x(t) \rangle + i \langle \sigma_i^y(t) \rangle\right)$. In the continuum limit, the complex field $u(t,x)$ satisfies (see Supplementary Material)
\begin{eqnarray}\label{eq:1magnon_pde}
    i \partial_t u + \frac{1}{2} \partial_{x}^2 u + \left(1 - \Delta \right)u + B(x) u + \mathcal{O}(\partial_x^4 u)=0,
\end{eqnarray}
where $B(x)$ is the continuous version of the magnetic fields $B_i$.
The gradient expansion in Eq.~(\ref{eq:1magnon_pde}) stems from the Taylor expansion of the tight-binding kinetic-energy operator $\cos(i\partial_x)=1+\partial_x^2/2 +\partial_x^4/24 + \ldots$.
Note that Eq.~(\ref{eq:1magnon_pde}) is valid for both ferromagnetic and antiferromagnetic XXZ models as long as the initial state corresponds to a superposition of a ferromagnetic product state and a single spin flip.   

In order to learn PDE (\ref{eq:1magnon_pde}), we construct a dataset by exactly solving the Schr\"odinger equation for the Hamiltonian (\ref{eq:xxz_ham_main}) in the single-magnon subspace, where the initial state is prepared by imposing a Gaussian envelope profile $f(n)$.
We limit our search to a set of ten candidate terms as follows: 
$u_t = F(1, u, \partial_{x}^n u, u\partial_{x}^n u)$, where $n=1,\dots,4$. Applying our PDE-learning algorithm to a quench problem with $B_i = 0$ for all $i$ and $\Delta = 0.5$, we arrive at the following equation: 
\begin{equation}\label{eq:discovered_1magnon_pde}
    i\partial_t u + 0.495\, u_{xx} + 0.499\, u=0.
\end{equation}
The inferred PDE~(\ref{eq:discovered_1magnon_pde}) corresponds to a fixed nonzero value of the penalty constant $\lambda_0=10^{-3}$, and when scanning across the values of the penalty parameter our algorithm finds a frontier of equations matching the exact gradient expansion in (\ref{eq:1magnon_pde}). 
Although the analytical derivation of Eq.~(\ref{eq:1magnon_pde}) is straightforward,
 we first discovered it using the PDE-learning algorithm and then retrospectively derived the equation.
A reader interested in more examples of PDE-learning in the context of single-magnon dynamics in the XXZ model---examples featuring a confining potential $B(x)$ and long-range interactions---can find them in the Supplementary Material.

\begin{figure}[t]
    \centering
    \includegraphics[scale=0.35]{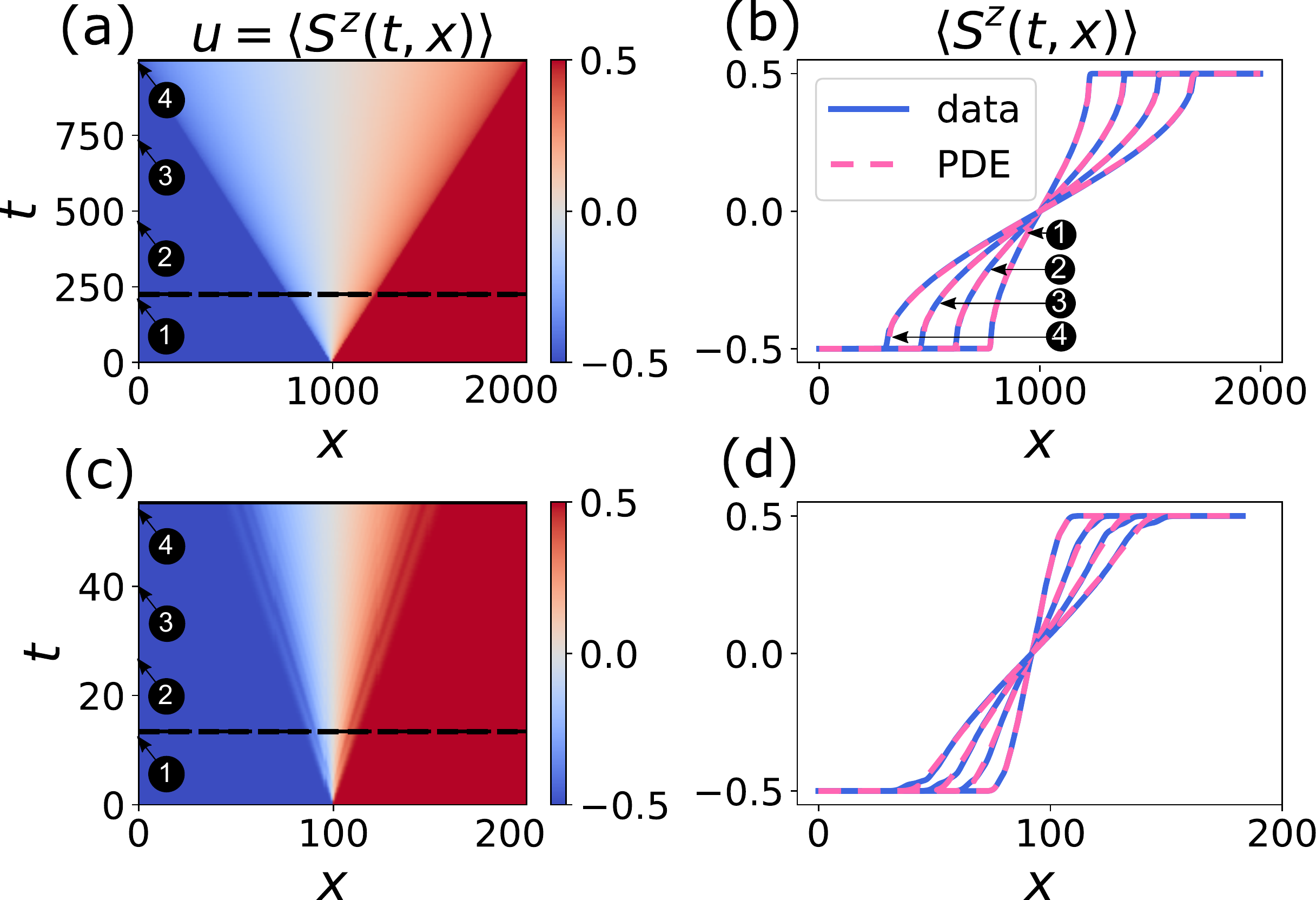}
    \caption{Quench evolution of the onsite magnetization $u(t,x) = \langle S^z(t, x)\rangle$  in XXZ model for the domain-wall initial state: (a, b) $\Delta=0$ (exact) and (c, d) $\Delta=0.5$ (TEBD simulations). 
    The solutions of the recovered PDE (\ref{eq:recovered_dw_taylor}) are shown with the red dashed line.
    The horizontal black dashed line at time $t_0$ in (a,c) separates the portion of the data $t\geq t_{0}$ that was used for PDE-reconstruction. 
    The four curves in panels (b,d) correspond to the four evolution times labeled in (a,c).
    }
    \label{fig:domain_wall}
\end{figure}

Now we turn to nontrivial test cases where the analytical derivation of evolution PDEs is more challenging. 
One such problem is the quench dynamics of onsite magnetization in the XXZ model for a domain-wall initial state $|\psi_0\rangle = |\downarrow\dots\downarrow\rangle\otimes |\uparrow\dots\uparrow\rangle$.
As it has been analytically shown, the magnetization dynamics in the continuum limit is governed by the following  PDE~\cite{collura2018analytic}
\begin{equation}
\label{eq:pde_collura}
u_t + \zeta_0 \sin{\left(\frac{2 \pi}{P} u\right)}   u_x = 0; \quad u(t, x_i) = \langle\psi(t)|S_i^z|\psi(t)\rangle,
\end{equation}
for the values of the anisotropy parameter satisfying  $\Delta = \cos{(\pi Q/ P)}$ with $Q,P\in\mathbb Z$ being coprime integers, and $\zeta_0 = \sqrt{1-\Delta^2}/\sin(\pi/P)$.
Equation (\ref{eq:pde_collura}) could be interpreted as the continuity equation representing conservation of total longitudinal magnetization.
To learn Eq.~(\ref{eq:pde_collura}) with $\Delta = 0$, we map the XXZ Hamiltonian to non-interacting fermions via the Jordan-Wigner transformation, allowing us to simulate the dynamics exactly for large system sizes.
On the other hand, in the interacting regime  ($\Delta \neq 0$), we apply the time evolving block decimation (TEBD) algorithm to generate data for systems up to a few hundreds of lattice sites.
Next, we apply our PDE reconstruction method with the penalty coefficient $\lambda_0=10^{-6}$ looking for an equation of the form $u_t=F(\partial_x^n u, u^n \partial_x u),\; n=1\ldots 5$. As a result, 
 from  data 
shown in Fig.~\ref{fig:domain_wall}(a, c),  we obtain
\begin{equation}
    u_t + a_1 u u_{x}+a_2u^3 u_x + a_3u^5 u_x  = 0.
\label{eq:recovered_dw_taylor}
\end{equation}
First, we consider the  non-interacting case, $\Delta\to 0$,
that results in the values of extracted coefficients equal to  $a_1 = 3.135$, $a_2 = -5.056$, and $a_3 = 1.92$. These values can be compared to corresponding values obtained from the Taylor expansion of Eq.~(\ref{eq:pde_collura}):
\begin{equation}
    u_t  = -\sin(\pi u)u_x \approx - \pi u u_x + \frac{\pi^3}{3!} u^3 u_x - \frac{\pi^5}{5!} u^5 u_x + \ldots,
\end{equation}
which is in an excellent agreement with the values of coefficients $a_i$ in (\ref{eq:recovered_dw_taylor}).
Second, if we use TEBD data from Fig.~\ref{fig:domain_wall}(c) corresponding to $\Delta=0.5$, we obtain
$a_1 = 2.1$, $a_2 = -1.67$, $a_3=0$, matching the values of coefficients from the Taylor expansion of the theoretically expected PDE $u_t + \sin{(2\pi/3 u)}u_x=0$, which follows from Eq.~(\ref{eq:pde_collura}).
The proposed method not only discovers the relevant terms in the PDEs, but also accurately identifies their coefficients, and hence can be used as a method for parameter estimation in cases when the theoretical form of the PDE is known.

Next, we focus on the problem where the analytical form of the PDE remains unknown.
Let us consider the problem where the initial state is the high-temperature domain wall state, which is qualitatively different from the zero-temperature case discussed above.
The initial state is prepared by combining left and right reservoirs  ($L$ and $R$) having opposite directions of the longitudinal magnetic field, so that the density matrix reads $\rho(t=0)=\rho_R \otimes \rho_L$, where  the thermal state for the right (left) subsystem is $\rho_{R(L)} \propto \exp{(\pm \mu \sum_{i\in {R (L)}} \sigma^z_{i})}$, and $\mu\ll 1$  is  the inverse temperature of the initial Gibbs state.
For the mixed initial state, the spin dynamics is ballistic in the gapless phase $\Delta<1$, superdiffusive at the isotropic point $\Delta=1$,  and diffusive in the gapped phase $\Delta>1$ ~\cite{bertini2021finite}. In contrast, for the zero-temperature  initial state,   the dynamics is frozen in the gapped phase. 
Following Ref.~\cite{ljubotina2017spin}, we define a rescaled longitudinal magnetization $u(t,x)=\langle S^z(t, x) \rangle/\mu$.
\begin{figure}
    \centering
    \includegraphics[scale=0.28]{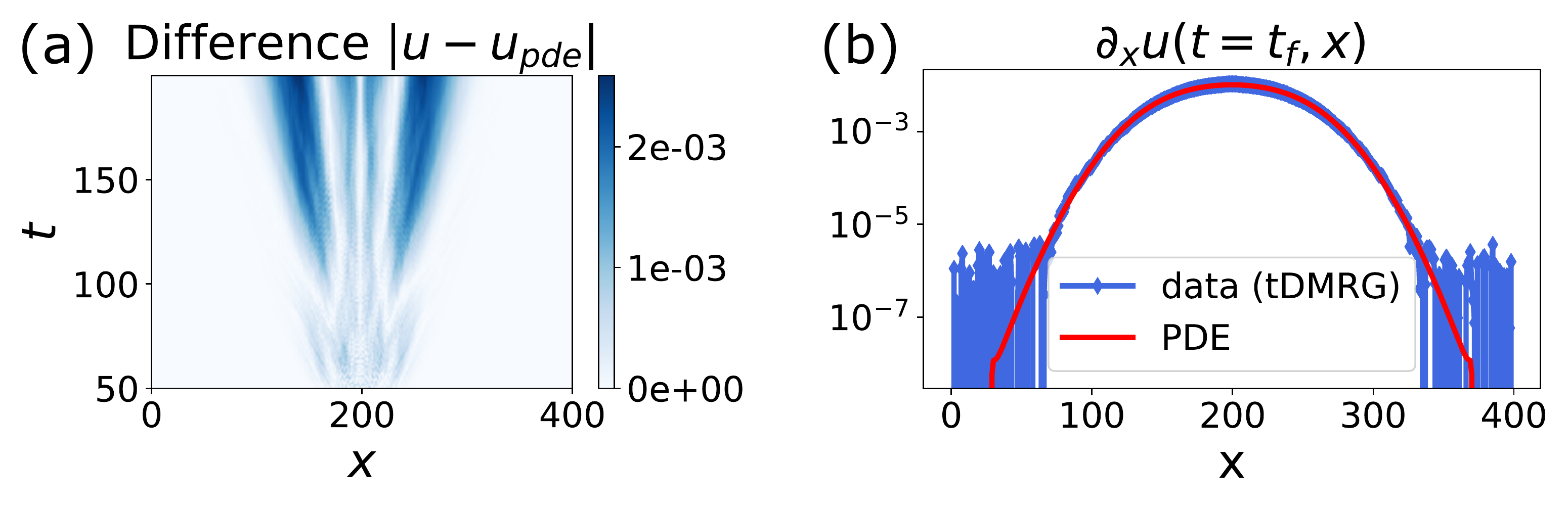}
    \caption{Evolution of a high-temperature domain-wall state in the XXZ model at the isotropic point, $\Delta=1$. Comparison between tDMRG data for local magnetization $u=\mu^{-1}\langle S^z(t,x) \rangle$ and the solution of the discovered PDE (\ref{eq:pde_delta=1_prosen}) denoted as $u_{\mathrm{pde}}(t,x)$.
    tDMRG data is taken from Ref.~\cite{ljubotina2019kardar}.
    }
    \label{fig:prosen_data_main}
\end{figure}
Using data from Ref.~\cite{ljubotina2019kardar},  our algorithm rediscovered an effective diffusion equation in the gapped phase, $u_t=D(\Delta) u_{xx}$, with the diffusion constant being a function of the anisotropy parameter $\Delta$; this form of the PDE agrees with the conclusions of Ref.~\cite{ljubotina2017spin}.
At the isotropic point $\Delta=1$,  we found the following deterministic PDE that matches the data with remarkable accuracy (see Fig.~\ref{fig:prosen_data_main}):
\begin{eqnarray}
\label{eq:pde_delta=1_prosen}
u_t + a u u_x = D u_{xx}, \quad a\approx 0.24,\; D\approx 1.90.
\end{eqnarray}
The discovered Eq.~(\ref{eq:pde_delta=1_prosen}) has the form of a viscous Burgers' equation and can be interpreted as a noise-averaged stochastic Burgers' equation, which is in turn equivalent to a noise-averaged Kardar-Parisi-Zhang (KPZ) equation for a field $h(t, x)$ after a variable substitution $u=h_x$ (using the notation of Ref.~\cite{kardar1986dynamic}).
Although KPZ scaling at the critical point  was empirically found in Ref.~\cite{ljubotina2019kardar}, the microscopic derivation of dynamical equations of KPZ-type  remains an outstanding theoretical challenge.

\begin{figure}
\centering
    \includegraphics[scale=0.35]{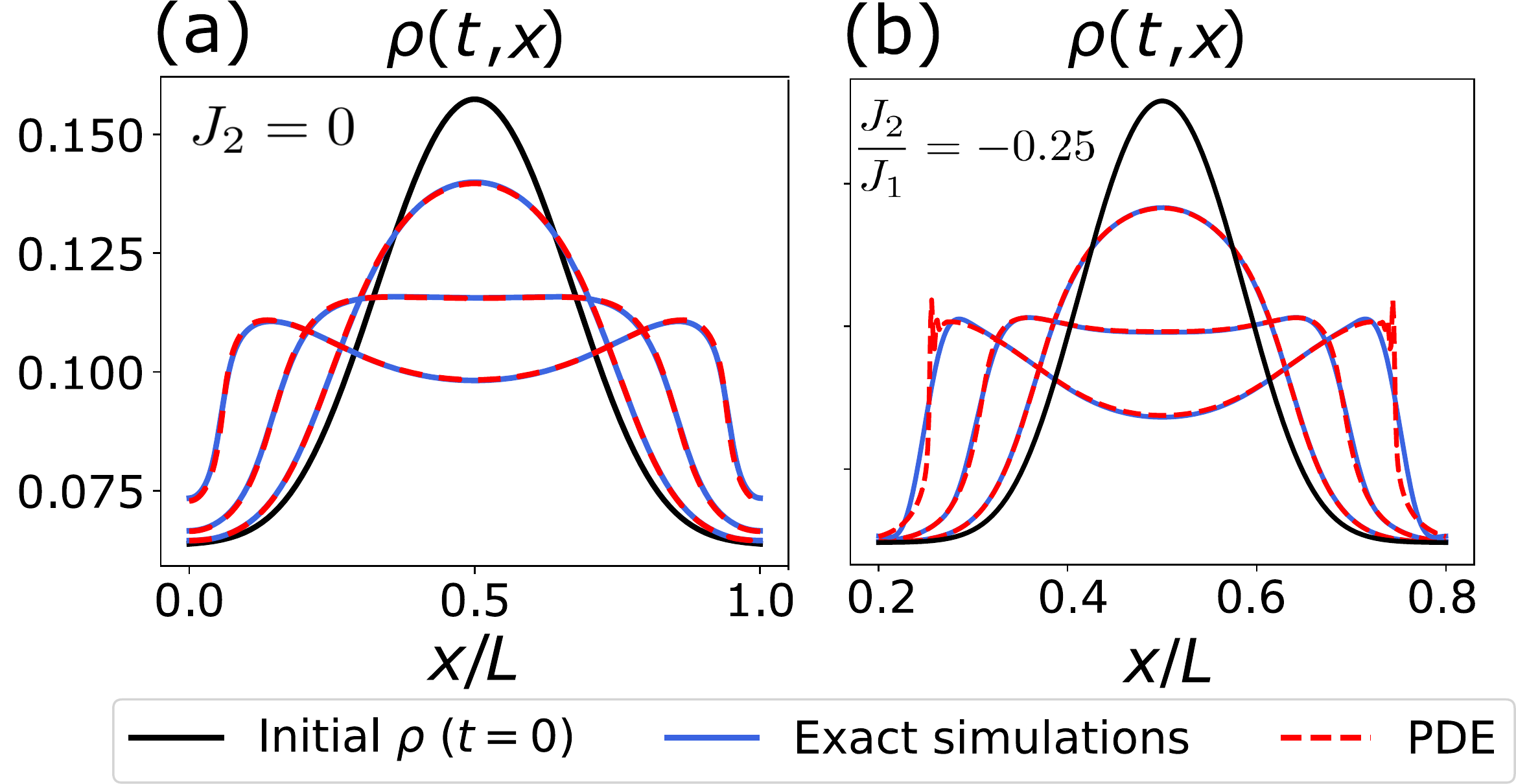}
    \caption{Hydrodynamics in a  non-interacting fermion gas: comparison of the exact evolution of fermion density $\rho(t,x)$  in the tight-binding model and the solutions of the recovered hydrodynamic system of PDEs [Eqs.~(\ref{eq:euler_recovered_v_main},~\ref{eq:vt_reconstr_j1j2_main})].  (a) Nearest-neighbor hopping only ($J_2=0$), (b)  fermion hydrodynamics in the $J_1-J_2$ model at the critical (Lifshitz) point corresponding to quartic dispersion $\varepsilon_k=  \beta k^4/4$.
    The evolution times are (a) 
    $J_1 t=(0, 125, 250, 375)$ and (b) $J_1 t=(0, 1250, 2500, 3750)$.
    Number of lattice sites is $L=1000$.
    }
    \label{fig:hydro_vs_exact_main}
\end{figure}

\textit{Fermion hydrodynamics}.---Another class of many-body quantum systems exhibiting hydrodynamic behavior is itinerant fermions. Let us first consider a system of free fermions described by the tight-binding Hamiltonian
\begin{equation}
\label{eq:Fermi-Hubbard_Hamiltonian}
    H_{\rm f} = - J\sum_{i} (c^\dag_i c_{i+1} + c^\dag_{i+1} c_i) - \mu \sum_i c^\dag_i c_i.
\end{equation}
The physical observables of interest are the particle density $\rho(t,i) = \langle c^\dag_ic_i\rangle$ and velocity $v(t,i) = 2J\Im \langle c^\dag_{i+1} c_i\rangle/\rho(t,i)$.
We assume that the density of fermions is small, so that the dispersion is well-approximated by  parabolic dispersion.
In the semiclassical approximation, the dynamics of free fermions with quadratic dispersion $\varepsilon_k = k^2/2m$ 
is governed by the system of hydrodynamic equations ~\cite{bettelheim2008quantum}:
\begin{equation}
\label{eq:pde_free_fermions_hydro}
\rho_t+(\rho v)_x = 0, \quad
v_t + v v_x  = - \frac{ 1}{m \rho} \partial_x P(\rho).   
\end{equation}
The first equation is the continuity equation for the density and represents the conservation of the total number of fermions. The second equation (for the velocity) has the form of a classical Euler equation for a barotropic compressible liquid flow, where the last term is given by the  ``Pauli pressure''  $P(\rho)=\pi^2 \rho^3/(3 m)$.
The key assumption behind the hydrodynamic model  (\ref{eq:pde_free_fermions_hydro}) is the   semiclassical approximation $|d\lambda_F(x)/dx|\ll 1$, where $\lambda_F(x)=2\pi/k_F(x)$ is the local de Broglie wavelength and $k_F(x)$ is the local Fermi-momentum. At the same time, the relative amplitude of the density perturbation is not required to be small for the validity of the hydrodynamic model. 
To study fermion dynamics we prepare the initial state as the ground state of  $H_{\mathrm{f}}+V(x)$, where $V(x)$ is a local Gaussian potential, and quench the potential to zero at $t>0$. By tuning the strength of the Gaussian potential, we intentionally set the amplitude of the hump in the density profile to be large in order to have a higher sensitivity to nonlinear effects. 
Considering a generic ansatz with candidate terms drawn from Table~\ref{table:t1},
we recover the following hydrodynamic PDEs for $J= 0.5$ (data are shown in Fig.~\ref{fig:collage}):
\begin{eqnarray}
\begin{cases}
\label{eq:euler_recovered_rho_main}
\rho_t + 1.006 \rho v_x + 1.0007 v \rho_x = 0,
\\
\label{eq:euler_recovered_v_main}
v_t + 0.97 v v_x + 9.45 \rho \rho_x = 0,
\end{cases}
\end{eqnarray}
which is in good agreement with the expected semiclassical equations (\ref{eq:pde_free_fermions_hydro}) for $m=1/(2J) = 1$ (note that $\pi^2\approx 9.869$). 
The recovered hydrodynamic model (\ref{eq:euler_recovered_v_main}) works surprisingly well  up to the time $t_c$ of the formation of the  ``gradient catastrophe''~\cite{mirlin2013, whitham2011linear}, see Fig.~\ref{fig:hydro_vs_exact_main}(a).
For $t \geq t_c$, the hydrodynamic model (\ref{eq:pde_free_fermions_hydro}) does not apply  because the semiclassical approximation breaks down~\cite{mirlin2013}.
The deviation of the recovered coefficients in the second line of Eq.~(\ref{eq:euler_recovered_v_main}) from theoretical values is due to  higher-order terms in the expansion of the tight-binding fermion dispersion, while the continuity equation in the first line of Eq.~(\ref{eq:euler_recovered_v_main}) is basically exact.
In addition to the identification of non-linear equations, our method is able to recover linearized approximations of exact equations, bypassing analytical derivations.
If we create an initial state with a small variation of the fermion density $\rho=\rho_0+\delta\rho$, where $\delta\rho/\rho_0\ll 1$, our algorithm recovers a linearized form of Euler equations, which could be reduced to the wave equation $\delta\rho_{tt}-v_F^2\delta\rho_{xx}=0$, where $v_F = \pi\rho_0$ is the wave speed, which coincides with the Fermi velocity.

\begin{center}
\begin{table}[h!]
\caption{Candidate terms for the rhs of Euler equation $v_t = G(\cdot)$. The library of candidate terms is constructed by considering an expansion in powers 
of fermion density, fermion velocity, and their spatial derivatives: we include terms up to quadratic order in $v$ ($v^m$ with $m\in [0,1,2]$) and up to fifth order in $\rho$ ($\rho^n$ with $n\in[0, 1, \ldots, 5]$), with spatial derivatives up to the second order. Only symmetry-allowed terms are selected, i.e.\ the last row, which contains terms that have the following signature with respect to $P$- and $T$-inversion transformations: $(P,T)=(-,+)$. Exploiting global symmetries in the PDE search is one of our novel contributions that leads to a significant reduction of the size of the candidate-terms library, thus facilitating the optimization problem.}
 \label{table:t1}
\begin{tabular}{|c|c|c|c|}
\hline
\textrm{Candidate terms} & $P$ & $T$ & \textrm{Select}\\
\hline
$\rho^{n}$, $\rho^n v^2$, $\rho^n \rho_{xx}$, $\rho^n v^2\rho_{xx}$, $\rho^n \rho_x^2$,  $\rho^n v_x^2$ & + & + & $\times$\\
$\rho^n v_{xx}$, $\rho^n v$, $\rho_x v_x$ & - & - & $\times$\\
$v_x$, $\rho v_x$, $\rho^2 v_x$, $\ldots$ & + & - & $\times$\\
$\rho^n \rho_x$, $\rho^n v v_x$, $v^2 \rho^n \rho_x$,
${\rho}_x/\rho$,\,$v^2 {\rho}_x/\rho$ & - & + & \checkmark\\
 \hline
\end{tabular}
\end{table}
\end{center}

Semiclassical equations of fermion hydrodynamics (\ref{eq:pde_free_fermions_hydro}) are derived within the assumption of a quadratic dispersion. Therefore, more accurate equations can be obtained by  accounting for higher order terms in the Taylor expansion of the tight-binding dispersion  $\varepsilon(k)\propto \cos{k}$. To the best of our knowledge, such corrections to the free-fermion Euler equation were not previously considered.
With CrossEntropy and  BruteForce algorithms, we were able to discover new correction terms: 
\begin{equation}
\label{eq:hydro_cos_corrections_main}
    v_t+v v_x + \frac{\pi^2}{m^2}\rho\rho_x \approx b_1 \rho^3\rho_x + b_2 v^2 \rho \rho_x + b_3 \rho^2 v v_x.
\end{equation}
Analytical derivation of the form of the correction terms and of the values of the corresponding coefficients is nontrivial (see Supplementary Material), while the algorithm readily discovers them. 
The coefficients $b_{i}$ are positive, and the terms on the rhs of Eq.~(\ref{eq:hydro_cos_corrections_main}) are in fact responsible for the shift of the values of extracted coefficients in (\ref{eq:euler_recovered_v_main}) relative to the Euler equation for free fermions with a parabolic dispersion, given in Eq.~(\ref{eq:pde_free_fermions_hydro}).
One should note that the discovery of subtle nonlinear corrections is possible only at the cost of increasing the precision of the input data by refining the spatiotemporal grid of the dataset.

For the next step, we extend the noninteracting tight-binding model (\ref{eq:Fermi-Hubbard_Hamiltonian}) by adding next-nearest neighbor hopping terms to yield the so-called $J_1-J_2$ model. Fermion hydrodynamics in such a model has not been previously studied.
The dispersion of fermions reads $\varepsilon_k = - 2J_1 \cos{(k)} - 2 J_2 \cos{(2k)}$. In the long-wavelength limit, we can perform expansion up to  fourth order in $k$:
$\varepsilon_k = \varepsilon_0 + \frac{\alpha k^2}{2} + \frac{\beta k^4}{4} + \mathcal{O}(k^6)$,
where $\alpha = 2(J_1+4 J_2)$, $\beta=-\left(\frac{J_1}{3}+\frac{16}{3}J_2\right)$.
By tuning the ratio of the hoppings $J_2/J_1$, one can set the coefficient $\alpha$ in front of the quadratic term to zero or even change the sign, while keeping $\beta$ positive.
The critical point $\alpha=0$ is a Lifshitz point where the Fermi surface changes the topology:  a single Fermi pocket at $\alpha>0$ splits into two pockets at $\alpha<0$~\cite{volovik2017topological}.
By taking symmetry-allowed terms, i.e.\ the last row of Table~\ref{table:t1}, the PDE-learning algorithm discovered the following equation for $J_1=0.5$  and $J_2=-0.125$ (corresponding to the Lifshitz critical point $\alpha = 0$):
\begin{equation}
\label{eq:vt_reconstr_j1j2_main}
    v_t + 4.98 v v_x + 225.7 \rho^5 \rho_x \approx 0. 
\end{equation}
The hydrodynamic equation at the Lifshitz critical point reads (see derivation in the Supplementary Material)
\begin{eqnarray}
\label{eq:vt_lifshitz_exact}
&v_t& + 5 v v_x - v^2(\log{\rho})_x + \beta^2 \pi^6 \rho^5 \rho_x \approx 0,
\end{eqnarray}
which, to the best of our knowledge, has not been previously reported.
Eq.~(\ref{eq:vt_lifshitz_exact})  is derived by performing an expansion in  $v/v_F$, where $v_F\propto \beta \rho^3$ is the Fermi velocity, and keeping only the leading terms.
The PDE discovery algorithm missed the term $v^2 (\log{\rho})_x$. However, this term is negligible in the regime of parameters considered, and it does not affect the solution of the PDE, see Fig. ~\ref{fig:hydro_vs_exact_main}(b).
As in the case of the fermionic gas with quadratic dispersion, the solution of the hydrodynamic PDE (\ref{eq:vt_lifshitz_exact})  develops a gradient catastrophe instability at large evolution times, that marks a breakdown of the semiclassical approximation.

\begin{figure*}[htbp]
    \centering
    \includegraphics[width=0.95\linewidth]{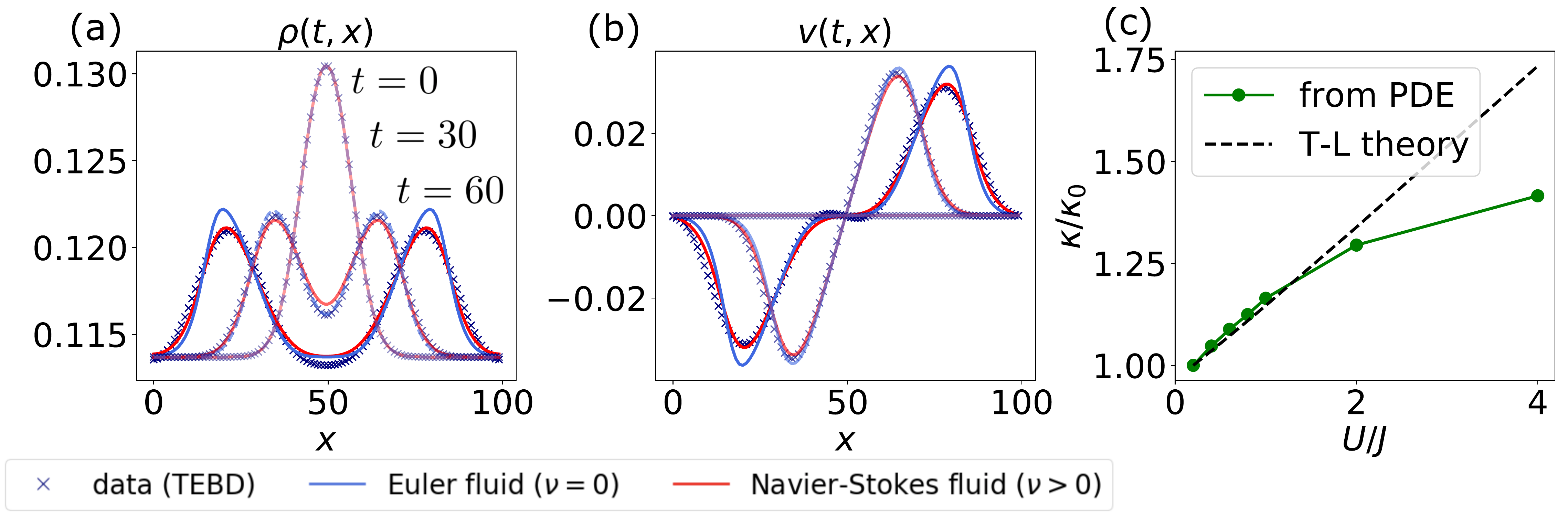}
    \caption{Hydrodynamics in the 1D Fermi-Hubbard model. (a, b)  Comparison between
    TEBD simulations (crosses, $U/J=2$, $U_2/J=2$) and solutions of the discovered hydrodynamic PDE~(\ref{eq:navier_stokes}). Solid blue line shows the solution of the Euler-fluid PDE ($\nu=0$), and solid red line corresponds to the solution of the viscous Navier-Stokes PDE ($\nu>0$). 
    (c) Dependence of the Pauli pressure renormalization parameter $\kappa(U)$ on the fermion-fermion interaction strength $U$ in  the Fermi-Hubbard model with  nearest-neighbor interactions ($U_2=0$); here $\kappa_0\equiv \kappa(U=0)$. Tomonaga-Luttinger (T-L)  theory  prediction~\cite{tsvelik2007quantum}  for $\kappa(U)$ is shown with the dashed black line.
    }
    \label{fig:navier_stokes}
\end{figure*}

Now we turn to the problem of the interacting Fermi-Hubbard model, adding to Eq.~\eqref{eq:Fermi-Hubbard_Hamiltonian} fermion-fermion interactions  in the form $V_{\rm int} = U\sum_i n_i n_{i+1}+U_2\sum_i n_in_{i+2}$, where $U$ and $U_2$ are the nearest-neighbor and next-nearest-neighbor couplings, respectively. Next-nearest-neighbor couplings break integrability of the Fermi-Hubbard model, resulting in generic hydrodynamic behaviour.
Similar to the case of non-interacting fermion systems, we prepare the initial state as the ground state of the Fermi-Hubbard model in the presence of an external localized potential. By employing our PDE-learning framework, we discover that fermion dynamics agrees with the Navier-Stokes-like equation for the velocity in the form
\begin{eqnarray}\label{eq:navier_stokes}
v_t + vv_x+\kappa\rho\rho_x = \nu v_{xx}
\end{eqnarray}
combined with the continuity equation for the fermion density, where $\kappa>0$ is the coefficient accounting for the renormalization of the pressure term, and $\nu\geq0$ is an emergent effective viscosity originating from short-range interactions. In Fig.~\ref{fig:navier_stokes}(a,b), we compare solutions of Eq.~(\ref{eq:navier_stokes}) with TEBD simulations.  We find that the extracted pressure term $\kappa(U)$ in the  model with nearest-neighbor interactions $(U_2=0)$ depends universally on the coupling constant, showing agreement with the Tomonaga-Luttinger theory~\cite{tsvelik2007quantum} in the region $U/J\lesssim 1$, see Fig.~\ref{fig:navier_stokes}(c).  Simultaneously, the role of viscosity is more complicated and depends on the evolution time. For short times, the system is well-described in terms of an ideal Euler liquid, $\nu=0$, which means that the algorithm does not favor the viscous term for the values of the penalty constant $\lambda_0\gtrsim 10^{-3}$ for evolution times $J t\lesssim 10$. In contrast, for longer times, the effect of viscosity becomes comparable to other terms, and $\nu$ saturates to a universal value at late times (see Supplementary Material).
Notably, a similar Navier-Stokes-like term was recently discovered in interacting 1D fermionic systems within the generalized hydrodynamics framework~\cite{de2018hydrodynamic}.

\textit{Analysis of experimental data.}---Our methodology for reconstructing PDEs can be directly applied to study quantum hydrodynamic regimes in quench-type experiments with systems including ultracold atoms, trapped ions, and superconducting circuits.   
In an experimental setting, it is quite common that some  physical observables cannot be directly measured. For example, in  ultracold-atom experiments, the evolution of atomic density is obtained via optical absorption measurements, but the velocity field is not directly accessible.
This limitation can be overcome by leveraging the continuity equation that provides a relationship between density and velocity. We reconstruct  velocity field data $v(t,\, x)$ from density data $\rho(t,x)$ by integrating the continuity equation:
\begin{equation}
    v(t, x) = - \frac{1}{\rho (t,x)} \int_{-\infty}^x\, dx' \partial_t \rho(t, x'). 
\end{equation}

\begin{figure}[h!]
    \centering
    \includegraphics[scale=0.4]{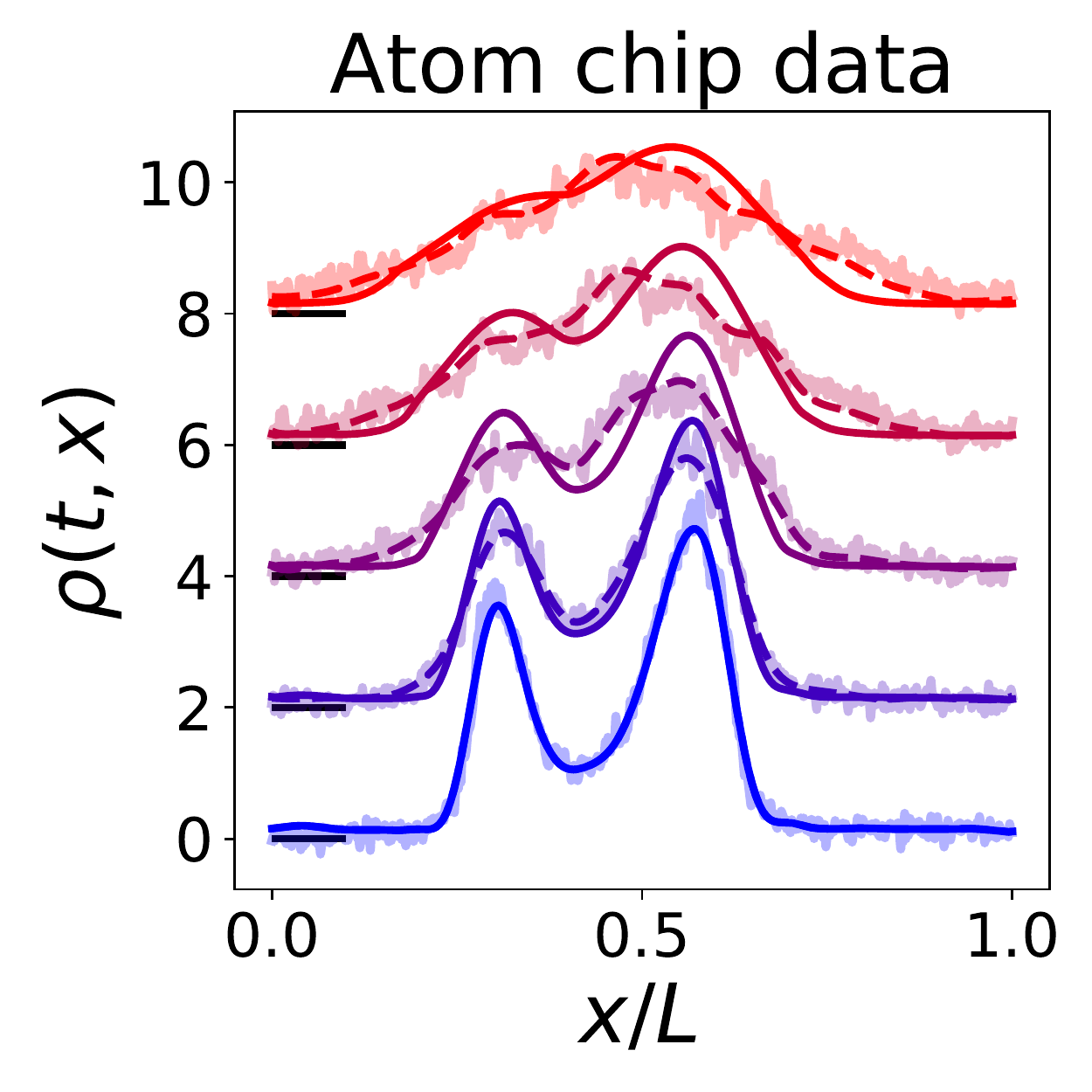}
    \caption{PDE-reconstruction from experimental data for the expansion of an interacting boson gas on an atom chip from a double-well potential~\cite{schemmer2019generalized}. Thick shadowed lines show experimental data for atom density (uniformly spaced within the time interval $[0, 45]$~ms), while solid lines correspond to the solution of the inferred PDE (\ref{eq:pde_atom_chip}) and the dashed lines show the post-processed (smoothed) data used for training our PDE-learning algorithm. We added vertical displacement to density profiles at each time step for visualization purposes (black horizontal dashes). The vertical scale for the density is in arbitrary units.}
    \label{fig:atom_chip}
\end{figure}

We test our PDE-learning algorithm using experimental data corresponding to the quench expansion of a 1D gas of interacting bosons on an atom chip~\cite{schemmer2019generalized}.
The atoms were confined in a double-well potential and, after releasing the potential, the evolution of gas density was measured. 
The system of bosons could be described by a Lieb-Liniger interacting gas with a contact repulsion~\cite{lieb1963exact}.
The original data has sufficient spatial resolution, but contains only a few time-points. We performed additional data preprocessing (noise filtering and interpolation) to obtain the necessary  resolution  to approximate derivatives with sufficient precision, see details in Supplementary Material.

The PDE discovery algorithm finds the following equation
\begin{eqnarray}
\label{eq:pde_atom_chip}
v_t + v v_x = - T(\log{\rho})_x + \nu\, v_{xx},
\end{eqnarray}
where we used $P$-symmetric candidate terms from Table~\ref{table:t1}.
The comparison between the experimental data and the solution of the recovered PDE (\ref{eq:pde_atom_chip}) are shown in Fig.~\ref{fig:atom_chip}.
The first term on the rhs of Eq.~(\ref{eq:pde_atom_chip}) has the form of a thermal pressure of an ideal Bolzmann gas, $P(\rho)=T \rho$ with temperature $T$. Hence, we conclude that the experimentally realized Lieb-Liniger gas behaves as an ideal thermal gas.
The viscosity-type contribution $\nu v_{xx}$ in Eq.~(\ref{eq:pde_atom_chip})  can be interpreted as an effective long-wavelength phenomenological term arising from short-range interactions between bosons.
The viscosity term regularizes the gradient catastrophe instability, commonly occurring in the  nonlinear Euler-type equations of quantum hydrodynamics  for ideal liquid flow~\cite{whitham2011linear}. Thus the viscosity term significantly extends the domain of validity of conventional hydrodynamics and results in a much better agreement with experimental data. 
Importantly, the reconstructed PDE (\ref{eq:pde_atom_chip}) is different from a widely-used conventional hydrodynamic model, based on the Gross-Pitaevskii equation, where the pressure term $P(\rho) \propto \rho^2$ originates from the $|\psi|^4$ interaction, rather than from thermal effects~\cite{Pitaevskii1999}. 
Based on the values of the extracted coefficient $T$, we  estimate the temperature of the boson gas: $T\approx 0.05 \mu K$.
We would like to note that  the predictions of conventional and generalized hydrodynamics coincide at short times, whereas, at long evolution times, the generalized hydrodynamics description should be more accurate.
In the future, our PDE-learning method could be extended to handle integro-differential equations, thus providing a direct connection to the generalized hydrodynamics  framework for quantum integrable systems~\cite{castro2016}.

Albeit the proposed PDE-learning method for the discovery of hydrodynamic equations in many-body quantum systems is very powerful, it has certain limitations.
The major problem is that closed-form PDEs  for the chosen set of physical observables might not exist at all. A well-known example is the Bogoliubov–Born–Green–Kirkwood–Yvon chain of kinetic equations for the $n$-point correlation functions. For generic interacting quantum systems, when starting with arbitrary initial states, this chain may continue indefinitely, involving higher-order correlation functions~\cite{stefanucci_leeuwen_2013}.
In this case, one can still approximately close the kinetic chain of equations, and the PDE-learning framework offers a new powerful tool to find such approximate closures.
Although a number of works investigated the closure of a hierarchical cumulant expansion in a similar context, they were based on rather ad-hoc assumptions on the properties of the quantum state~\cite{proukakis2001self,casteels2016truncated,colussi2020cumulant}, and as a result are not applicable in non-perturbative regimes. In contrast, the PDE-learning method proposed in the current paper could be used to approximately solve the closure problem for strongly-interacting systems, and we reserve this for future work.

\textit{Conclusions and outlook}.---In the present paper, we developed a new framework for symbolic regression-based PDE-learning  and applied it to a variety of non-equilibrium quantum problems. Our algorithm is able to find analytical forms of dynamical PDEs directly from raw data by discovering the long-wavelength limit of exact or approximate semiclassical equations, thus circumventing their analytical derivation.
First, we benchmarked our method on problems where exact evolution PDEs were known. Second, we discovered new PDEs that we were then able to derive analytically. Third, we discovered new PDEs whose analytical derivation is still an open problem. 

While we demonstrated our method in 1D, it applies in any dimension. Our work thus opens up new avenues for machine-assisted discovery of hydrodynamic-type equations in a wide range of many-body quantum systems, including ultracold atoms~\cite{gross2017quantum,bernien2017probing}, trapped ions~\cite{monroe2021programmable}, superconducting circuits~\cite{aleiner2020accurately},  critical phenomena in solids~\cite{sachdev2007quantum}, and hydrodynamics in graphene~\cite{levitov2016electron}.
In particular, our method can be used to find approximate solutions to the closure problem for a hierarchy of quantum kinetic equations,
discover conservation laws in many-body quantum systems, and uncover PDEs in open systems with a complex bath environment.
Closed-form PDEs for physical observables naturally arise in a semiclassical limit, such as the large spin limit of the XXZ model. Thus the PDE-learning framework can be fruitful for studying the quantum-classical correspondence.
The presented PDE-learning approach is especially powerful in the case of strongly interacting quantum systems, where theoretical tools based on perturbative calculations are no longer applicable.
Symbolic PDE-learning approaches can serve as a guide for theorists deriving effective long-wavelength descriptions.
Furthermore, experimentalists could utilize this approach to find the best-matching hydrodynamic model to describe the observed dynamics in a many-body quantum experiment or enhance parameter estimation in case a theoretical PDE for the dynamics is already known. Finally, it would be interesting to prove rigorous classical and quantum complexity results on finding hydrodynamic equations for any given Hamiltonian with and without access to data.

\textit{Acknowledgements}.---We are grateful to M. Ljubotina for providing  high-precision tDMRG data for the evolution for a thermal domain wall initial state in the XXZ model.
We thank J.\ Dubail and M.\ Schemmer for sharing experimental data for the Bose gas hydrodynamics on an atom chip. We thank Samuel Rudy and Brayden Ware for useful discussions and comments. Y.K., O.S., P.B., and A.V.G. acknowledge funding by the DoE ASCR Accelerated Research in Quantum Computing program (award No.~DE-SC0020312), U.S.~Department of Energy Award No.~DE-SC0019449, NSF PFCQC program, DoE ASCR Quantum Testbed Pathfinder program (award No.~DE-SC0019040), DoE Quantum Systems Accelerator, AFOSR, AFOSR MURI, ARO MURI, DARPA SAVaNT ADVENT.
A.S., M.V.R. and M.H. were sponsored by
ARO W911NF-15-1-0397,W911NF2010232, AFOSR FA9550-19-1-0399, QSA-DOE, Simons foundation. A.S. is supported by a Chicago Prize Postdoctoral Fellowship in Theoretical Quantum Science.

\let\oldaddcontentsline\addcontentsline
\renewcommand{\addcontentsline}[3]{}

\bibliography{references}
\bibliographystyle{Science}

\let\addcontentsline\oldaddcontentsline
\newpage


\renewcommand{\bibnumfmt}[1]{[S#1]}
 \setcounter{equation}{0}
    \setcounter{figure}{0}
    \setcounter{table}{0}
    
\renewcommand{\theequation}{S\arabic{equation}}
\renewcommand{\thefigure}{S\arabic{figure}}

\renewcommand{\thetable}{S\Roman{table}}

\widetext
\newpage 
\addtocontents{toc}{\protect\setcounter{tocdepth}{-1}}

\begin{center}
\textbf{\large Supplementary Material for ``Discovering hydrodynamic equations of many-body quantum systems''}
\end{center}
\setcounter{equation}{0}
\setcounter{figure}{0}
\setcounter{table}{0}
\makeatletter
\renewcommand{\theequation}{S\arabic{equation}}
\renewcommand{\thefigure}{S\arabic{figure}}
\renewcommand{\bibnumfmt}[1]{[S#1]}

\tableofcontents

\section{Details of sparse regression algorithms}\label{sec:sparse_regr}
In this Section of the Supplementary Material, we discuss algorithms for the optimization of the non-convex  objective function for the sparse regression problem:
\begin{equation}
    \mathcal{L} = ||\mathbf{U}_t-\Theta(\mathbf{U}, \mathbf{U}_x, \ldots)\cdot\xi||_2+\lambda_0 ||\xi||_0.
    \label{eq:object_func_l0}
\end{equation}
In Sections~\ref{sec:brute_force} and \ref{sec:cem},  we provide additional details, including the pseudocode, of the brute-force algorithm and the cross-entropy algorithm, respectively. In Sections~\ref{sec:stridge} and~\ref{sec:lasso}, we briefly discuss two other popular sparse-regression methods: Sequential Thresholding and Ridge regression (STRidge) and least absolute shrinkage and selection operator (LASSO) regression. 

\subsection{BruteForce algorithm}\label{sec:brute_force}
The brute-force algorithm (BruteForce) for PDE-learning consists of two stages (see Algorithm~\ref{alg:brute_force} below): (i) looping over all possible combinations of terms from the dictionary, (ii) for the selected terms, reconstruct coefficients via linear regression and evaluate the objective function (\ref{eq:object_func_l0}). Finally, the algorithm returns coefficients that minimize the objective function $\mathcal{L}$. Although  this algorithm has exponential complexity when  increasing the number of candidate terms, it could still be used in practice in a lot of cases. The largest problem instance we were able to solve with BruteForce contained $M=20$ candidate terms (see Table~\ref{table:summary}, Problems \#9, \#10).
\begin{algorithm}[H]\label{alg:brute_force}
\caption{BruteForce algorithm for sparse selection of PDE terms}
\begin{algorithmic}[1]
\Function{BruteForceL0}{$\mathbf{U}_t$, $D_x^n\mathbf{U}$, $\lambda_0$}
    \For{nonzero indexes $\in$ all $2^M$ combinations of $M$ terms in the dictionary $\Theta$}\Comment{Iterate over term combinations}
        \State $\tilde{\Theta} = \Theta[:, \textrm{nonzero indexes}]$\Comment{Select columns in the dictionary matrix}
        \State $\xi = (\tilde\Theta^\dag \tilde\Theta)^{-1} \tilde\Theta^\dag \mathbf{U}_t$\Comment{Perform linear regression to extract nonzero coefficients}
        \State $\mathcal{L} \gets ||\mathbf{U}_t-\tilde\Theta(U, D_x^n U)\cdot \xi||_2 + \lambda_0 ||\xi||_0$ \Comment{Evaluate objective function}
        \If {$\mathcal{L}<\mathcal{L}_{best}$} \Comment{Objective function improved}
        \State $\mathcal{L}_{best} \gets \mathcal{L}$
        \State $\xi_{best} \gets \xi$
        \EndIf
    \EndFor
\State \textbf{return} $\xi_{best}$
\EndFunction
\end{algorithmic}
\end{algorithm}

\subsection{CrossEntropy algorithm}\label{sec:cem}
As a scalable alternative to the BruteForce method, we propose a sampling-based algorithm which we call CrossEntropy, see Algorithm~\ref{alg:cem}.
CrossEntropy is conceptually similar to BruteForce, but, instead of performing an exhaustive search over $2^M$ combinations of terms, it relies on the Cross-Entropy method (CEM)~\cite{rubinstein1999cross, de2005tutorial} as a subroutine for combinatorial optimization (term selection) of a ``black-box'' function $\mathcal{L}$. CEM is a heuristic method that shows reliable practical performance for hard optimization problems (e.g.~the travelling salesman problem), is computationally efficient and is relatively simple in implementation. The CEM algorithm is analogous to a derivative-free evolutionary algorithm with a Monte-Carlo-like update rule.
The key steps in the algorithm are
\begin{itemize}
    \item Initialize a weights vector $\vec{W}=(W_1,\ldots, W_M)$ with zero values. The weights define the probability of a term being present  via the Bolzmann distribution (SoftMax policy).
    \item Create a population of weights vectors, independently update vector elements in the population by adding i.i.d. Gaussian fluctuations.
    \item In order to estimate the value of the objective function $\mathcal{L}$ in each population, we perform a series of rollouts for a given vector of SoftMax weights. 
    In each rollout, the indexes of nonzero terms are sampled using the SoftMax policy. The coefficients $\xi$ of non-zero terms are recovered via linear regression and then used for the evaluation of the objective function $\mathcal{L}$. 
    \item Select top performing (``elite'') candidates in the population (e.g.~top 1\%) according to the objective function $\mathcal{L}$.
    \item Update the current weights vector by taking element-wise mean of the elite weights array.
\end{itemize}
The largest problem we were able to solve with the CrossEntropy algorithm contained $M=45$ terms (see Table~\ref{table:summary}, Problems \#11, \#12).
Typical values of hyperparameters we used in our PDE-learning experiments are: $\texttt{number of rollouts} = 100$, $\texttt{batch size} = 100$, $\texttt{elite fraction} = 1\%$. 
\begin{algorithm}[H]
\caption{Sparse selection algorithm based on the Cross Entropy method for combinatorial optimization}
\label{alg:cem}
\begin{algorithmic}[1]
\Function{SoftMaxPolicy}{$\vec W$}
    \For{i \textbf{in} $1,\ldots, M$}
    \State $p_i\gets \exp{(W_{i})}/\sum_{j=1}^M \exp{(W_j)}$ \Comment{Get term probabilities from weights vector $\vec{W}$}
    \State $indx[i]\sim Bernoulli(p_i)$  \Comment{Sample indices of nonzero terms from Bernoulli distribution}
    \EndFor
    \State \textbf{return} $indx$ \Comment{Return vector of indexes of nonzero terms}
\EndFunction
\Function{EstimateBestLossAndCoefs}{$\vec{W},\mathbf{U}_t, D_x^n\mathbf{U}, \, num\; rollouts, \lambda_0$}
\For{rollout \textbf{in} $1,\ldots, num\; rollouts$}\hspace{10pt}\Comment{Perform sampling of indexes of non-zero terms and estimate min loss for a fixed vector of SoftMax weights  }
\State $indx \gets \textrm{SoftMaxPolicy}(\vec{W})$
\State $\tilde{\Theta} = \Theta[:,  indx]$\Comment{Select columns in the dictionary matrix}
\State $\xi[indxs] = (\tilde\Theta^\dag \tilde\Theta)^{-1} \tilde\Theta^\dag \mathbf{U}_t$\Comment{Perform linear regression to extract nonzero coefficients}
\State $\xi[\sim indx] \gets 0$\Comment{Assign zero values to the remaining coefficients}
\State $\mathcal{L} \gets ||\mathbf{U}_t-\tilde\Theta(U)\cdot\xi||_2 + \lambda_0 ||\xi||_0$ \Comment{Compute current loss function}
\If {$\mathcal{L}<\mathcal{L}_{best}$} \Comment{Check if the objective function has improved}
\State $\mathcal{L}_{best} \gets \mathcal{L}$
\State $\xi_{best} \gets \xi$
\EndIf
\EndFor
\State \textbf{return} $\mathcal{L}_{best}$, $\xi_{best}$
\EndFunction

\Function{TrainCEM}{$elite\, frac$, $batch\, size$, $num\, rollouts$}
\State $\vec{W}_{popul}\gets [batch\, size \times M]$ \Comment{Initialize array of weights for the CEM population}
\For{iter \textbf{in} $1,\ldots, niter$}
\For{b \textbf{in} $1, \ldots, batch\, size$}
\State $\vec{dW} \sim  \mathcal{N}(0, \sigma_{W})$ \Comment{Update weights in each batch by adding vector of i.i.d.~Gaussian variables}
\State $\vec{W}_{popul}[b] = \vec{W}+\vec{dW}$
\State $\mathcal{L}_{popul}[b],\, \xi \gets  \textrm{EstimateBestLossAndCoefs}(\vec{W}_{popul}[b],\, num\, rollouts)$
\EndFor
\State $indx_{elite} \gets  argsort(\mathcal{L}_{popul})[:elite\; frac \times batch\; size]$ \Comment{Select elite weights, e.g.~$1\%$ of top performing weights samples from the population}
\State $\vec{W}_{elite}\gets \vec{W}_{popul}[indx_{elite}]$
\State $\vec{W} \gets mean(\vec{W}_{elite})$
\State $\sigma_{W} \gets std(\vec{W}_{elite})$
\EndFor
\State \textbf{return} $\vec{W}$
\EndFunction
\Function{CrossEntropyL0}{$\mathbf{U}_t$, $D_x^n\mathbf{U}$, $\lambda_0$, $elite\, frac=0.01$, $num\, rollouts$, $batch\, size$, $niter$}
    \State $\vec{W}\gets \textrm{TrainCEM}(elite\; frac, batch\; size, num\; rollouts)$
    \State $\mathcal{L}_{best},\, \xi_{best} \gets \textrm{EstimateBestLossAndCoefs}(\vec{W},\mathbf{U}_t, D_x^n\mathbf{U},\, num\; rollouts, \lambda_0)$
\State \textbf{return} $\xi_{best}$
\EndFunction
\end{algorithmic}
\end{algorithm}

\subsection{Sequential Thresholding and Ridge regression (STRidge)}\label{sec:stridge}
STRidge is a heuristic algorithm for the least-squares sparse regression problem in the presence of  $L_0$ and $L_2$ penalty terms and is based on an annealing-like schedule for thresholding of non-zero regression coefficients. See description and pseudocode in Ref.~\onlinecite{rudy2017data}.\\

\subsection{LASSO regression}\label{sec:lasso}
A commonly used approach to promote sparsity is to consider convex relaxation of the original problem (\ref{eq:object_func_l0})  by using $L_1$ regularization  instead of $L_0$. This method is  known as LASSO regression:
$\mathcal{L} = ||\mathbf{U}_t-\Theta(\mathbf{U}, \mathbf{U}_x, \ldots)\cdot\xi||_2^2+\lambda_1 ||\xi||_1$.
However, LASSO tends to have difficulty finding a sparse basis when the data matrix $\Theta$ has high correlations between columns (which could be the case for nonlinear terms in $\Theta$), which results in a poor PDE reconstruction quality~\cite{rudy2017data}.

\subsection{Summary of PDE-reconstruction results for various sparse selection algorithms}

In this subsection, we present a short  summary (see table~\ref{table:summary})  of the PDE-learning problems considered in the main text and the performance of three algorithms for term selection: BruteForce, STRidge, and CrossEntropy.  
\bgroup
\def\arraystretch{1.5}
\begin{center}
\begin{table}[H]
\caption{Performance of sparse selection algorithms on the problems considered in the main text. Successful reconstruction of an entire sequence of PDEs is marked as  $(\checkmark)$, failure is marked as $(\times)$, and partial success---when only some PDEs depending on the value of $\lambda_0$ were correctly identified---is marked as $(\pm)$. In the problem list column, ``fermion hydro.'' stands for ``fermion hydrodynamics'', while ``extended lib.'' refers to an extended library of candidate terms.}\label{table:summary}
\begin{tabularx}{\textwidth}{|c|c|c|c|c|X|}
\hline
&\textbf{Problem} & BruteForce & STRidge & CrossEntropy & Candidate Terms \\
\hline
1&Single magnon, $B(x)=0$ & $\checkmark$ & $\checkmark$ & $\checkmark$ & \noindent\parbox[c]{\hsize}{
$1$, $\partial_x^n u$, $u \partial_x^n u$,\\  $n\in[1,..,4]$ }\\
 \hline
2&Single magnon, $B(x)=B_0(x-x_0)^2$ & $\checkmark$ & $\checkmark$ & $\checkmark$& \noindent\parbox[c]{\hsize}{
$1$,  $\partial_x^n u$,  $(x-x_0)^n u$,
$n\in[1,..,4]$ }\\
 \hline
3& \begin{tabular}[c]{@{}c@{}}
Domain wall, XXZ $(\Delta=0)$,\\
zero temperature
\end{tabular}
& $\checkmark$ & $\checkmark$ & $\checkmark$ &
\noindent\parbox[c]{\hsize}{
$\partial_x^n u$, $u^m \partial_x u$  $n\in[1,..,4]$, $m\in [1,..,5]$ }\\
 \hline 
4&\begin{tabular}[c]{@{}c@{}} Domain wall, XXZ $(\Delta=0)$, \\
zero temperature (extended lib.) 
\end{tabular}
& $\checkmark$ & $\times$ & $\checkmark$ & \noindent\parbox[c]{\hsize}{
$\partial_x^n u$, $u^m \partial_x u$, $\sin{(2\pi u/P)} u_x$ \\ $n\in[1,..,4]$,
$m\in [1,..,5]$, $P\in [1,..,10]$ }\\
 \hline 
5&\begin{tabular}[c]{@{}c@{}}
Domain wall, XXZ $(\Delta/J=0.5)$,\\ zero temperature
\end{tabular}& $\checkmark$ & $\pm$ & $\checkmark$ & 
\noindent\parbox[c]{\hsize}{
$\partial_x^n u$, $u^m \partial_x$, $n\in[1,..,4]$, $m\in [1,..,5]$ }\\
 \hline
6&
\begin{tabular}[c]{@{}c@{}}Domain wall, XXZ $(\Delta/J=0.5)$,\\ zero temperature (extended lib.)
\end{tabular}
 & $\checkmark$ & $\pm$ & $\checkmark$ & \noindent\parbox[c]{\hsize}{
$\partial_x^n u$, $u^m \partial_x u$, $\sin{(2\pi u/P)} u_x$ \\ $n\in[1,..,4]$, $m\in [1,..,5]$, $P\in [1,..,10]$ }\\
 \hline 
7& 
\begin{tabular}[c]{@{}c@{}}
Domain wall, XXZ $(\Delta/J=1)$,\\ high-temperature state
\end{tabular}
& $\checkmark$ & $\checkmark$ & $\checkmark$ & 
\noindent\parbox[c]{\hsize}{
$u_t=-\partial_x \mathcal{J}(u)$,\,\\ $\mathcal{J}(u)$: $u^n$, $u^n \partial_x u$, $u^n \partial_x^2 u$, $n\in[1,..,5]$  }\\
 \hline 
8&
\begin{tabular}[c]{@{}c@{}}
Domain wall, XXZ $(\Delta/J=2)$, \\ high-temperature state
\end{tabular}
 & $\checkmark$ & $\checkmark$ & $\checkmark$ & \noindent\parbox[c]{\hsize}{
$u_t=-\partial_x \mathcal{J}(u)$,\,\\ $\mathcal{J}(u)$: $u^n$, $u^n \partial_x u$, $u^n \partial_x^2 u$, $n\in[1,..,5]$}\\
 \hline  
9&\begin{tabular}[c]{@{}c@{}}
Fermion hydro., \\
$U=0$ ($J_1=0.5$, $J_2=0$) 
\end{tabular}& $\checkmark$ & $\times$ & $\checkmark$ & 
\noindent\parbox[c]{\hsize}{
Table~\ref{table:t2} $(P, T)=(-,+)$ }\\
 \hline  
10&\begin{tabular}[c]{@{}c@{}}
Fermion hydro., \\
$U=0$ ($J_1=0.5$, $J_2=-0.125$) 
\end{tabular}& $\checkmark$ & $\times$ & $\checkmark$ & 
\noindent\parbox[c]{\hsize}{
Table~\ref{table:t2} $(P, T)=(-,+)$}\\
\hline
11&\begin{tabular}[c]{@{}c@{}}Fermion hydro. (extended lib.), \\
$U=0$ ($J_1=0.5$, $J_2=0$)
\end{tabular}& not tractable & $\times$ & $\checkmark$ & \noindent\parbox[c]{\hsize}{ all terms from Table~\ref{table:t2}}\\
 \hline  
12&\begin{tabular}[c]{@{}c@{}}
Fermion hydro. (extended lib.),\\ $U=0$ ($J_1=0.5$, $J_2=-0.125$)
\end{tabular}& not tractable & $\times$ & $\checkmark$ & \noindent\parbox[c]{\hsize}{ all terms from Table~\ref{table:t2} }\\
 \hline 
13&\begin{tabular}[c]{@{}c@{}}
Fermion hydro., \\
$U/J=4$ ($J_1=0.5$, $J_2=0$) 
\end{tabular}& not tractable & $\times$ & $\checkmark$ &  \noindent\parbox[c]{\hsize}{ Table~\ref{table:t2} $(P, T)=(-,\cdot\,)$ }\\
 \hline 
\end{tabularx}
\end{table}
\end{center}
\egroup


\section{PDE-learning of quench dynamics in the XXZ model: analytical derivations and additional examples}
In this Section, we derive closed-form PDEs presented in the main text describing long-wavelength dynamics of excitations in the low-energy sector of the XXZ model. We provide additional details of PDE-learning methodology and discuss numerical schemes to calculate spatiotemporal derivatives from the data.

We consider the following benchmarking cases: (i) single-magnon dynamics in the nearest-neighbor XXZ model with/without an external magnetic field [Section~\ref{sec:single_magnon_xxz}], (ii)  non-local PDEs for single-magnon dynamics in the long-range XXZ model [Section~\ref{sec:long_range}], (iii) evolution of a domain-wall initial state corresponding to a zero temperature product state and to a high-temperature Gibbs state [Section~\ref{sec:domain_wall}].
Cases (i) and (iii) were considered in the main text, whereas, for case (ii), we introduce a new model---the long-range interacting XXZ spin chain---and show how our PDE-learning method can be extended to systems with power-law-decaying interactions.

\subsection{Magnon dynamics in the nearest-neighbor XXZ model}\label{sec:single_magnon_xxz}

In this subsection, we consider quench dynamics of the XXZ spin chain in the single-magnon excitation sector and provide an analytical derivation of Eq.~(6) from the main text.

The Hamiltonian of the XXZ model reads
\begin{equation}
\label{eq:xxz_ham}
    H =  \sum_{i} \left[J\left(S^x_i S^x_{i+1} + S^y_i S^y_{i+1}\right) + \Delta  S^z_i S^z_{i+1} +  B_i S_i^z\right],
\end{equation}
where $S^\mu_i=\sigma^\mu_i/2$ are spin operators, $\sigma^\mu_i$ are standard Pauli operators associated with the $i$th spin polarization, and coefficients $J$, $\Delta$, and $B$ are real  parameters. In our simulations we set periodic boundary conditions in the Hamiltonian (\ref{eq:xxz_ham}).

The initial state $|\psi_0\rangle$ is prepared as a wave packet in the single-magnon excitation sector over the ferromagnetic product state:
\begin{equation}\label{eq:1p_psi_0}
    |\psi_0\rangle  = \sum_{n}  f(n)\, \mathcal U(\theta_n , \phi_n) \ket{\downarrow}_{n} \prod_{j \neq n} \ket{\downarrow}_{j} = \frac 1{\sqrt{\pi\sigma^2}} \sum_{n} e^{-(n - x_0)^2/\sigma^2 + ik_0 n} | \theta_n, \phi_n \rangle_n \prod_{j \neq n} \ket{\downarrow}_{j} ,
\end{equation} 
where $f(n)$ is Gaussian wave-packet envelope function corresponding to momentum $k_0$ and centered around coordinate $x_0$. Here $\mathcal U(\theta_n, \phi_n)$ is an $SU(2)$ unitary rotation operator acting as
\begin{equation}
| \theta_n, \phi_n \rangle  =  \mathcal U(\theta_n , \phi_n)\ket{\downarrow}_n = \cos{(\theta_n/2)}\ket{\uparrow}_n + \sin{(\theta_n/2)} e^{i \phi_n}\ket{\downarrow}_n.    
\label{eq:wave_packet_psi}
\end{equation}

We introduce the following complex-valued function $u(t, x)$. At the sites of the spin chain, $x_i \equiv i a$, where $a$ is the lattice spacing, we set  the value of the function to
\begin{equation}
u(t, x_i) = \langle S^+_i(t) \rangle = \frac{1}{2}[\langle \sigma^x_i(t)\rangle + i\langle \sigma^y_i(t)\rangle],
\end{equation}
where $S^+_i=S^x_i+iS^y_i = \frac{1}{2} (\sigma^x_i + i \sigma^y_i)$ is the spin raising operator, $O(t) = \exp{(iHt)}O\exp{(-iHt)}$ is the time-dependent operator in the Heisenberg picture, and $\langle O \rangle \equiv \langle\psi_0|O|\psi_0\rangle$ is the expectation value taken in the initial state.

To derive the equations of motion, we use the canonical commutation relations for the spin operators,
\begin{equation}
    [S^+_i(t), S^-_j(t)] = 2\delta_{ij}S^z_i(t), \quad  [S^+_i(t), S^z_j(t)] = -\delta_{ij}S^+_i(t).
\end{equation}
Calculating the time derivative of the observable of interest in the Heisenberg representation, we obtain
\begin{equation}
\label{eq:s_plus_t}
    i\partial_t \langle S^+_i(t)\rangle = \langle [ S^+_i(t), H] \rangle =  J \Bigl(\langle S^+_{i-1}(t)S^z_i(t)\rangle+\langle S^+_{i+1}(t)S^z_i(t) \rangle\Bigl)- \Delta \Bigl(\langle S^+_i(t)S^z_{i+1}(t)\rangle+\langle S^+_i(t)S^z_{i-1}(t) \rangle\Bigl) - B_i \langle S_i^+(t) \rangle.
\end{equation}
The right-hand side of Eq.~(\ref{eq:s_plus_t}) depends on two-point same-time correlation functions of the type $\langle S^+_i(t) S^z_j(t) \rangle$. Therefore, for a generic initial state, the time derivative could not be expressed via $u(t,x_i)$ only.
However, in the case of initial states in the form of Eq.~(\ref{eq:1p_psi_0}), i.e.~a superposition of a zero-magnon state and a one-magnon state---$|\psi_0\rangle = |\psi_{0,m=0}\rangle + |\psi_{0,m=1}\rangle$ where $|\psi_{m=0}\rangle = \ket{\downarrow}^{\otimes L}$---the equation can be simplified.
Projecting the r.h.s.~terms in Eq.~(\ref{eq:s_plus_t}) onto the span of $|\psi_{0,m=0}\rangle$ and $|\psi_{0,m=1}\rangle)$, we obtain
\begin{eqnarray}
   && \langle \psi_{0,m=1}| S^+_i(t)| \psi_{0,m=0}\rangle = u(t,x_i),\\
  && \langle S^+_i(t) S^z_j(t) \rangle  = \langle \psi_{0,m=1} |S^+_i(t) S^z_j(t)|\psi_{0,m=0} \rangle = \frac 12 u(t, x_i).
\end{eqnarray}
As a result, we arrive at the following closed equation:
\begin{equation}
\label{eq:fin_diff_xxz_hz(x)}
    i \partial_t u(t, x_i) = \frac{J}{2} \Bigl(u(t, x_{i+1})+u(t, x_{i-1})\Bigl) -\Delta u(t, x_i) - B_i u(t, x_i).
\end{equation}
Due to the linearity of dynamical equations, there is no dependence on the choice of the envelope function $f(n)$ for the initial state [see Eq.~(\ref{eq:1p_psi_0})].

We would like to note that the simple closed form of Eq.~(\ref{eq:fin_diff_xxz_hz(x)}) is due to the specific choice of observable $u = \langle S^+_i(t)\rangle$.
Another natural choice of initial condition and observable is 
\begin{equation}\label{eq:psi0_altern}
|\psi_0 \rangle = \sum_{n} f(n) | \uparrow \rangle_{n} \prod_{j \neq n}|\downarrow \rangle_{j}, \qquad \tilde u(t, i) = \langle S^z_i(t) \rangle.
\end{equation}
The initial state in Eq.~(\ref{eq:psi0_altern}) has a conventional form of a single-magnon excitation, whereas Eq.~(\ref{eq:1p_psi_0}) corresponds to a superposition of a single-magnon and a ferromagnetic ground state.
However, in the former case, the onsite $z$-magnetization alone does not contain enough information to predict its evolution at later times, hence, for this choice of observable $\tilde u(t,x)$,  a simple self-contained PDE does not exist.

Now we consider the long-wavelength limit of Eq.~(\ref{eq:fin_diff_xxz_hz(x)}). We assume that $u(t,x)$ is a smooth interpolation of integer-valued points. We also consider $B(x)$ as a smooth interpolation for the local magnetic field such that $B(x_i) = B_i$.
The continuous form of Eq. (\ref{eq:fin_diff_xxz_hz(x)}) reads 
\begin{equation}\label{eq:cos_q_exact}
i \partial_t u = J\cos{(i\partial_x)}u-\Delta u - B(x) u.
\end{equation}
Next, we assume that the magnetic field and the observables of the spin system change slowly in space, with the smallest-scale variations characterized by a length-scale $\lambda\gg 1$, implying that $|\partial_x^n u|,\, |\partial^n_x B| \leq \mathcal O(\lambda^{-n})$. Then the dynamics of the complex function $u(t,x)$ can be approximated as
\begin{eqnarray}
\label{eq:pde_xxz_hz(x)}
i\partial_t u =\frac{J}{2} \partial^2_x u + (J-\Delta) u - B(x) u + \mathcal{O}(\lambda^{-4}).
\end{eqnarray}
Notably, Eq.~(\ref{eq:pde_xxz_hz(x)}) has the form of the single-particle Schr\"odinger equation in an external potential generated by the longitudinal magnetic field $B(x)$. Although formally the derivation of Eq.~(\ref{eq:pde_xxz_hz(x)}) does not require the magnetic field profile $B(x)$ to have small spatial gradients, such a condition could be important to guarantee smoothness of the solution $u(t,x)$ during the evolution.

\begin{figure}[h!]
    \includegraphics[scale=0.3]{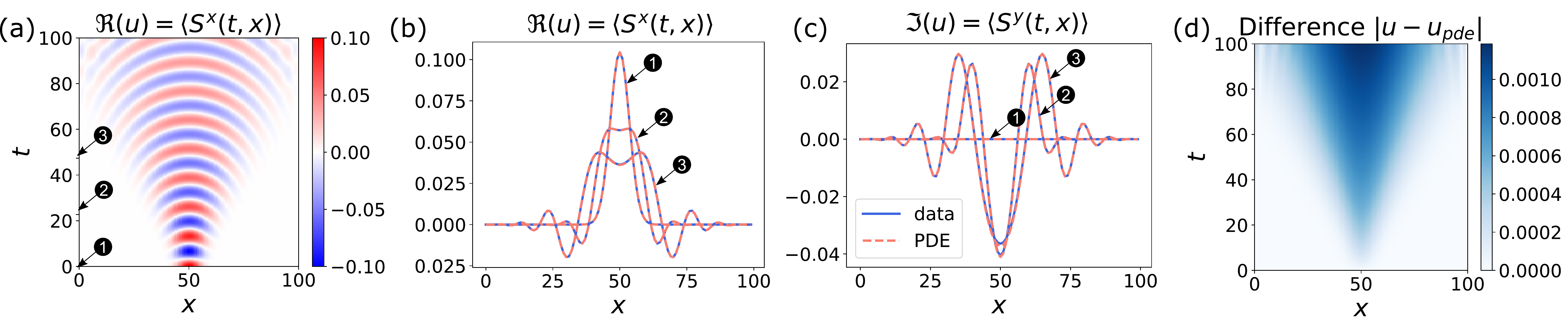}
    \caption{Propagation of a wave packet in the XXZ spin chain, with $\Delta/J=0.5$  (exact diagonalization). The initial state $|\psi_0\rangle$ corresponds to a superposition of a single-magnon excitation and a ferromagnetic state: $|\psi_0\rangle   = A \sum_{n} e^{-(n - x_0)^2/\sigma^2 }| + \rangle_n \prod_{j \neq n} \ket{\downarrow}_{j}$,  where $|+\rangle_n = \frac{1}{\sqrt{2}}(\ket{\uparrow}_n + \ket{\downarrow}_n)$. Periodic boundary conditions are imposed. The parameters are: total number of lattice sites $L=100$, $J=-1$, $\sigma = 5$, and number of time steps $N_t=2000$. Panels (a, b) correspond to $\Re[u]=\langle S^x(t,x)\rangle$, while panel (c) corresponds to $\Im[u]=\langle S^y(t,x) \rangle$.  Solid lines display the exact evolution, while dashed lines show the solution of the PDE (\ref{eq:pde_xx_num}). The evolution times in (b, c) are labeled in panel (a). (d) Difference between the exact solution and the solution of the inferred PDE.}
    \label{fig:1p_wave_packet}
\end{figure}

\begin{figure}[h!]
    \includegraphics[scale=0.33]{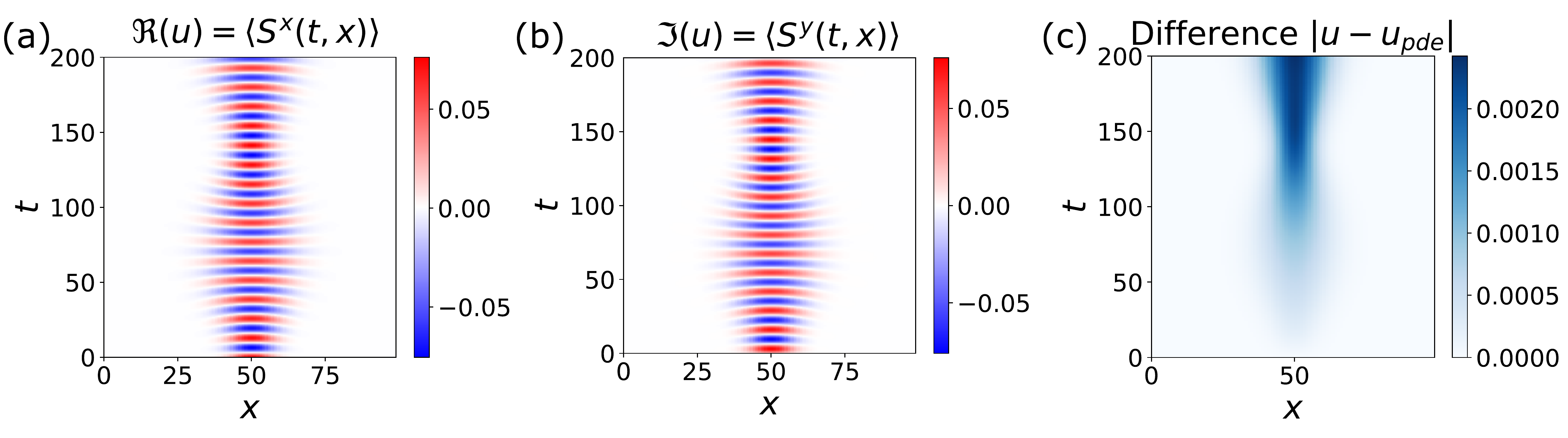}
    \caption{Propagation of a wave packet in the XXZ spin chain,  with $\Delta/J=0.5$, in the presence of a parabolic longitudinal magnetic field $B_i = B_0\left(i - x_0\right)^2$  (exact diagonalization). The initial state is the same as in Fig.~\ref{fig:1p_wave_packet}, periodic boundary conditions are imposed, $B_0 = 5\times 10^{-4}$, and the number of time steps is  $N_t=4000$. Panels (a, b) correspond to the input data for our algorithm: $\Re[u]=\langle S^x(t,x)\rangle$ and $\Im[u]=\langle S^y(t,x)\rangle$. (c) Difference between the exact solution and the solution of the recovered PDE (\ref{eq:extract_pde_xxz_hz(x)}).}
    \label{fig:1p_wave_packet_hz}
\end{figure}
First, we consider dynamics of a single magnon in the XXZ model in the case of zero magnetic field, $B_i=0$.
For instance, if we choose the following library of candidate terms, 
\begin{equation}
    u_t = F(1, u, u_x, u_{xx}, u_{xxx}, u_{xxxx}, u^2, u u_x, u u_{xx}, u u_{xxx}, u u_{xxxx}),
    \label{eq:1p_F_dict}
\end{equation}
and, using the data shown in Fig.~\ref{fig:1p_wave_packet},
we obtain the following PDE  with the BruteForce algorithm for the case $\Delta/J=0.5$ and the penalty constant $\lambda_0=10^{-3}$:
\begin{equation}
    i u_t + 0.4999 u_{xx} + 0.4997 u = 0.
    \label{eq:pde_xx_num}
\end{equation}
The temporal and spatial derivatives in Eq.~(\ref{eq:pde_xx_num}) were computed from data using the second-order finite-difference scheme, see details in Sec.~\ref{sec:num_schemes}.
We included nonlinear terms up to the second order in $u$  to the candidate terms dictionary (\ref{eq:1p_F_dict}) in order  to perform  a consistency check of the sparse selection algorithm.

\begin{figure}
    \centering
    \includegraphics[scale=0.5]{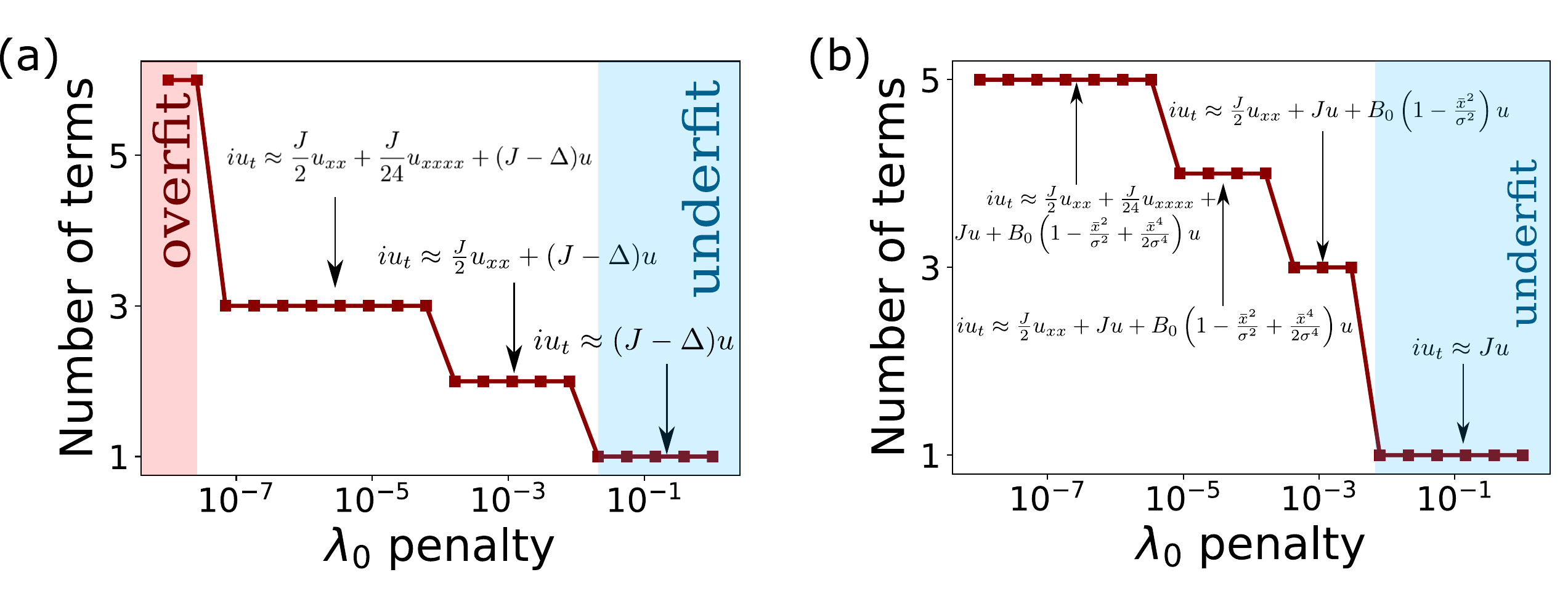}
    \caption{Number of  terms on the rhs of the reconstructed PDE $u_t=F(\cdot)$, Eq.~(\ref{eq:1p_F_dict}), found with BruteForce algorithm vs the $L_0$ penalty constant $\lambda_0$. (a) Quench dynamics in the XXZ model in the single-magnon sector, with $\Delta/J=0.5$, for (a) $B(x)=0$, see Fig.~\ref{fig:1p_wave_packet}, and (b)  inhomogeneous magnetic field $B(x)=B_0 e^{-(x-x_0)^2/\sigma^2}$.
    The ``underfit'' region corresponds to the range of $\lambda_0$, where the number of terms in the inferred PDE is underestimated, whereas in the ``overfit'' region our algorithm finds spurious terms that are not present in the true PDE.
    Spatial derivatives were calculated  using the spectral method [see Section \ref{sec:num_schemes}].}
    \label{fig:num_terms_1magnon}
\end{figure}

Now we consider single-magnon dynamics in the presence of an external longitudinal magnetic field. We impose a parabolic magnetic field   $B_i= B_0 (x_i-x_0)^2$, where $x_0=L/2$.
Post-quench dynamics is confined by the trapping potential, and the evolution of the observable $u(t,x)$ is shown in Fig.~\ref{fig:1p_wave_packet_hz}. 
Recovering the  PDE from the following ansatz,
\begin{equation}
\label{eq:1p_F_dict_hz}
u_t = F(u, u_x, u_{xx}, u_{xxx}, u_{xxxx}, \bar x, \bar x^2, \bar x^3, \bar x^4, \bar x u, \bar x^2 u, \bar x^3 u, \bar x^4 u), \qquad  \bar x = x-x_0, 
\end{equation}
using data corresponding to Fig.~\ref{fig:1p_wave_packet_hz} ($\Delta/J=0.5$, $B_0 = 5\cdot 10^{-4}$) with the BruteForce, CrossEntropy, and STRidge algorithms, we obtain
\begin{equation}
\label{eq:extract_pde_xxz_hz(x)}
i u_t = - 0.4998 u_{xx} - 0.4999 u + 4.998\cdot 10^{-4}\left(x-x_0\right)^2 u.
\end{equation}
The extracted PDE in Eq.~(\ref{eq:extract_pde_xxz_hz(x)}) matches with high precision the expected Eq.~(\ref{eq:pde_xxz_hz(x)}).
In Eq.~(\ref{eq:extract_pde_xxz_hz(x)}), we again used the finite-difference approximation of the derivatives, see Sec.~\ref{sec:num_schemes}. 
The frontiers of reconstructed PDEs as a function of penalty parameter $\lambda_0$ corresponding to the cases of single-magnon dynamics with/without the confining magnetic field are shown in Fig.~\ref{fig:num_terms_1magnon}. 

\subsection{Numerical schemes for the approximation of derivatives}\label{sec:num_schemes}
In this subsection, we discuss approximation schemes for computing derivatives from data and comment on how these numerical schemes affect recovered PDEs.

For the purposes of reconstructing Eq.~(\ref{eq:pde_xx_num}), we employed the standard second-order finite difference scheme when
calculating temporal $\partial_t u$ and spatial $\partial_x^n u$ derivatives from data:
\begin{eqnarray}
\label{eq:ut_finite_diff_scheme}
&& u_t(t, x) = \frac{u(t+dt, x)-u(t-dt, x)}{2dt} + \mathcal O(dt^2),\\
\label{eq:ux_finite_diff_scheme}
&& u_{x}(t, x) = \frac{u(t, x+dx)-u(t, x-dx)}{2dx} + \mathcal O(dx^2),\\
\label{eq:uxx_finite_diff_scheme}
&& u_{xx}(t, x) = \frac{u(t, x+dx)+u(t, x-dx)-2 u(t,x)}{dx^2} + \mathcal O(dx^2).
\end{eqnarray}
One can notice that the coefficients in the inferred PDE in Eq.~(\ref{eq:pde_xx_num}) are very close to the exact theoretical values. Such high precision of the recovered coefficients could be surprising at first glance, given that the PDE in Eq.~(\ref{eq:pde_xxz_hz(x)}) contains corrections with higher-order spatial derivatives.
In fact, when using the second-order finite difference scheme (\ref{eq:uxx_finite_diff_scheme}), the finite difference discretization of PDE (\ref{eq:pde_xxz_hz(x)})  coincides with the exact differential-difference equation (\ref{eq:fin_diff_xxz_hz(x)}). The reconstruction error of the coefficients, when considering the second-order finite difference scheme, could be estimated as $\delta\xi \sim \mathcal O (dt^2 + dx^2)$.

The spectral (Fourier) method for the calculation of spatial derivatives $u_x$ and $u_{xx}$ could be used as an alternative method to the finite difference schemes (\ref{eq:ux_finite_diff_scheme},~\ref{eq:uxx_finite_diff_scheme}) when periodic boundary conditions are imposed:
\begin{equation}
\label{eq:fourier_method}
    \partial_x^n u(t,x) = \mathrm{iFFT}\left[(iq)^n \hat{u}(t, q)\right], \qquad \hat{u}\left(t,q_m=\frac{2\pi m}{L}\right) = \mathrm{FFT}(u) = \sum_{j=0}^{L-1} e^{i q_m j} u(t, x_j),
\end{equation}
where FFT (iFFT) denotes Fast Fourier Transform (inverse Fast Fourier Transform).
Taylor expansion of the ``kinetic term``  $\cos{q} = 1-\frac{q^2}{2!}+\frac{q^4}{4!} + \ldots$ in Eq.~(\ref{eq:cos_q_exact}) results in the following correction to the evolution PDE:
\begin{equation}
    i u_t \approx \frac{J}{2}u_{xx}+\frac{J}{24}u_{xxxx} + (J-\Delta) u - B(x)u.
\end{equation}
Applying the spectral method for the calculation of spatial derivatives from data [shown in  Fig.~\ref{fig:1p_wave_packet}] resulted in the following reconstructed PDE (parameters of the XXZ model are $J=-1$, $\Delta/J=0.5$, $B(x)=0$):
\begin{equation}
i u_t + 0.495 u_{xx} + 0.5\, u = 0.
\end{equation}
The STRidge algorithm turned out to be insensitive to the forth-order derivative term  $u_{xxxx}$ and missed it during reconstruction.
Performing a full search over all possible combinations of $M=10$ terms in $F(\cdot)$ and scanning across a range of values for the $L_0$ penalty factor $\lambda_0$, we were able to recover, at $\lambda_0=10^{-4}$, the expected form of the PDE that includes the forth-order derivative term:
\begin{equation}
 i u_t + 0.4996 u_{xx} + 0.041 u_{xxxx} + 0.499 = 0,   
\end{equation}
where we again used the spectral method to compute spatial derivatives from data.
As displayed in Fig.~\ref{fig:num_terms_1magnon}, as we decrease the strength of the $L_0$ penalty term, we obtain a ``staircase'' of  PDEs, which reproduces the gradient expansion of the exact PDE (\ref{eq:cos_q_exact}). Note, that each additional term in the inferred PDE persists over a finite range of $\lambda_0$ values.
By increasing the precision of the input dataset (refining the spatiotemporal grid), it is possible in principle to recover higher-order derivative terms originating from the tight-binding dispersion $\propto\cos{(i\partial_x)}$. 

Generally, the reconstructed PDE 
could be slightly sensitive to the choice of the numerical scheme used for the calculation of temporal and spatial derivatives, as shown in the examples above.
However, such dependence will mostly appear in the high-order gradient terms.
It is worth noting that finite difference schemes could be used to recover differential-difference equations instead of PDEs [e.g.~Eq.~(\ref{eq:fin_diff_xxz_hz(x)})] even when the envelope function is not smooth and the continuous approximation is not valid.

\subsection{Magnon dynamics in the long-range XXZ model}\label{sec:long_range}

In this subsection, we consider the one-dimensional XXZ model with power-law-decaying spin-spin interactions,
\begin{equation}
\label{eq:H_lr}
    H = -\sum_{i > j}  \frac{1}{|i-j|^\alpha} \Bigl(J\left(S^x_i S^x_j+ S^y_i S^y_j\right) + \Delta S^z_i S^z_j\Bigl),
\end{equation}
where $\alpha$ is a power-law exponent, and spin operators are defined as in Eq.~\eqref{eq:xxz_ham}. We will assume that $\alpha>1$, so that the Hamiltonian (\ref{eq:H_lr}) has a well-defined thermodynamic limit.

The phase diagram for the model in Eq.~(\ref{eq:H_lr}) for $J=1$ was obtained in Ref.~\onlinecite{maghrebi2017continuous}.
Depending on the value of $\Delta$, the ground state of the model (\ref{eq:H_lr}) can be in (i) the ferromagnetic phase for $\Delta>1$, or (ii) the antiferromagnetic phase for large $\alpha$-dependent values $\Delta<0$, or (iii) the XY phase (Tomonaga-Luttinger liquid) with algebraically decaying correlations (and characterized by the conformal charge $c=1$), or (iv) the continuous symmetry breaking phase  for intermediate values of $\Delta$ and small power-law exponents $\alpha$. 
The continuous symmetry breaking phase, which is generally forbidden by the Mermin-Wagner theorem in the case of low dimensional systems with local interactions, arises as a consequence of the long-range interactions.
The phase boundary between the ferromagnetic phase and either the XY or the continuous symmetry breaking phase corresponds to a first-order phase transition.
Here, we will be considering only excitations in the ferromagnetic phase. 

The exact form of evolution equations for the observable $u(t,x_i)=\langle S^+_i(t)\rangle$ has the form
\begin{equation}
    i \partial_t u(t,x_i) = -\frac{J}{2}\sum_{j\neq i} \frac{1}{|i-j|^\alpha} u(t,x_j) + \frac \Delta 2u(t,x_i)\sum_{j \neq i} \frac{1}{|i-j|^\alpha}.
\end{equation}
In the continuous limit, the evolution PDE reads~\cite{gong2016kaleidoscope}
\begin{equation}
    i u_t = J \mathcal D(i \partial_x) u + c u, \quad \mathcal D(\hat q) := \sum_{n=1}^{\infty} \frac{1-\cos{(\hat q n)}}{n^\alpha},
\end{equation}
where the constant $c=(\Delta-J)\sum_{n=1}^\infty n^{-\alpha} = (\Delta-J) \zeta(\alpha)$, where $\zeta(\alpha)$ is the Riemann zeta function.
It is convenient to formulate the PDE-learning problem in the time-momentum $(t, q)$ representation instead of the $(t, x)$ representation by considering the Fourier components
\begin{equation}
    \hat u(t,q) := \sum_{j=0}^{L-1} u(t, x_j) e^{-i q x_j}.
\end{equation}
Then, the equation of the Fourier component has the form
\begin{equation}
    i \hat u_t = J \mathcal D(q) \hat u(t,q) + c \hat u(t,q),
\end{equation}
where the operator $\mathcal D(q)$ for non-integer $\alpha$ has the long-wavelength expansion
\begin{equation}
\label{eq:Dq_expansion}
    \mathcal D(q) = M_\alpha -\Gamma (1-\alpha) \cos{\left[\frac{\pi}{2}(\alpha-1)\right]}|q|^{\alpha-1} + \frac{1}{2!}\zeta(\alpha-2) q^2 - \frac{1}{4!}\zeta(\alpha-4)q^4 + \mathcal{O}(q^6),
\end{equation}
where
\begin{equation}
M_\alpha = 
\begin{cases}
\frac{(-1)^{n+1}}{(2n)!}q^{2n}\log{|q|}, \quad &\alpha = 2n+1,\quad n\in\mathbb Z, \quad n\geq 1,\\
0, &\text{other } \alpha>1.
\end{cases}
\end{equation}
In the case of integer $\alpha$, new additional logarithmic terms will appear $\sim |q|^{\alpha-1} \log{|q|}$ (for odd integer $\alpha=3,5,\ldots$) in the expression in Eq.~(\ref{eq:Dq_expansion}).
Indeed, logarithmic terms at odd integer values of $\alpha$ appear when accounting for the singularity of the zeta function $\zeta(1+\epsilon)=\frac{1}{\epsilon}+\gamma_E +\mathcal{O}(\epsilon)$ and the Gamma function $\Gamma(-n+\epsilon)=\frac{(-1)^n}{n!}\left(\frac{1}{\epsilon}+\psi_1(n+1) + \mathcal{O}(\epsilon)\right)$, where $\psi_1(z)=\Gamma'(z)/\Gamma(z)$ is the digamma function, $\psi_1(n)=\sum_{k=1}^{n-1}\frac{1}{k}-\gamma_E$. Considering $\alpha=2n+1+\epsilon$ and taking the limit $\epsilon\to 0$  gives the following contribution $ \frac{(-1)^{n+1}}{(2n)!}q^{2n}\log{|q|}$ that comes from the singular $\frac{1}{\epsilon}$ terms.

We define the candidate terms library as 
\begin{equation}
\label{eq:uq_ansatz}
\partial_t \hat u = F(\hat u, q^2\hat u, q^4\hat u,  |q|^\mu \hat u, \log{|q|}\hat u, q^2 \log{|q|}\hat u, q^4\log{|q|}\hat u),    
\end{equation}
where $\mu$ is a free parameter subject to tuning.
Next, we sequentially perform optimization of the $L_2+L_0$ loss function  using a three-step procedure: (1) perform sparse selection of the most relevant candidate terms (e.g.~brute force search with an $L_0$ penalty term or STRidge algorithm), (2) get coefficients for each term in the library via least-squares regression, (3) run several steps of optimization (using the Powell line search algorithm) to find the best value for the parameter $\beta$. Steps (1), (2), and (3) are repeated in a loop.

\begin{figure}
    \centering
    \includegraphics[scale=0.35]{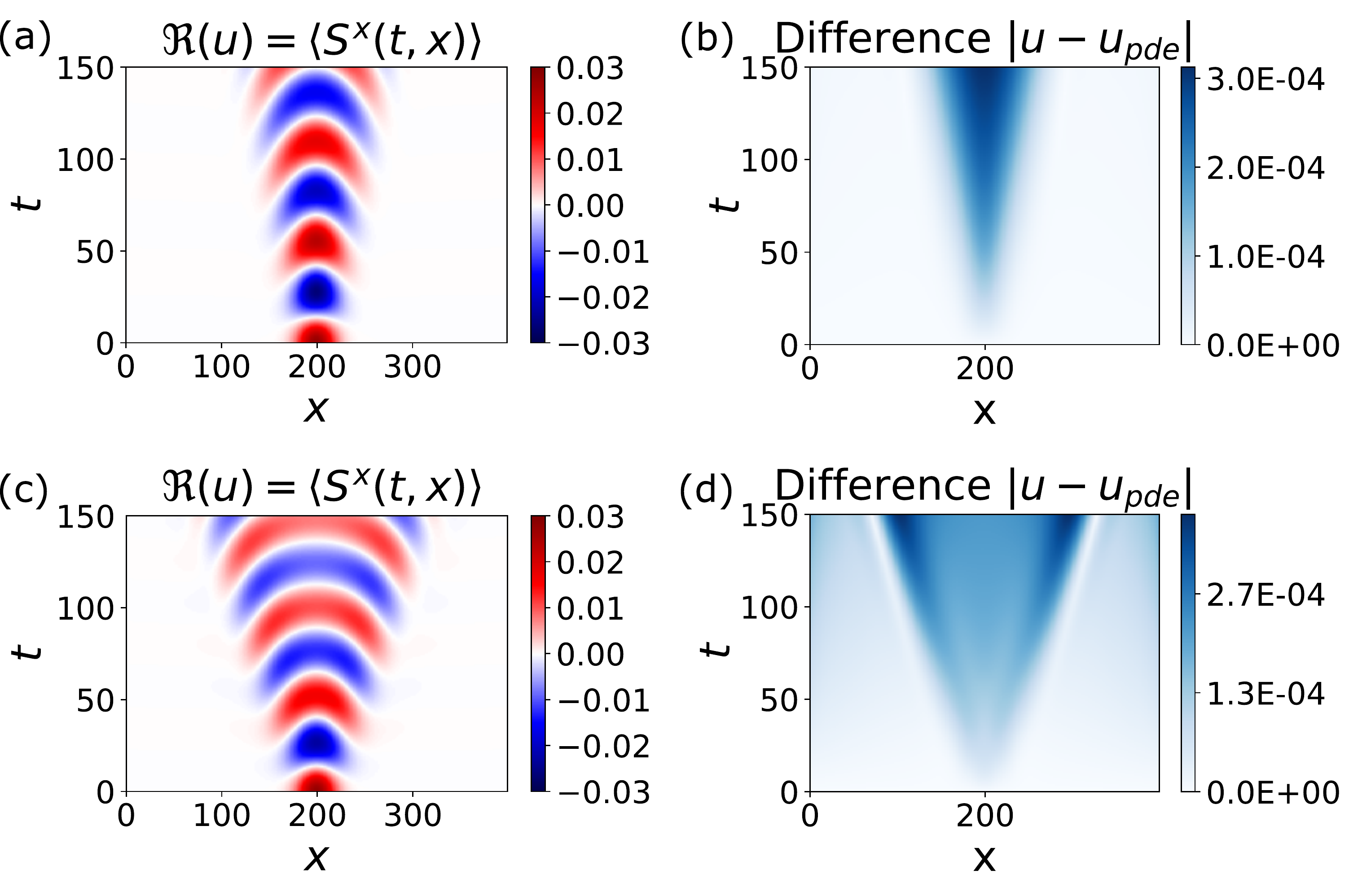}
    \caption{Magnon wave packet propagation in the long range XXZ model ($L=400$, $\Delta/J=0.9$), exact vs reconstructed evolution: (a, b) $\alpha=3$, (c, d) $\alpha=2.5$. Panels (a,c) display the  real part $\Re{[u(t,x)]}$ of the input dataset, while panels (b,d) show the difference between the exact solution and the recovered non-local PDEs (\ref{eq:pde_lr_alpha=3}, \ref{eq:pde_lr_alpha=2.5}).  
    }
    \label{fig:long_range_xxz}
\end{figure}

The above-described algorithm, for the case $\alpha=3$, $\Delta/J=0.9$, $J=-1$ and initial conditions corresponding to Fig.~(\ref{fig:long_range_xxz}), results in the following reconstructed equation:
\begin{equation}
    \label{eq:pde_lr_alpha=3}
    i  u_t(t,x) +  \int_{-\infty}^\infty dq\int_{-\infty}^\infty dx' \Bigl(0.752\, |q|^{2.02}+ 0.505 \, q^2\log{|q|}\Bigl) e^{iq(x-x')}u(t,x')  - 0.12  u(t,x) = 0,
\end{equation}
which is in good agreement with the theoretically predicted equation up to a $\mathcal{O}(q^4 \hat u) $ correction term:
\begin{equation}
\label{eq:pde_lr_alpha=3_theory}
    i u_t(t,x) + \int_{-\infty}^\infty dq\int_{-\infty}^\infty dx' \Bigl(\frac{3}{4} q^2 + \frac 12 q^2\log{|q|}\Bigl) e^{iq(x-x')}u(t,x') - 0.1 \zeta(3)u(t,x)= 0,
\end{equation}
where $\zeta(3)\approx 1.202..$.
Equations (\ref{eq:pde_lr_alpha=3},~\ref{eq:pde_lr_alpha=3_theory}) are written in the integro-differential form since we performed conversion from the momentum representation to the coordinate representation in Eq.~(\ref{eq:uq_ansatz}).

In the case of non-integer $\alpha=2.5$, $\Delta/J=0.9$, $J=-1$, the reconstruction results in the equation
\begin{equation}
\label{eq:pde_lr_alpha=2.5}
i  u_t + 1.68\int_{-\infty}^\infty dq\int_{-\infty}^\infty dx'  |q|^{1.505} e^{iq(x-x')}u(t,x') + 0.72 u_{xx}(t,x) - 0.134  u(t,x) = 0,
\end{equation}
that should be compared to the theoretically expected one from Eq.~(\ref{eq:Dq_expansion}):
\begin{equation}
i u_t  +\frac{1}{\sqrt{2}}\Gamma\left(-\frac{3}{2}\right)\int_{-\infty}^\infty dq\int_{-\infty}^\infty dx'|q|^{3/2}e^{iq(x-x')}u(t,x')  -\frac{1}{2}\zeta{\left(\frac{1}{2}\right)} u_{xx}(t,x) + 0.1\zeta(5/2) u(t,x),
\end{equation}
where $\frac{1}{\sqrt{2}}\Gamma\left(-\frac{3}{2}\right)\approx 1.671...$, $\frac{1}{2}\zeta{\left(\frac{1}{2}\right)}\approx -0.7301...$, and $\zeta(5/2) \approx 1.341...$

As an additional application of the reconstruction algorithm, from the inferred PDEs~(\ref{eq:pde_lr_alpha=2.5}, \ref{eq:pde_lr_alpha=3}) for the observable $u(t,x)$, one can extract physical parameters of the long-range XXZ model, including the power-law exponent $\alpha$, by comparing coefficients of the reconstructed PDE with the theoretical values.
Hydrodynamic behavior in a trapped-ion quantum simulator was recently measured experimentally~\cite{joshi2021observing}.

\subsection{Dynamics of a domain-wall initial state in the XXZ spin chain}\label{sec:domain_wall}

In the main text, we showed how the PDE-learning approach allows one to recover evolution equations of a domain-wall initial state in the XXZ spin chain with nearest-neighbor couplings.  
In the present subsection, we provide additional details regarding our results  for both the zero-temperature and the high-temperature initial states. We also elaborate on previously known theoretical results.

The conservation of total magnetization along the $z$ axis implies the continuity equation of the form 
\begin{equation}
\label{eq:spin_current}
\partial_t u  + \partial_x \mathcal{J}^{z}(u)=0, \quad u(t,x) \equiv \langle S^z(t,x) \rangle.    
\end{equation}
The exact solution for the ``domain-wall'' initial state $|\psi_0\rangle =  \ket{\downarrow}^{\otimes L/2} \ket{\uparrow}^{\otimes L/2}$ in the thermodynamic limit $L\to\infty$ is given by~\cite{collura2018analytic}
\begin{align}
\label{eq:sz_dw}
\left\langle S^{z}\right\rangle&=\frac{1}{2 \pi / P} \arcsin \left(\frac{\zeta}{\zeta_{0}}\right), \\ \mathcal{J}^{z} &=\frac{1}{2 \pi / P} \zeta_{0}\left[\sqrt{1-\frac{\zeta^{2}}{\zeta_{0}^{2}}}-\cos \left(\frac{\pi}{P}\right)\right],
\label{eq:jz_dw}
\end{align}
where $\zeta=x/t$ is the lightcone coordinate. Here the coefficients are given by  $\gamma = \arccos{(\Delta/J)}$, $\zeta_0 = \sin{(\gamma)}/\sin{\left(\frac{\pi}{P}\right)}$, and $\gamma=\pi Q/P$, where $Q$ and $P$ are coprime integers.
Formally, the derivation of Eqs.~(\ref{eq:sz_dw}, \ref{eq:jz_dw}) is restricted to the specific values of the anisotropy parameter, such that $Q/P=\frac{1}{\pi}\arccos{(\Delta/J)}\in \mathbb{Q}$ is a rational number. However, if $\frac{1}{\pi}\arccos{(\Delta/J)}$ is an irrational number, the ratio $Q/P$ can be tuned to approximate the irrational number with a desired precision.
The overall sign in the expression for the current (\ref{eq:jz_dw}) assumes the specific choice of  boundary conditions at infinity: $\langle S^z \rangle \to \pm 1/2$ for $x\to \pm \infty$.
The evolution PDE could be simplified to the form
\begin{equation}
\label{eq:ut_zeta}
u_t + \zeta_0 \sin{\left(\frac{2 \pi}{P} u\right)}   u_x = 0.
\end{equation}

\begin{figure}
    \includegraphics[scale=0.4]{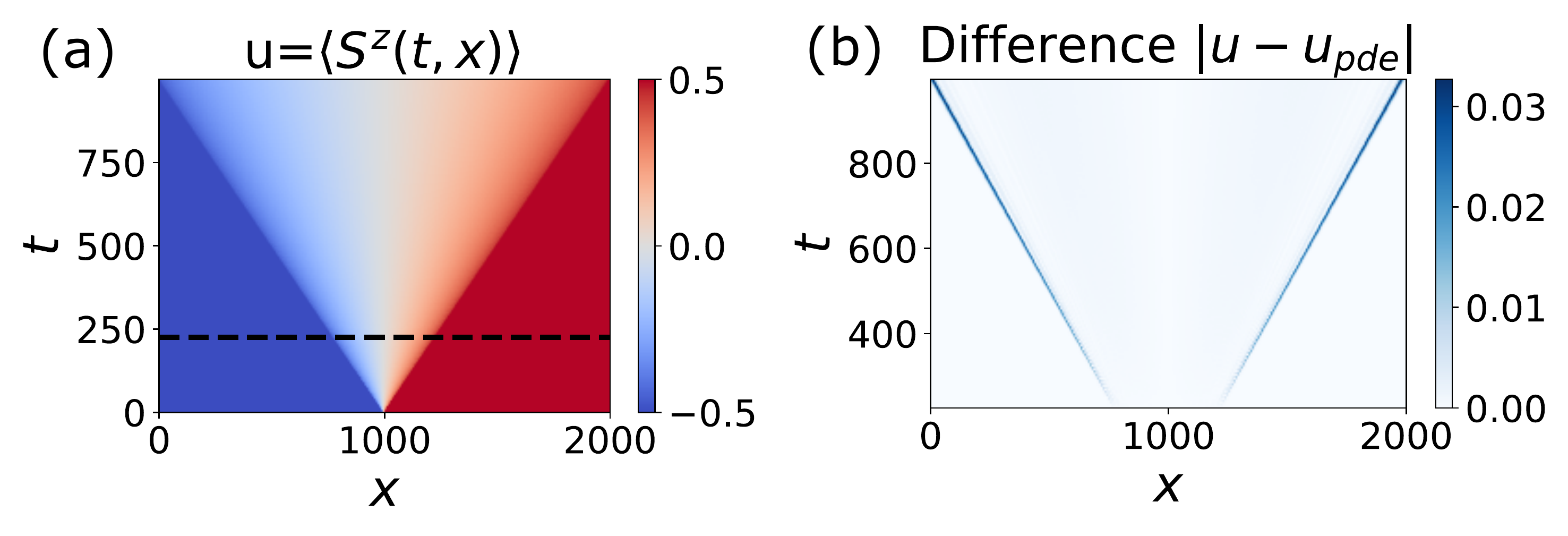}
    \caption{(a) Evolution of a domain-wall initial state in the 1D XX model, $\Delta=0$ (exact). 
    (b) Difference between the exact evolution of $u(t,x)=\langle S^z(t,x)\rangle$  and the solution of the inferred PDE, Eq.~(\ref{eq:xx_dw_2terms}).
    Total number of sites is $L=2000$. Dashed line shows the starting time  used for PDE reconstruction. 
    }
    \label{fig:xx_domain_wall}
\end{figure}

Using data obtained from numerical simulations of the dynamics of the XXZ spin chain, we perform PDE reconstruction.
Specifying, using the  library of terms
\begin{equation}
\label{eq:xx_dictionary}
   u_t  = F(u_x, u_{xx}, u_{xxx}, u_{xxxx}, u u_x, u^2 u_{x}, u^3 u_{x}, u^4 u_x, u^5 u_x)
\end{equation}
and the  data presented in Fig.~\ref{fig:xx_domain_wall}---obtained from numerical computation of the exact evolution for the case of the XX spin chain ($\Delta=0$---we obtain the following  PDEs: 
\begin{eqnarray}
  \label{eq:xx_dw_2terms}
  & u_t& + 3.12 u u_{x} - 4.49  u^3 u_x = 0, \quad (\lambda_0 = 10^{-4})\\
  & u_t& + 3.135 u u_{x} - 5.056  u^3 u_x + 1.92 u^5 u_x  = 0, \quad (\lambda_0 = 10^{-6}).
  \label{eq:pde_extract_dw_delta=0}
\end{eqnarray}
The functional form (\ref{eq:xx_dictionary}) is consistent with spin-current conservation (\ref{eq:spin_current}).
Coefficients in Eq.~(\ref{eq:ut_zeta}) are quite close to the theoretically expected ones, obtained via Taylor expansion of the $\sin(\cdot)$ term  up to the 5th order:
$u_t+\pi u - \frac{\pi^3}{3!}u^3 u_x + \frac{\pi^5}{5!}u^5 u_x\approx0$.

\begin{figure}
    \includegraphics[scale=0.4]{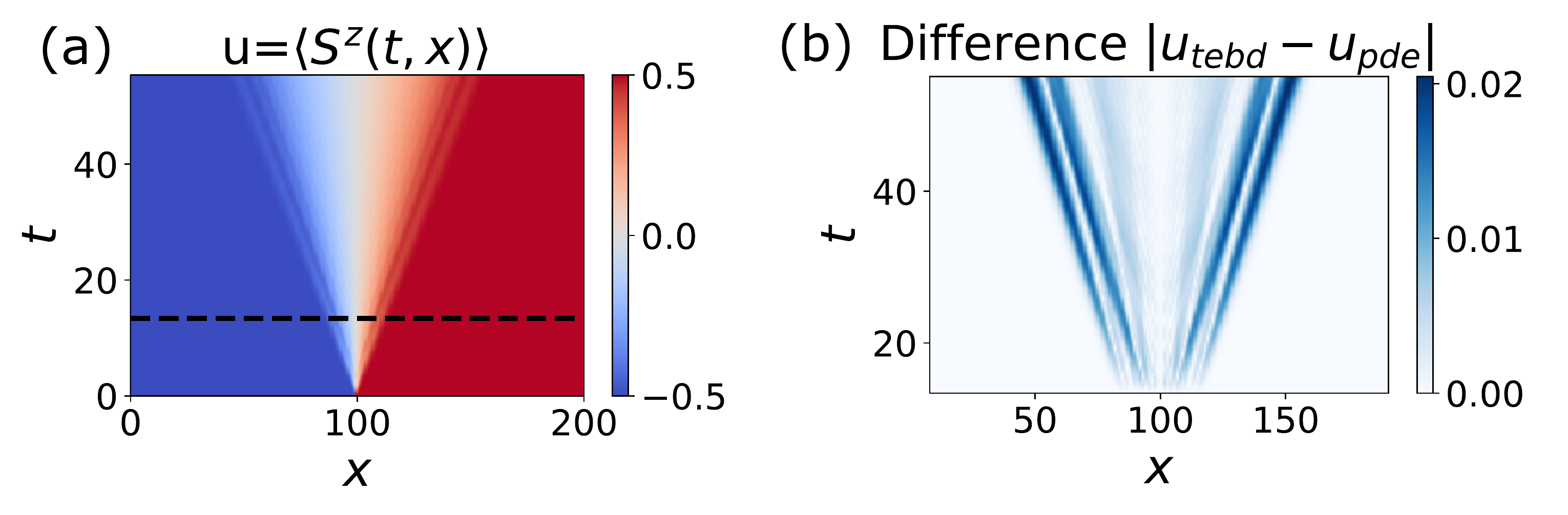}
    \caption{(a) Evolution of a domain wall initial  state in the XXZ model, $\Delta/J=0.5$ (TEBD). 
    (b) Difference between exact evolution of magnetization and the solution of the inferred PDE, Eq.~(\ref{eq:dw_pde_ab}).
    The total number of sites is $L=200$.  Original TEBD data $u(t,x)$ was smoothed along the $x$ dimension by applying the Savitsky-Golay filter. The MPS bond dimension was  $\chi=200$.}
    \label{fig:xxz_domain_wall}
\end{figure}

For the XXZ model with  $\Delta/J = 2$, we get $(P, Q) =(3, 1)$, $\zeta_0=1$, and the corresponding PDE in the limit $t\to \infty$ reads
\begin{equation}
    u_t + \sin{\left(\frac{2\pi}{3} u \right)} u_x = 0.
\end{equation}
PDE reconstruction from TEBD data shown in Fig.~\ref{fig:xxz_domain_wall}
gives the following equation:
\begin{equation}
\label{eq:dw_pde_ab}
    u_t + a u u_x - b u^3 u_x = 0,
\end{equation}
where $a\approx 2.1$, $b\approx 1.67$.
Knowing parameters $(a,b)$ of the PDE allows one to extract the Hamiltonian parameter $\Delta/J=\cos{\gamma}$ directly from data:
\begin{equation}
\label{eq:dw_pde_zeta_p}
    u_t \approx -\zeta_0 \left(\frac{2\pi}{P}\right)  u u_x + \frac{\zeta_0}{3!}\left(\frac{2\pi}{P}\right)^3 u^3 u_x .
\end{equation}
Comparing Eqs.~(\ref{eq:dw_pde_ab}) and (\ref{eq:dw_pde_zeta_p}), we obtain
\begin{equation}
    P =  2\pi \sqrt{\frac{a}{6b}} \approx 2.9, \qquad \zeta_0 = \frac{a P}{2 \pi} \approx 0.96, \qquad \Delta/J = \sqrt{1-\zeta_0^2 \sin^2{\left(\frac{\pi}{P}\right)}} \approx 0.52.
\end{equation}

Motivated by the theoretically expected form of the evolution equations (\ref{eq:sz_dw}) and  (\ref{eq:jz_dw}), we could also try to search for a PDE of the form
\begin{equation}
    u_t = F\left(u_x, u_{xx},  u u_x, u^2 u_x, u^3 u_x, u^4 u_x, u^5 u_x, \sin{\left(\frac{2\pi}{P_1} u\right)} u_x,\, \sin{\left(\frac{2\pi}{P_2} u\right)} u_x, \ldots \right),
\end{equation}
 where $P_i$ are integers. 
The goal of such a test is to see if the PDE-learning algorithm would be able to identify a concise form of the equation and find the correct value of $\Delta$.
We set the integer parameter to be in the range $P_i \in \{1, 2, \ldots, 10\}$.
The BruteForce and CrossEntropy algorithms were able to recover the theoretically expected equation from $M=19$ terms:
\begin{eqnarray}
u_t + 0.994 \sin{\left(\frac{2\pi}{3}u\right)}u_x = 0, \quad (\lambda_0 = 10^{-3}),
\end{eqnarray}
which immediately gives $\Delta/J\approx 0.5$.
The algorithm finds a sparse solution and favors a compact form with the $\sin{(\cdot)}$ term on the rhs, as opposed to the truncated Taylor expansion for the same  expression.
Interestingly, the STRidge algorithm was not able to find the correct solution for any value of the penalty parameter $\lambda_0$.
This shows that STRidge, although reliable in most test cases, sometimes fails.

\begin{figure}
    \includegraphics[scale=0.36]{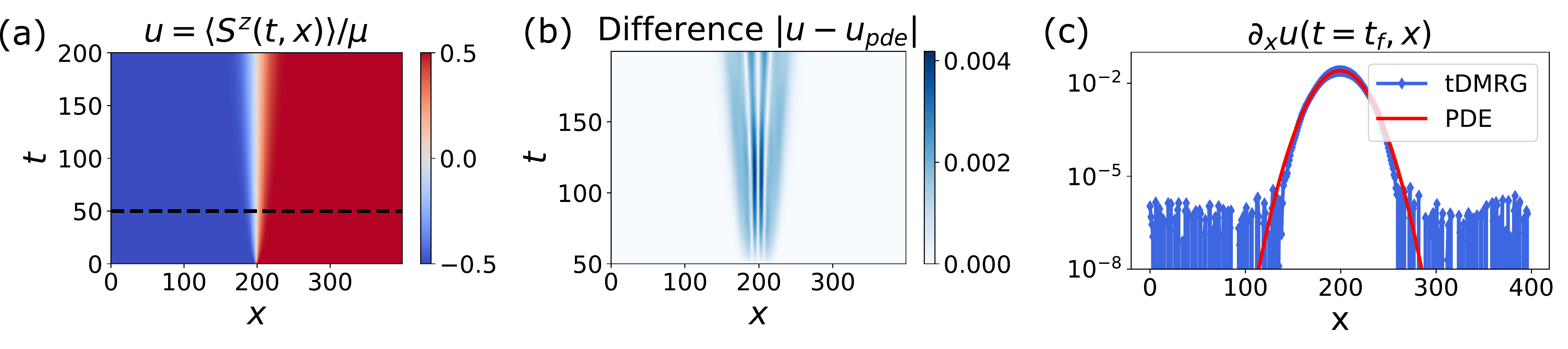}
    \caption{Evolution of a high-temperature domain-wall initial state in the XXZ spin chain in the gapped phase $\Delta/J=2$. (a) Data (high-precision tDMRG) is reproduced from Ref.~\onlinecite{ljubotina2017spin}. 
    The density matrix of the initial state is given by Eq.~(\ref{eq:rho_thermal}). The dashed horizontal line, $t_0$, separates the portion of the data $t\geq t_0$ used for PDE-learning. (b) Difference between tDMRG data and the solution of the recovered PDE (\ref{eq:pde_delta=2}). (c) Comparison between spatial derivatives $\partial_x u(t, x)$, where $u(t,x)$ corresponds to data or to the solution of the PDE at the maximum evolution time, $t_f=200$.
        }
    \label{fig:prosen_data}
\end{figure}

The example considered above corresponds to the spreading of the domain-wall initial state in the gapless phase of the XXZ model at zero temperature.
Interestingly, in the gapped phase, $\Delta/J>1$, equations~(\ref{eq:sz_dw}, \ref{eq:jz_dw})
are not valid: domain-wall evolution in the XXZ model freezes and the domain-wall spreading stops.
As a result, the PDE reconstruction is problematic in this case.
On the other hand, for high-temperature mixed initial states, the spin dynamics is qualitatively different. The initial high-temperature state is prepared by combining two reservoirs with the opposite direction of the longitudinal magnetic field,  
\begin{equation}\label{eq:rho_thermal}
\rho(t=0) = \frac{\exp{(\mu \sum_{i\in L} \sigma^z_{i})}}{Z_L} \otimes  \frac{\exp{(-\mu \sum_{j\in R}\sigma^z_j)}}{Z_R},
\end{equation}
where $0<\mu \ll 1$.

Using tDMRG data from Ref.~\onlinecite{ljubotina2017spin}, we perform PDE reconstruction for $\Delta/J=2$ and $\Delta/J=1$.
In the gapped phase ($\Delta/J=2$) presented in Fig.~\ref{fig:prosen_data}, using ansatz 
\begin{equation}
u_t = F(u_x, u_{xx}, u_{xxx}, u_{xxxx}, u u_x, u^2 u_{x}, u^3 u_{x}, u^4 u_{x}, u^5 u_{x}),
\end{equation}
we obtain the following equation for the rescaled magnetization $u=\mu^{-1}\langle S^z \rangle$:
\begin{eqnarray}
&u_t& = D u_{xx}, \quad D \approx 0.64,
\label{eq:pde_delta=2}
\end{eqnarray}
which agrees with the self-similar scaling law in the gapped phase, $u(t,x)=f(x/\sqrt{t})$,  observed numerically in Ref.~\onlinecite{ljubotina2017spin}. The value of the diffusion coefficient is close to the theoretically predicted value $D=0.76$ at infinite temperature for $\Delta/J=2$.~\cite{gopalakrishnan2019}
In Fig.~\ref{fig:prosen_data}, we compare the tDMRG data and the solution of the reconstructed PDE, Eq.~(\ref{eq:pde_delta=2}); the agreement is excellent. 

In spite of a number of recent papers on the topic~\cite{gopalakrishnan2019, de2018hydrodynamic, de2020superdiffusion}, the full theoretical explanation of the properties of spin dynamics at the isotropic point $\Delta/J=1$ is still lacking. 
A superdiffusion behaviour at large times $t\gg 1$ was empirically observed in Ref.~\onlinecite{ljubotina2017spin},  $u(x,t) \propto f(x/t^{\eta})$, with an anomalous scaling exponent  $\eta\approx 2/3$.
Moreover, it was shown in Ref.~\onlinecite{ljubotina2019kardar} that the shape of the profile of the magnetization $f(y=x/t^{\eta})$ asymptotically approaches  the KPZ scaling function, thus revealing a connection between the KPZ equation and the effective dynamics of magnetization in the Heisenberg model.
Following our PDE reconstruction methodology, we are interested in finding a closed-form evolution equation for $u(t,x)$, where the rhs $u_t=F(\cdot)$ does not have an explicit time dependence.
Using BruteForce, STRidge, and CrossEntropy algorithms (the list of candidate terms is shown in Table~\ref{table:t2}), we found the following equation that describes data with high precision:
\begin{eqnarray}
\label{eq:pde_delta=1_2terms}
&u_t& + a u u_x = D u_{xx}, \quad a\approx 0.24,\; D\approx 1.90 \quad (\lambda_0=10^{-2}),
\end{eqnarray}
which is known as Burgers' equation.
A similar diffusion-type term was recently predicted in Ref.~\onlinecite{de2018hydrodynamic} for integrable 1D models  based on a generalized hydrodynamics approach.

\begin{figure}
    \centering
    \includegraphics[scale=0.5]{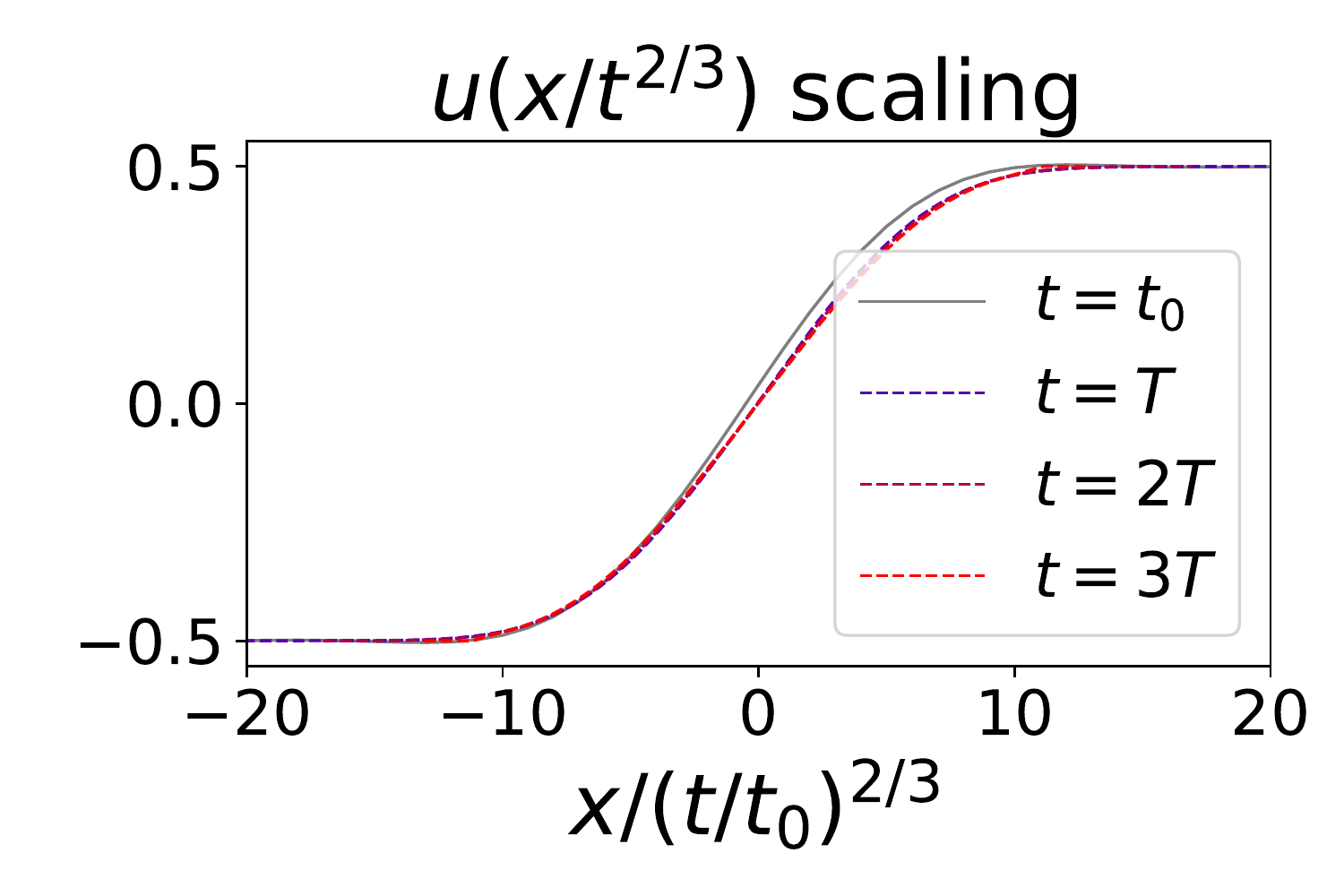}
    \caption{KPZ-scaling relation  $u=u(x/t^{2/3})$ of the solution of the inferred deterministic PDE (\ref{eq:pde_delta=1_2terms}) for the magnetization dynamics in the high-temperature Heisenberg model ($\Delta/J=1$) at late evolution times. The KPZ scaling law has been observed numerically~\cite{ljubotina2019kardar} and is anticipated theoretically.~\cite{gopalakrishnan2019}
    The initial condition for the evolution of the PDE corresponds to tDMRG data at time $t_0=50$ (dashed horizontal line in Fig.~\ref{fig:prosen_data}). $T=200$ is the total evolution time [$\max(t)$] in the tDMRG data. We observe that the scaling relation for the solution of the PDE still holds for times $t\in [T, 3T]$,  which is beyond the time range presented in the dataset $t\in[0, T]$. } 
    \label{fig:kpz_scaling}
\end{figure}

It is natural to interpret the discovered equation (\ref{eq:pde_delta=1_2terms}) as a noise-averaged stochastic Burgers' equation: 
\begin{equation}
u_t + a u u_x = D u_{xx} + \partial_x \eta (x, t),   
\end{equation}
where $\eta(x, t)$ represents uncorrelated Gaussian noise, $\langle\eta (x,t)\rangle=0$.
The stochastic Burgers' equation is closely related to the 1D KPZ equation
\begin{equation}
    h_t +\frac{a}{2}(h_x)^2  = D h_{xx} + \eta(x,t)
\end{equation}
via the substitution $u(t,x)=h_x(t,x)$. Therefore, our equation (\ref{eq:pde_delta=1_2terms}) also  demonstrates a connection between magnetization dynamics in the Heisenberg model and the KPZ physics. 
Interestingly, we found that the solution of the Burgers' equation (\ref{eq:pde_delta=1_2terms}) obeys a KPZ-type scaling law $u(x/t^{2/3})$ for late evolution times, see Fig.~\ref{fig:kpz_scaling}. Although the KPZ scaling for the inferred deterministic equation (\ref{eq:pde_delta=1_2terms}) is exhibited numerically with high accuracy, we were not able to prove analytically whether or not the solution of Burgers' equation (\ref{eq:pde_delta=1_2terms}) with the initial condition given by data admits asymptotic scaling $u(x/t^{2/3})$ at $t\to\infty$.

\newpage
\section{PDE-learning of hydrodynamic equations in fermionic systems: additional details}\label{sec:fermions}

In this Section of the Supplementary Material, we provide details of PDE-learning in fermionic systems: the 1D non-interacting fermion gas and the strongly interacting Fermi-Hubbard model.
In Section~\ref{sec:fermion_hydro}, we give a quick  overview of the analytical derivation of hydrodynamic equations describing dynamics in the free fermion gas in the semiclassical approximation. 
In Section~\ref{sec:cos_corrections}, we derive  correction terms to the hydrodynamic equations for the free fermion gas---terms that stem from the non-parabolic (tight-binding) dispersion---both analytically and using our PDE-learning algorithm.
In Section~\ref{sec:fermion_hydro_quartic}, we consider hydrodynamics of the non-interacting fermion gas in the vicinity of a Lifshitz critical point. The salient feature of the Lifshitz critical point is the quartic fermion dispersion at small momenta, which results in an unusual hydrodynamic equation.
In Section \ref{sec:particle_current} we derive expression for the particle current and velocity a tight-binding model with additional next-nearest-neighbour hopping terms. 
In Section~\ref{sec:global_symmetries}, we discuss the global symmetry properties of the hydrodynamic equations and show how leveraging of these symmetries significantly reduces the size of the search space of candidate PDEs. In Section~\ref{sec:partial_observ}, we propose a method to perform PDE-reconstruction from partial observations, when only data for fermion density evolution (but not velocity) is available.  
In Section~\ref{sec:single_pde_rho}, we derive a single second-order-in-time PDE that describes the evolution of density in a gas of free fermions.
In Section~\ref{sec:inter_ferm_hydro}, we provide supplementary details on PDE-learning of hydrodynamics in the spinless Fermi-Hubbard model and summarize our findings from the main text. In particular, we discuss in more detail the connection between the discovered effective Euler equation and  the Tomonaga-Luttinger theory. Finally, in Section~\ref{sec:viscosity}, we discuss the emergent Navier-Stokes equation and  the role of the discovered viscosity term. 

\subsection{PDE-learning of bosonization equations:  Semiclassical regime of  hydrodynamics of non-interacting fermions}\label{sec:fermion_hydro_parabol}
\label{sec:fermion_hydro}
In this subsection, we provide details of the derivations of semiclassical hydrodynamic equations for a free-fermion gas.

We consider a 1D non-interacting gas of spinless fermions on a lattice described by the tight-binding Hamiltonian
\begin{equation}
\label{eq:H_tight_binding}
    H = -J \sum_i (c^\dag_i c_{i+1} + c^\dag_{i+1} c_i) + \sum_i V_i c^\dag_i c_i,
\end{equation}
where $c_i$ ($c^\dag_i$) are fermion annihilation (creation) operators at lattice site $i$, $J$ is the hopping parameter, and $V_i$ is the external potential. The energy dispersion of free fermions  in the low-density limit could be well-approximated as parabolic: $\varepsilon_k/2J = -\cos{(k)} \approx -1 + \frac{k^2}{2} + \mathcal{O}(k^4)$.
The dynamics of the fermion gas with parabolic dispersion $H=\sum_k \frac{k^2}{2 m} c^\dag_k c_k$
in the Wentzel–Kramers–Brillouin (WKB) approximation  could be described by
hydrodynamic equations~\cite{bettelheim2008quantum}:
\begin{eqnarray}\label{eq:euler_rho}
&&\rho_t + (v \rho)_x =0,\\
&&\label{eq:euler_v}
v_t + v v_x  = - \frac{1}{m\rho} \partial_x P(\rho) - \partial_x V(x), \qquad P(\rho)  = \frac{\pi^2}{3 m}\rho^3.
\end{eqnarray}
Eq.~(\ref{eq:euler_rho}) is the continuity equation, which describes the conservation of the total number of fermions. Eq.~(\ref{eq:euler_v}) is the Euler equation describing barotropic compressible fluid flow. 
Here $P(\rho)$ is the Pauli pressure that could be derived from the textbook thermodynamic relation
$P(\rho) = \rho \partial_\rho \epsilon(\rho) - \epsilon(\rho)$, where
 $\epsilon(\rho)$ is the specific energy of the fermion gas (energy per unit volume):
\begin{equation}
    \rho(x) = \int_{-k_F}^{k_F}\frac{dk}{2\pi} = \frac{k_F(x)}{\pi}, \qquad
    \epsilon(\rho) = \int_{-k_F}^{k_F}\frac{dk}{2\pi} \frac{k^2}{2m} =  \frac{k^3_F(x)}{6\pi m} = \frac{\pi^2\rho^3}{6m},
\end{equation}
where $k_F(x)$ is the local Fermi momentum.
The system of hydrodynamic equations ~(\ref{eq:euler_rho}, \ref{eq:euler_v}) can be diagonalized by introducing two Riemann invariants $k_{R,L} = mv \pm \pi \rho$ corresponding to the local momenta of right- and left-movers: 
\begin{equation}\label{eq:J_PDE}
\partial_t k_R +  \frac{k_R}{m} \partial_x k_R=0, \qquad
\partial_t k_L +  \frac{k_L}{m} \partial_x k_L=0,
\end{equation}
where the effective velocity of the right (left) movers is given by the group velocity of right (left) moving fermions $v_{gr}(x)=\partial_k \varepsilon_k=k_{R,L}(x)/m$. 
Eqs.~(\ref{eq:J_PDE}) are known as the Riemann-Hopf equations (or the inviscid Burgers' equations). 
The Riemann-Hopf equations (\ref{eq:J_PDE}) form a shock-wave singularity at finite time $t_{c}$, which is also known as the ``gradient catastrophe''. The semiclassical hydrodynamic equations remain valid only for evolution times $t\leq t_{c}$.
The collapse time $t_c$ depends on the density profile of the initial state: a larger amplitude of the density hump corresponds to a shorter $t_c$ due to higher non-linearity. The value of $t_c$ can be computed by solving the Riemann-Hopf Eqs.~(\ref{eq:J_PDE}) separately for left- and right-moving modes using the method of characteristics, see e.g.~Ref.~\onlinecite{whitham2011linear}.
For example, for a given initial condition $k_R(t=0, x)=f(x)$, the 
 collapse time $t_c$ corresponds to the minimal (over all choices of $\xi$) positive value of the expression $t_c = -\frac{m}{ f'(\xi)}$, which is $t_c=-\frac{m}{\min{f'(\xi)}}$ (assuming that $f'(\xi)$ takes negative values). For the equilibrium initial state (zero initial velocity), the local Fermi momenta of right- and left-movers are proportional to the local fermion density, $k_{R,L}(t=0,x)=\pm \pi\rho_0(x)$. 
 Thus, the singularity formation time is inversely proportional to the amplitude of the  density hump in the initial state, $t_c=-\frac{m}{\pi} [\min(\partial_x\rho_0(x))]^{-1}$, where $\rho_0(x)$ is the initial density profile.

After the formation of the shock wave, the fermion density profile $\rho(t,x)$ develops quantum ripples, which are not captured by semiclassical equations~\cite{mirlin2013}. 
However, at $t>t_c$ the envelope of the density profile, after averaging over quantum oscillations, can  still can be computed from semiclassical equations by using specialized PDE solvers (e.g.~Riemann solvers~\cite{godunov1959difference}), which allow one to propagate solutions beyond the shock-wave formation time.

It is instructive to provide an alternative derivation of the hydrodynamic system ~(\ref{eq:euler_rho}, \ref{eq:euler_v}), which will be straightforward to generalize to other dispersion relations.   
Taking the transport equation (\ref{eq:J_PDE}) as a starting point, we can cast it in the form of hydrodynamic equations for the fermion density $\rho$ and the velocity $v$ by expressing $\rho$ and $v$ in terms of the local Fermi momenta  of the left- and right-moving modes: $\rho=\rho(k_R, k_L)$, $v=v(k_R,k_L)$. The fermion density and the current read
\begin{equation}
\label{eq:rho_v_defin_from_k}
    \rho = \int_{k_L}^{k_R}\frac{dk}{2\pi} = \frac{k_R-k_L}{2\pi}, \quad     j \equiv\rho v = \int_{k_L}^{k_R} \frac{dk}{2\pi} \partial_k \varepsilon(k) = \frac{1}{2\pi}\left(\varepsilon(k_R)-\varepsilon(k_L)\right).
\end{equation}
Substituting the parabolic dispersion $\varepsilon(k)=k^2/2m$ into Eq.~(\ref{eq:rho_v_defin_from_k}), solving for the momenta of the left- and right-moving modes in terms of the fermion density and velocity, $k_{R,L}(\rho, v)=mv\pm \pi\rho$, and substituting into the transport equation (\ref{eq:J_PDE}), we obtain
\begin{equation}
\label{eq:pm_kR_kL}
    \partial_t \left(mv\pm \pi\rho\right) + \frac{1}{m}\left(mv\pm \pi\rho\right)\partial_x\left(mv\pm \pi\rho\right)=0.
\end{equation}
By adding and subtracting the $(\pm)$ equations in (\ref{eq:pm_kR_kL}), we obtain the continuity and Euler equations, Eqs.~(\ref{eq:euler_rho}, \ref{eq:euler_v}).

Now let us consider the quench dynamics where the initial state is prepared by applying a smooth localized potential, e.g. 
\begin{equation}
\label{eq:V0_gauss}
V(x) = V_0 e^{-(x-x_0)^2/\sigma^2},
\end{equation}
and then setting the potential to zero at $t>0$.
The initial density profile in the Thomas-Fermi approximation reads
\begin{equation}
    \rho_{TF}(t=0, x) = \frac{1}{\pi}\sqrt{2(E-V(x))}, \qquad E = \frac{k_F^2(\infty)}{2m},
\end{equation}
which can be explicitly obtained from  Eq.~(\ref{eq:euler_v}) when setting $v\to 0$ and integrating the rhs over $x$.

\begin{figure}[h!]
\centering
    \includegraphics[scale=0.3]{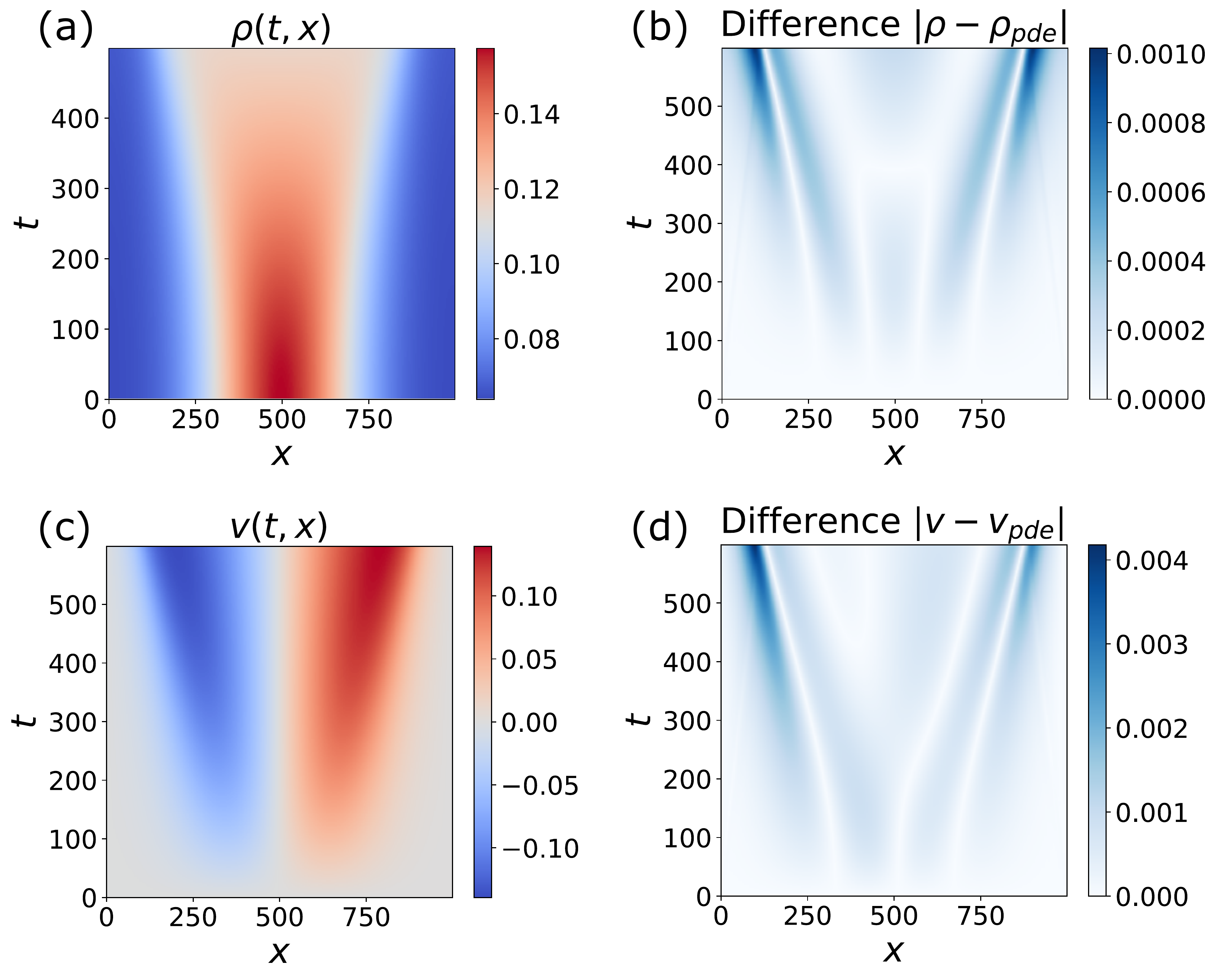}
    \caption{Hydrodynamics in the 1D free-fermion system. (a,c) Exact evolution of fermion density $\rho(t,x)$ and velocity $v(t,x)$ in the tight-binding model. (b, d) The difference between the data and the solutions of recovered hydrodynamic PDEs ~(\ref{eq:euler_recovered_rho}, \ref{eq:euler_recovered_v}). Parameters of simulations: number of lattice sites $L=1000$, filling factor $\nu=0.1$, and initial potential $V(x) = V_0 e^{-(x-x_0)^2/\sigma^2}$ with $V_0/J=-0.2$, $\sigma = 0.2 L$.  Periodic boundary conditions were imposed. The data for the density and velocity is the same as in Fig.~1 of the main text. }
    \label{fig:hydro_vs_exact}
\end{figure}

Our goal is to reconstruct hydrodynamic equations describing the evolution of $\rho(t, x)$ and $v(t, x)$ directly from data obtained via numerical simulations.
We search for hydrodynamic equations of the form
\begin{eqnarray}
\label{eq:rho_dict_hydro}
\rho_t = F(1, \rho, \rho_x, \rho_{xx}, v, v_x, v_{xx}, \rho v_x, v \rho_x, v v_x, \rho \rho_x, \ldots),\\
\label{eq:v_dict_hydro}
v_t = G(1, \rho, \rho_x, \rho_{xx}, v, v_x, v_{xx}, \rho v_x, v \rho_x, v v_x, \rho \rho_x, \ldots).
\end{eqnarray}
From data for the fermion density and velocity presented in Fig.~\ref{fig:hydro_vs_exact}, we reconstruct the system of hydrodynamic PDEs. Both the BruteForce algorithm and the STRidge algorithm result in
\begin{eqnarray}
\label{eq:euler_recovered_rho}
&\rho_t& + 1.006 \rho v_x + 1.0007 v \rho_x = 0,
\\
\label{eq:euler_recovered_v}
&v_t& + 0.97 v v_x + 9.45 \rho \rho_x = 0,
\end{eqnarray}
 which is very close to the expected Eqs.~(\ref{eq:euler_rho}, \ref{eq:euler_v}).
In Fig.~\ref{fig:hydro_vs_exact}, we also compare 
the data  and the solution of the inferred system of PDEs~(\ref{eq:euler_recovered_rho}, \ref{eq:euler_recovered_v}).

\begin{figure}[H]
     \centering
     \includegraphics[scale=0.3]{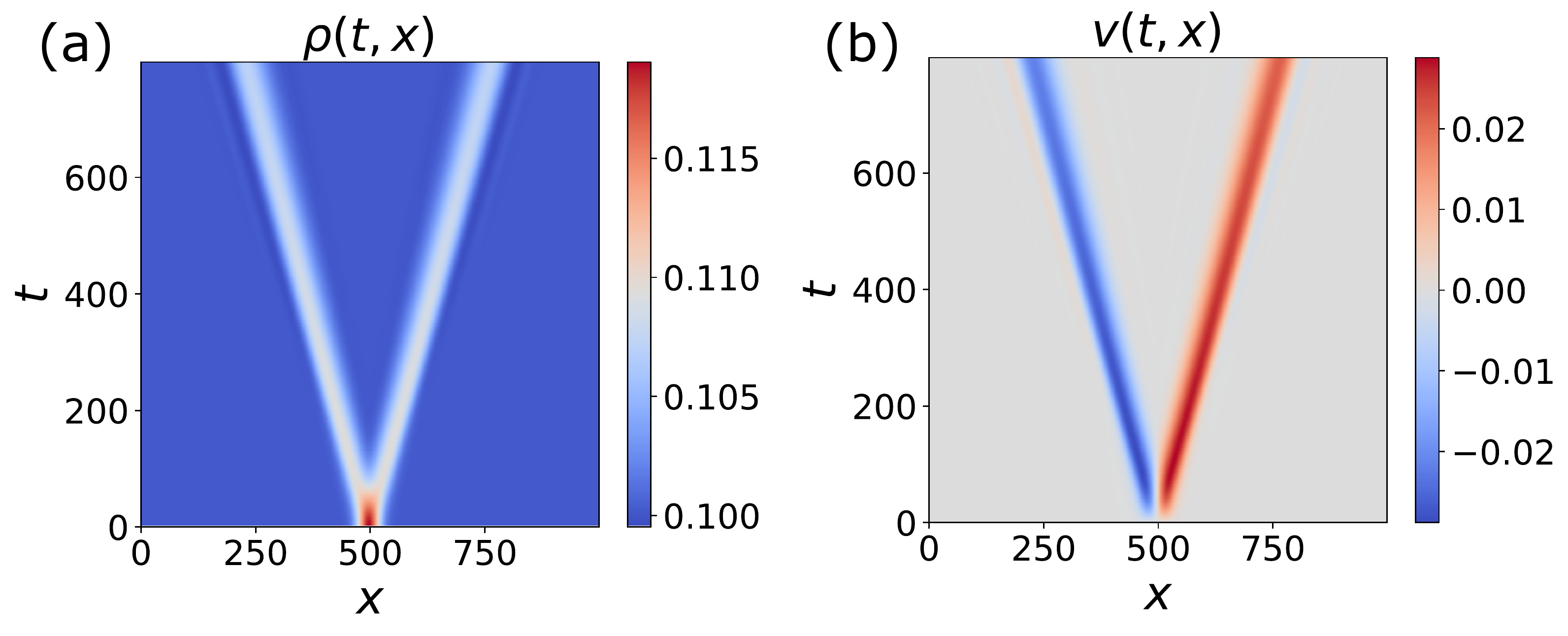}
     \caption{Regime of linearized hydrodynamics ($\delta\rho/\rho_0 \ll 1$) in non-interacting fermion gas: data for the evolution of (a) fermion   density and (b) fermion velocity. PDE reconstruction results in the system of linearized hydrodynamic equations~(\ref{eq:lin_hydro_reconstr}). The linearized system is equivalent to the  wave equation $\rho_{tt}-v_F^2 \rho_{xx}=0$, with the wave speed given by the Fermi velocity $v_F=\pi\rho_0$. The amplitude of the external potential is $V_0/J=-0.02$. }
     \label{fig:fermion_wave_eq}
 \end{figure}

Interestingly, in the case of a small amplitude of the initial density hump, $\delta\rho/\rho_0\ll 1$, our PDE reconstruction algorithm recovers the correct form of linearized  Euler equations:
 \begin{equation}
 \label{eq:linearized_euler}
     \begin{cases}
      \rho_t + \rho_0 v_x \approx 0,\\
       v_t + \pi^2\rho_0 \rho_x \approx 0.
    \end{cases}
 \end{equation}
These equations, in turn, imply the wave equation $\rho_{tt}=v_F^2 \rho_{xx}$, with the wave speed equal to the Fermi velocity, $v_F=\pi\rho_0$. For example, for the data presented in Fig.~(\ref{fig:fermion_wave_eq}), the BruteForce algorithm yields
 \begin{equation}
     \rho_t \approx -0.109 v_x, \quad v_t \approx -1.036 \rho_x,
     \label{eq:lin_hydro_reconstr}
 \end{equation}
which is in perfect agreement with the linearized system (\ref{eq:linearized_euler}) for $\rho_0\approx 0.1$.
 
\subsection{Corrections to hydrodynamic equations due to the tight-binding dispersion}\label{sec:cos_corrections}
In this subsection, we consider how hydrodynamic equations (\ref{eq:euler_rho}) and (\ref{eq:euler_v}) are modified due to corrections generated by subleading terms in the expansion of the dispersion relation $\varepsilon_k=-2J \cos{(k)}=2J\left(-1+\frac{k^2}{2}-\frac{k^4}{4!}+\ldots\right)$ and analytically derive these correction terms, which we discovered with our PDE-learning algorithm [see main text].

Following the steps from  Section~\ref{sec:fermion_hydro_parabol}, we first express the momenta of left- and right-movers via fermion density and velocity:
\begin{eqnarray}\label{eq:k_RL}
\frac{k_R-k_L}{2} = \pi \rho, \qquad
\frac{k_R+k_L}{2} = \sin^{-1}{\left[ \frac{  \pi \,\rho\, m\,v}{  \sin{\pi\rho}}\right]},
\end{eqnarray}
where $m=1/(2J)$. 
Since we are interested in finding corrections stemming from the deviation of the tight-binding dispersion from the parabolic dispersion in the limit $k_{R,L}\ll 1$, we assume that the fermions occupy the bottom of the band, $\rho\ll 1$. Furthermore, given that, to the leading order (i.e.~in parabolic approximation), the momenta of right- and left- movers are $k_{R,L}= mv\pm \pi\rho$, the condition $k_{R,L}\ll 1$ also implies that the fermion velocity should be small, $mv\ll 1$. 

Solving for $k_{R,L}(\rho, v)$ from Eq. (\ref{eq:k_RL}), we expand the solution in powers of $\rho$ and $v$. Keeping the terms $\propto v^{n_1} \rho^{n_2}$ with $n_1+n_2\leq 5$, we obtain: 
\begin{equation}
    k_{R,L} = \pm \pi\rho + \sin^{-1}{\left[ \frac{   \pi \rho\, m\,v}{ \sin{\pi\rho}}\right]} = \pm \pi\rho + m v + \frac{(mv)^3}{6} + \frac{3m^5}{40}v^5 + \frac{\pi^2}{6 } m v\rho^2 + \frac{\pi^2 m^3}{12}v^3\rho^2 + \frac{7 \pi^4 }{360}mv\rho^4 + \ldots.
\end{equation}
Dynamical equations for the momenta of the left- and right-movers read
\begin{eqnarray}\label{eq:kR_t}
&&\partial_t k_{R} + \frac{1}{m}\sin{(k_R)}\, \partial_x k_R = 0,\\
&&\partial_t k_{L} + \frac{1}{m}\sin{(k_L)}\, \partial_x k_L = 0.
\label{eq:kL_t}
\end{eqnarray}

First, taking the difference of Eqs.~(\ref{eq:kR_t}, \ref{eq:kL_t})
and using the definition of particle current (\ref{eq:rho_v_defin_from_k}),
\begin{equation}\label{eq:rho_v}
\rho v = - \frac{1}{2 m \pi } \left(\cos{k_R}-\cos{k_L}\right),
\end{equation}
one can recover  the exact continuity equation $\rho_t+(\rho v)_x= 0$, which remains valid to all orders in perturbation theory.   
Second, multiplying both sides of Eqs.~(\ref{eq:kR_t}, \ref{eq:kL_t}) by $\sin{(k_{R,L})}$, respectively, we obtain
\begin{eqnarray}\label{eq:cos_kRL_t}
-\partial_t \cos{(k_\alpha)} +\frac{1}{m}\partial_x \left(\frac{k_\alpha}{2}-\frac{1}{4}\sin{(2 k_\alpha)}\right)  = 0, \quad \alpha = R, L.
\end{eqnarray}
Subtracting the two Eqs.~(\ref{eq:cos_kRL_t}) for right- and left- movers and using (\ref{eq:rho_v}), we arrive at the dynamical equation for the density of momentum:
\begin{eqnarray}
(\rho v)_t &=& - \frac{1}{2  \pi m^2}\partial_x \left(\frac{k_R-k_L}{2}-\frac{\sin{(2 k_R)}-\sin{(2 k_L)}}{4}\right)  =\nonumber \\ 
&&- \frac{1}{2 \pi m^2 } \partial_x \left(2m^2\pi v^2 \rho + \frac{2\pi^3}{3}\rho^3 - \frac{2\pi^3}{3}m^2v^2\rho^3 - \frac{2\pi^5}{15}\rho^5 + \ldots\right) = 0,
\end{eqnarray}
where we performed Taylor expansion of $\sin{(\cdot)}$ up to the fifth order.
Substituting  $\rho_t=-(
\rho v)_x$ from the continuity equation, 
we obtain a modified Euler equation with additional correction terms:
\begin{eqnarray}
&&v_t + v v_x + \frac{\pi^2}{m^2}\rho\rho_x = \frac{2\pi^2}{3}\rho^2 v v_x + \pi^2 v^2 \rho \rho_x + \frac{\pi^4}{3 m^2}\rho^3 \rho_x + \ldots.
\label{eq:v_t_correct}
\end{eqnarray}

If we set $\rho_0 \approx 0.1$ and $m=1$ ($J=0.5$), which corresponds to the parameter regime used in Fig.~\ref{fig:hydro_vs_exact},
the signs and the magnitude of the corrections are in full agreement with the PDE (\ref{eq:euler_recovered_v}) extracted from direct simulations, as we will now discuss.
One can perform dimensional analysis of equation (\ref{eq:v_t_correct}) by noticing that, for our choice of units (we set the lattice spacing to unity), we have $[v]=J$, $[\rho]=1$, $[x]=1$, $[t]=J^{-1}$.
We performed a BruteForce search of corrections to the rhs of the Euler equation by considering the symmetry allowed terms, $(P,T)=(-,+)$, from Table~\ref{table:t2}:
\begin{equation}
    v_t + v v_x + \pi^2\rho\rho_x = f(\rho^2\rho_x, \rho^3\rho_x, \ldots, \rho^5\rho_x, \rho v v_x, \rho^2 v v_x, v^2\rho_x, v^2\rho \rho_x, \ldots,  (\log{\rho})_x, v^2 (\log{\rho})_x ).
\end{equation}
We found the following corrections using data shown in Fig.~\ref{fig:hydro_vs_exact}(a,c) [the dataset has the spatiotemporal resolution $N_t,N_x=(10^3, 10^3)$]:
\begin{eqnarray}
v_t+v v_x +\pi^2\rho\rho_x \approx  1.003\, \frac{2\pi^2}{3} \rho^2 v v_x + 1.18\, \pi^2 v^2\rho \rho_x + 0.967\, \frac{\pi^4}{3} \rho^3 \rho_x, 
\qquad (\lambda_0 = 10^{-5}),
\end{eqnarray}
which are in excellent agreement with the analytical result, Eq.~(\ref{eq:v_t_correct}).
Moreover, starting from the generic ansatz $v_t=G(\cdot)$, our algorithm was able to recover the entire series of terms, including the leading terms ($v v_x$, $\rho\rho_x$) and the subleading corrections ($\rho^2 v v_x$, $v^2\rho \rho_x$, $\rho^3 \rho_x$):
\begin{equation}
\label{eq:vt_corrections_reconstr}
    v_t = - 1.008\, v v_x - 0.9993\, \pi^2 \rho \rho_x = 
    1.10\, \frac{2\pi^2}{3} \rho^2 v v_x+
    1.05\, \pi^2 v^2 \rho\rho_x + 0.96\, \frac{\pi^4}{3} \rho^3\rho_x, \qquad (\lambda_0 = 10^{-6}).
\end{equation}
The high spatiotemporal resolution of the data results in a high accuracy of the reconstructed coefficients of the hydrodynamic equation (\ref{eq:vt_corrections_reconstr}).

\subsection{Fermion hydrodynamics at the Lifshitz transition}
\label{sec:fermion_hydro_quartic}
In this subsection, we provide the analytical derivation of semiclassical hydrodynamic equations in a fermion gas with  quartic dispersion and present additional details of  PDE-learning.

Now we extend the tight-binding model (\ref{eq:H_tight_binding}) by adding next-nearest-neighbor hopping terms:
\begin{equation}
\label{eq:H_j1_j2}
     H = -J_1 \sum_i (c^\dag_i c_{i+1} + c^\dag_{i+1} c_i) - J_2 \sum_i (c^\dag_i c_{i+2} + c^\dag_{i+2} c_i).
\end{equation}
Fermion dispersion is $\varepsilon_k = - 2J_1 \cos{(k)} - 2 J_2 \cos{(2k)}$. In the long-wavelength limit, we can perform an expansion up to the fourth order in $k$:
\begin{equation}
    \label{eq:p2_p4_dispersion}
    \varepsilon_k = \varepsilon_0 + \frac{\alpha k^2}{2} + \frac{\beta k^4}{4} + \mathcal{O}(k^6),
\end{equation}
where $\alpha = 2(J_1+4 J_2)$ and $\beta=-\left(\frac{J_1}{3}+\frac{16}{3}J_2\right)$.
\begin{figure}[H]
    \centering
    \includegraphics[scale=0.3]{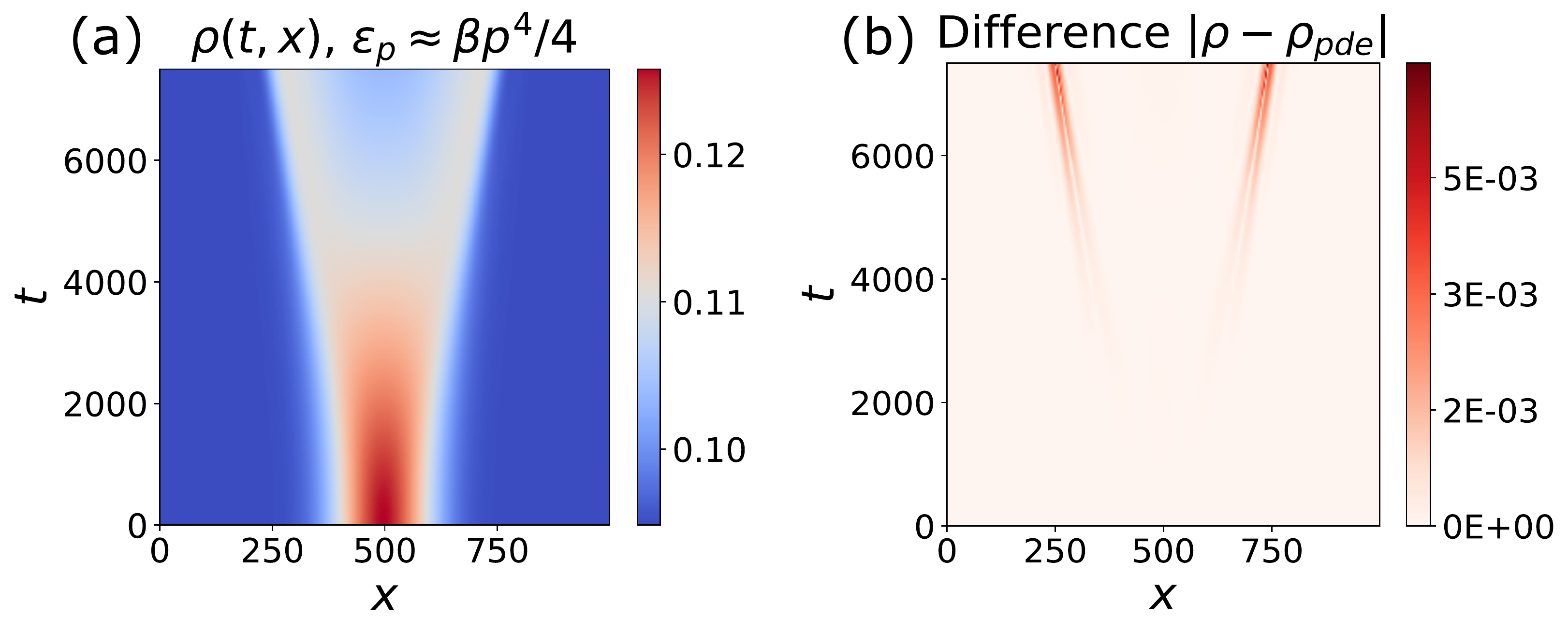}
    \caption{Fermion hydrodynamics at the Lifshitz critical point ($J_2/J_1=-0.25$), which is characterized by the quartic dispersion, $\varepsilon_k \approx  \frac{\beta k^4}{4}$. Panel (a) displays exact simulations of evolution of fermion density, while panel (b) shows the difference between the data and the solution of the recovered PDE (\ref{eq:vt_reconstr_j1j2_main}). The initial state corresponds to the ground state in the Gaussian-shaped potential Eq.~(\ref{eq:V0_gauss}), with amplitude $V_0/J=-4\cdot 10^{-3}$ and width $\sigma=0.1 L$. At large times $t \gtrsim 6\cdot 10^{3}$, a shock wave starts to form, and the semiclassical hydrodynamic approximation breaks down.}
    \label{fig:hydro_p4}
\end{figure}
The hydrodynamic equation for the generalized dispersion with quadratic and quartic terms (\ref{eq:p2_p4_dispersion}) can be derived in analogy with Eq.~(\ref{eq:J_PDE}), considering separately right- and left-movers:
\begin{equation}
\label{eq:k_RL_p2_p4}
\partial_t k_R + (\alpha k_R + \beta k_R^3)\partial_x k_R = 0, \quad \partial_t k_L + (\alpha k_L + \beta k_L^3)\partial_x k_L = 0.
\end{equation}
To derive the hydrodynamic equations, we proceed by analogy with Section~\ref{sec:fermion_hydro_parabol}. 
The expressions for the fermion density and current read:
\begin{equation}
\label{eq:rho_v_defin_j1j2}
    \rho(t,x) = \int_{k_L}^{k_R}\frac{dk}{2\pi} = \frac{k_R-k_L}{2\pi}, \quad    \rho(t,x) v(t,x) = \int_{k_L}^{k_R} (\alpha k + \beta k^3) \frac{dk}{2\pi} = \frac{\alpha}{4\pi} (k^2_R - k^2_L) +  \frac{\beta}{8\pi} (k_R^4-k^4_L).
\end{equation}
From Eq.~(\ref{eq:rho_v_defin_j1j2}) and Eq.~(\ref{eq:k_RL_p2_p4}), we obtain the exact continuity equation:
\begin{equation}
    \rho_t + (\rho v)_x = 0.
\end{equation}
To express $k_{R,L}$ in terms of $\rho$ and $v$, we need to solve the following system of equations:
\begin{equation}
\label{eq:rho_v_j1_j2}
 \rho = \frac{k_R-k_L}{2\pi}, \quad
 v = \frac{1}{2}\left[\alpha + \frac{\beta}{2}(k_R^2+k_L^2)\right](k_R+k_L).
\end{equation}
Unfortunately, in order to express $k_{R,L}$ in terms of hydrodynamic variables $(\rho, \, v)$, one has to solve a qubic equation (\ref{eq:rho_v_j1_j2}). 
To simplify the problem, we consider the limit $\alpha\to 0$, corresponding to the quartic dispersion $\varepsilon_k = \beta k^4/4$.
Solving Eq.~(\ref{eq:rho_v_j1_j2}) for $k_{R,L}$ and expanding the solution as a Taylor series in powers of $v/v_F$, we obtain
\begin{equation}
\label{eq:k_RL_expansion}
    k_{R,L} = \pm \pi \rho + \frac{v}{\beta \pi^2 \rho^2} + \mathcal{O}(v^3/v_F^3).
\end{equation}
Here $v_F=\partial_k \varepsilon_k|_{k=k_F}=\beta \pi^3  \rho^3$ is the Fermi velocity of a Fermi gas with quartic dispersion.
Substituting Eq.~(\ref{eq:k_RL_expansion}) into Eq.~(\ref{eq:k_RL_p2_p4}) and keeping terms up to the second-order in $v$, we obtain
\begin{eqnarray}
&v_t& + 5 v v_x - v^2\frac{\rho_x}{\rho} + \beta^2 \pi^6 \rho^5 \rho_x = 0. 
\end{eqnarray}
The initial density profile could be derived in the Thomas-Fermi approximation by imposing a condition that the Fermi energy is  constant:
\begin{equation}
    \rho_{TF}(t=0, x) = \frac{1}{\pi}\left[\frac{4}{\beta}(E - V(x))\right]^{\frac{1}{4}}.
\end{equation}
We perform PDE reconstruction of the equation  using the following dictionary of candidate terms: 
\begin{equation}
\label{eq:v_dict_pde_j1j2}
v_t = G(\rho, \rho_x, \rho\rho_x, \rho^2\rho_x, \ldots, \rho^5\rho_x, \rho_{xx}, v, v_x, v_{xx}, v v_x, v^2 \rho_x, (\log{\rho})_x, v^2 (\log{\rho})_x, \ldots ),    
\end{equation}
which contains terms up to order $v^2$. In order to constrain the search space, we remove terms that do not satisfy the $P$- and $T$-inversion symmetry constraint $(P,T) = (-,+)$.
Performing such preselection, we composed the dictionary $G(\cdot)$ consisting of $M=20$ candidate terms marked with a check in Table~\ref{table:t2}. 
The PDE recovered from data presented in Fig.~\ref{fig:hydro_p4}(a)  using the BruteForce and the CrossEntropy algorithms reads
\begin{equation}
\label{eq:vt_reconstr_j1j2}
    v_t \approx -4.98 v v_x - 225.7 \rho^5 \rho_x, \quad  (J_1=0.5,\, J_2=-0.125,\, \beta=0.5).
\end{equation}
The  term $\sim v^2 \rho_x/\rho$ is missing in the recovered Eq.~(\ref{eq:vt_reconstr_j1j2}); however, this term turns out to be negligible in the regime of parameters considered here.
In Fig.~\ref{fig:hydro_p4}(b), we compare the solution of the inferred PDE (\ref{eq:vt_reconstr_j1j2}) with the original data. 
We find that adding the $L_2$ regularization term $\lambda_2$ to the loss function $\mathcal L = ||U_t - \Theta \cdot \xi||_2 + \lambda_0 ||\xi||_0 + \lambda_2 ||\xi||_2^2$ stabilizes the regression problem in the presence of highly nonlinear terms. Without $L_2$ regularization,  we obtain extremely large values of the regression coefficients $\xi$. On the other hand, even very small value for the penalty constant, such as $\lambda_2=10^{-12}$, suffices. Interestingly, the STRidge algorithm  was not able to identify a correct PDE.

\subsection{Fermion current and velocity in lattice simulations}\label{sec:particle_current}
In this subsection we derive the expression for the particle current and velocity for the 1D tight-binding Hamiltonian with/without interactions.

Let us consider a subsystem cut between the lattice sites $i=i_0-1$ and $i=i_0$. The current between these two parts is defined via the particle number conservation, i.e.
\begin{equation}\label{eq:j_ham}
j:=-\left\langle \frac{d N_{L}}{d t}\right\rangle=-i\left\langle\left[H, N_{L}\right]\right\rangle
\end{equation}
where $N_{L}=\sum_{i<i_0} c_{i}^{\dagger} c_{i}$ is the number of particles to the left from the cut, $H$ is the system's Hamiltonian, and $\langle\ldots\rangle$ is the expectation value. 
We consider the tight-binding Hamiltonian (\ref{eq:H_j1_j2}) with the nearest-neighbour and next-nearest-neighbour hopping terms. 
The particle current can be expressed as
\begin{equation}\label{eq:j_commutator}
j(t, i_0)=-i \sum_{i<i_0, n} J_{1}\left\langle\left[c_{n}^{\dagger} c_{n+1}, c_{i}^{\dagger} c_{i}\right]\right\rangle+J_{1}\left\langle\left[c_{n+1}^{\dagger} c_{n}, c_{i}^{\dagger} c_{i}\right]\right\rangle+J_{2}\left\langle\left[c_{n}^{\dagger} c_{n+2}, c_{i}^{\dagger} c_{i}\right]\right\rangle+J_{2}\left\langle\left[c_{n+2}^{\dagger} c_{n}, c_{i}^{\dagger} c_{i}\right]\right\rangle. 
\end{equation}
The expectation value of the commutator reads
\begin{equation}
\sum_{i<i_0}\left\langle\left[c_{a}^{\dagger} c_{b}, c_{i}^{\dagger} c_{i}\right]\right\rangle 
=\sum_{i<i_0}\left(\delta_{i b}-\delta_{i a}\right)\left\langle c_{a}^{\dagger} c_{b}\right\rangle=\left\{\begin{array}{ll}
-\left\langle c_{a}^{\dagger} c_{b}\right\rangle, & a<i_0,\, b \geq i_0, \\
\left\langle c_{a}^{\dagger} c_{b}\right\rangle, & b<i_0,\, a \geq i_0, \\
0, & \text { otherwise.}
\end{array}\right.
\end{equation}
Thus, the current at the cut between the sites $i-1$ and $i$ has the following form 
\begin{equation}\label{eq:j_definit}
j(t,i)=i J_{1}\left[\mathcal{G}_{i,i-1}(t)-\mathcal{G}_{i-1,i}(t)\right]+i J_{2}\left[\mathcal{G}_{i,i-2}(t)-\mathcal{G}_{i-2,i}(t)\right]+i J_{2}\left[\mathcal{G}_{i+1,i-1}(t)-\mathcal{G}_{i-1,i+1}(t)\right],
\end{equation}
where $\mathcal{G}_{a b}(t):=\left\langle c_{b}^{\dagger}(t) c_{a}(t)\right\rangle$ is the single-particle density matrix. The fermion velocity is related to the particle current as $v(t,i)=j(t,i)/\rho(t,i)$, where $\rho(t,i)=\left\langle c^\dag_i c_i \right\rangle$ is the on-site fermion density. We employ the expression~(\ref{eq:j_definit}) to construct datasets for the fermion velocity field for the PDE-learning algorithm both in the main text and throughout Section~\ref{sec:fermions} of the Supplementary Material.
Note, that the expression for the current~(\ref{eq:j_definit}) remains unchanged in the presence of the Fermi-Hubbard-type interactions, e.g. $V_{\rm int}=U \sum_i n_i n_{i+1}$, since the interaction term $V_{\rm int}$ commutes with the fermion number operator $n_i=c^\dag_i c_i$,
and therefore does not contribute to the particle current in~(\ref{eq:j_ham}). 
We use this fact later in Sections~\ref{sec:inter_ferm_hydro} and~\ref{sec:viscosity} when performing PDE-learning of hydrodynamics of interacting fermions.

 
\subsection{Global symmetries and term preselection}\label{sec:global_symmetries}

Prior knowledge of global symmetries, such as invariance with respect to time-reversal $(T)$/spatial-inversion $(P)$ transformations,  provides a powerful method to significantly reduce the number of candidate terms in the dictionary.
The transformation properties of the fermionic density and velocity are:
\begin{equation}
    P(\rho)=1, \quad T(\rho)=1, \quad P(v)=-1, \quad T(v)=-1.
\end{equation}
The summary of the preselection procedure for fermionic systems obeying $P$- and $T$-inversion symmetries  is presented in Table~\ref{table:t2}.
\begin{center}
\begin{table}[htbp]
\caption{Example of candidate terms for the rhs of the Euler equation $v_t = G(\cdot)$, Eq.~(\ref{eq:v_dict_pde_j1j2}). 
Selected terms (see last column) have the following signature with respect to $P$- and $T$-inversion: $(P,T)=(-,+)$.}
 \label{table:t2}
\begin{tabular}{|c|c|c|c|}
\hline
\textrm{Terms} & $P$ & $T$ & \textrm{Select}\\
\hline
$const$, $\rho$, $\rho^2$, $\rho^3$, $\ldots$, $\rho^5$,  $v^2$ & + & + & $\times$\\
$\rho_x$, $\rho \rho_x$, $\ldots$ $\rho^5\rho_x$ & - & + & \checkmark\\
$\rho_{xx}$, $\rho \rho_{xx}$,  $\rho^2 \rho_{xx}$, $\ldots$, $v^2 \rho_{xx}$ & + & + & $\times$\\
$v^2\rho$, $v^2 \rho_{xx}$ & + & + & $\times$\\
$\rho_x v_x$, $v_{xx}$, $\rho v_{xx}$, $\rho^2 v_{xx}$& - & - & $\times$\\
$\rho_x^2$, $v_x^2$ & + & + & $\times$\\
$v$, $\rho v$, $\rho^2 v$, $\rho^3 v$ & - & - & $\times$\\
$v_x$, $\rho v_x$, $\rho^2 v_x$ & + & - & $\times$\\
$v v_x$, $\rho v v_x$, $\rho^2 v v_x$, $\ldots$, $\rho^5 v v_x$ & - & + & \checkmark\\
$v^2 \rho_x$, $v^2 \rho \rho_x$, $\ldots$, $v^2 \rho^5 \rho_x$, $(\log{\rho})_x$, $v^2 (\log{\rho})_x$ & - & + & \checkmark\\
 \hline
\end{tabular}
\end{table}
\end{center}

\subsection{Learning of hydrodynamic PDEs from partial observations}\label{sec:partial_observ}
In an experimental setting, it is quite common that only some  physical observables can be directly measured. In this subsection, we propose two approaches for PDE-learning of hydrodynamic equations from partial observations, when only data for the evolution of density (but not velocity) is available.
Such a situtation is common in ultracold-atom experiments, since it is relatively easy to measure the density of the atomic cloud via optical absorption, but it is hard to directly measure the velocity field of the atomic cloud in situ.

The evolution of density between different time snapshots could be considered as a ``movie'' that contains information about particle velocity at each spatial point.
The velocity field can then be extracted by integrating  the continuity equation, Eq.~(\ref{eq:euler_rho}):
\begin{equation}\label{eq:v_hidden}
    v(t,x) = -\frac{1}{\rho(t,x)} \partial_t \left[\int_{-\infty}^x dy\, \rho(t, y)\right].
\end{equation}
The right-hand side of Eq.~(\ref{eq:v_hidden}) can be directly evaluated from the data at spatiotemporal points of interest. 
After the extraction of the velocity field $v(t,x)$, we can proceed with the standard PDE-reconstruction procedure of the hydrodynamic equation for the velocity, described in Section~\ref{sec:fermion_hydro}. 
Applying the method described above to the $\rho(t,x)$ data shown in Fig.~\ref{fig:hydro_vs_exact}, we reconstruct the Euler equation from the library of candidate terms $v_t=G(\cdot)$, Eq.~(\ref{eq:v_dict_hydro}),  using the BruteForce, CrossEntropy, and STRidge algorithms (all three algorithms leading to the same result),
\begin{equation}
    v_t + 0.97 v v_x + 9.42 \rho \rho_x = 0,
\end{equation}
which is in good agreement with the theoretically expected equation (\ref{eq:euler_v}) and nearly identical to Eq.~(\ref{eq:euler_recovered_v}), where both density and velocity data were used for PDE learning.

The method presented above efficiently solves the problem of reconstructing the Euler equation for the velocity $v_t=G(\cdot)$ from partial observations (only from the density data $\rho(t,x)$).
However, this method has a few drawbacks. (i) 
The velocity reconstruction procedure via Eq.~(\ref{eq:v_hidden}) introduces additional numerical errors due to the finite-difference computation of the time-derivative $\partial_t [\ldots]$. (ii) The situation worsens in the regions when the density approaches zero, $\rho\to 0$, resulting in a vanishing  denominator in Eq.~(\ref{eq:v_hidden}), that amplifies numerical errors.
(iii) The velocity reconstruction trick (\ref{eq:v_hidden}) works only if the continuity equation is valid. In problems where the total number of particles is not conserved (e.g.~in the presence of three-body loss in cold-atom experiments), the continuity equation is no longer exact and has to be modified with appropriate loss terms. 

Problem (ii) can be partially alleviated by the considering hydrodynamic equation for the particle current,
\begin{equation}
    (\rho v)_t + (\rho v^2)_x = - \partial_x P(\rho),
\end{equation}
so that the we search for an unknown equation of the form $(\rho v)_t = G(\cdot)$. We utilized this method  in the main text for extracting PDEs from experimental data (boson gas expansion on an atom chip).

In addition, we propose a  modified PDE learning method to address problems (ii) and (iii).
We assume that  
\begin{eqnarray}
&\rho_t&  = F(\xi_1; \rho, v, \rho_x, \rho_{xx}, v_x, v_{xx}, \ldots),\\
&v_t& = G(\xi_2; \rho, v, \rho_x, \rho_{xx}, v_x, v_{xx}, \ldots),
\end{eqnarray}
where $\xi_{1,2}$ are the coefficients that parametrize functions $F$ and $G$.
We define the objective function as 
\begin{equation}\label{eq:obj_func_pdesolve}
\mathcal{L}(\xi_{1,2}; \rho^*, \lambda_0) = \sum_{t_k, x_i} \left|\rho^{*}(t_k, x_i) - \textrm{PDESolve}( t_k, x_i, \rho_0, v_0, F(\xi_1; \cdot), G(\xi_2; \cdot), BC\right| + \lambda_0 ||\xi_2||_0,
\end{equation}
where $\rho_0$ and $v_0$ are the initial conditions at $t=0$, $BC$ represents a set of boundary conditions, and $\rho^*(t_k, x_i)$ are  the data points for the density evaluated at the spatiotemporal grid $\{t_k, x_i\}$. $\textrm{PDESolve}(\ldots)$ denotes a PDE solver that takes initial and boundary conditions and the coefficients $\xi_{1,2}$ parametrizing the unknown function $f(\xi_{1,2}; \cdot)$ and outputs a solution $\rho_{pde}$ at the grid points $\{t_k, x_i\}$.
We assume that we know the initial conditions for the velocity [in quench experiments with ultracold atoms, one usually has $v(t=0, x)=0$].
The goal is to find a set of coefficients $\xi_{1,2}$ that minimize the objective function:
\begin{equation}\label{eq:xi_12_solver}
    \xi_{1,2}  = \argmin_{\xi_{1,2}} (\mathcal{L}(\xi_{1,2}; \rho^*, \lambda_0)).
\end{equation}
Optimization of Eq.~(\ref{eq:xi_12_solver}) is more computationally costly compared to the sparse-regression methods discussed in Sec.~\ref{sec:sparse_regr}. 
The former approach requires  (i) discrete optimization to find optimal combinations of non-zero terms, (ii) continuous optimization to find optimal values of PDE coefficients $\xi_{1,2}$ (e.g.~by gradient-descent algorithm), (iii) solution of the PDE forward in time for each optimization step in order to evaluate the current value of the objective function (\ref{eq:obj_func_pdesolve}).
Computational cost of this algorithm can be significantly improved by combining PDE solvers and reverse mode automatic differentiation~\cite{yashchuk2020bringing}.

\subsection{Could the evolution of the density of non-interacting fermions be described by a single PDE?}\label{sec:single_pde_rho}
In this subsection, we address the following question: is it possible to rewrite the coupled system of hydrodynamics equations describing fermion dynamics,
\begin{equation}
    \begin{cases}
    \rho_t+(\rho v)_x=0,\\
    v_t + v v_x + \pi^2 \rho \rho_x = 0,
    \label{eq:euler_system}
    \end{cases}
\end{equation}
as a single closed form PDE for the fermion density? The answer is positive, although surprisingly we did not find such an equation in the literature.
We introduce an auxiliary variable $w$ as:
\begin{eqnarray}
    &&\rho v = - w_t,
    \label{eq:w_t}
    \\
    &&\rho = w_x,
    \label{eq:w_x}
\end{eqnarray}
so that the continuity equation is automatically satisfied.
Solving Eq.~(\ref{eq:w_x}), we obtain $w(t,x) = \int_0^x dx'\, \rho(t,x') + g(t)$, where it is easy to show that $g(t)=const$ due to fixed boundary conditions at $x\to \pm \infty$.
The physical meaning of $w(t,x)$ is the number of fermions to the left of coordinate $x$. 
Expressing velocity from Eqs.~(\ref{eq:w_t},~\ref{eq:w_x}) as $v=-w_t/w_x$ and substituting the result into the Euler equation (\ref{eq:euler_system}), we obtain
\begin{equation}
\label{eq:w_final}
    -\partial_t \left(\frac{w_t}{w_x}\right) + \frac{w_t}{w_x}\left(\frac{w_t}{w_x}\right)_x + \pi^2 w_x w_{xx} = 0.
\end{equation}
The resulting PDE (\ref{eq:w_final}) is second-order in time and depends only on the fermion density $\rho(t,x)$ via $w$.
Unfortunately, Eq.~(\ref{eq:w_final}) has no transparent physical interpretation, to our knowledge. Although PDE-learning methodology allows one to reconstruct a second-order-in-time equation $w_{tt}=F(w_t, w_x, w_{tx}, \ldots)$,  such an equation would be hard to interpret, compared to a conventional hydrodynamic system of equations for the density and the velocity.

\subsection{Hydrodynamics of interacting fermions: emergent Euler and Navier-Stokes equations}\label{sec:inter_ferm_hydro}

While in previous subsections we considered hydrodynamics in a non-interacting fermion gas, here we focus on the case of interacting fermions described by the 1D spinless Fermi-Hubbard model:
\begin{equation}\label{eq:F-H_model}
    H = - J\sum_{i} (c^\dag_i c_{i+1} + c^\dag_{i+1} c_i) + U\sum_{i} n_i n_{i+1}  - \mu \sum_i c^\dag_i c_i,
\end{equation}
where $U$ is the interaction constant, and $\mu$ is the chemical potential of the fermion gas.

\begin{figure}[h]
    \centering
    \includegraphics[scale=0.4]{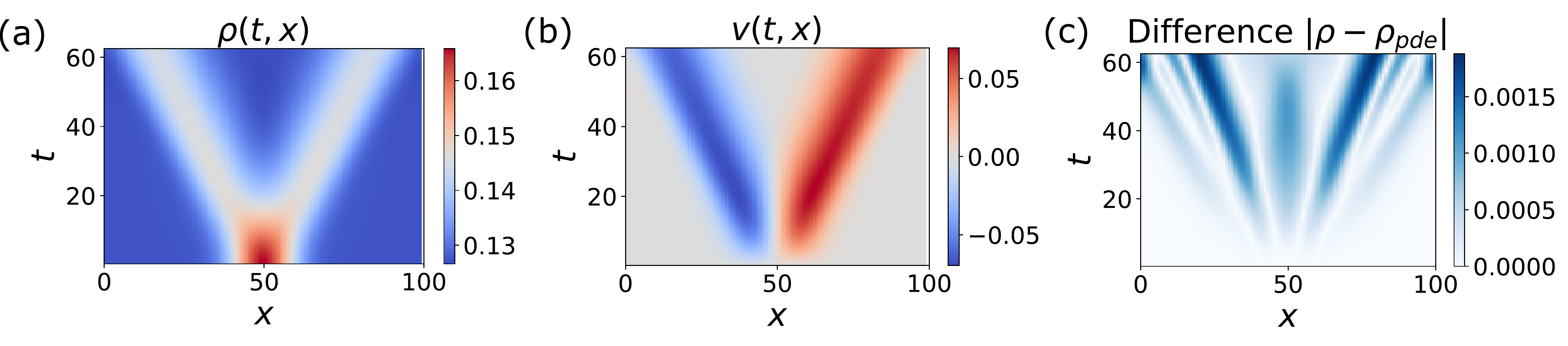}
    \caption{(a, b) TEBD data for the evolution of fermion density and velocity in the spinless Fermi-Hubbard model (\ref{eq:F-H_model}).
    (c) Difference between the solution of the reconstructed hydrodynamic PDE (``Navier-Stokes model'', Eq.~(\ref{eq:kappa_nu}))  and the TEBD data ($U/J=4$).
    In our simulations, we fixed the value of the chemical potential at $\mu/J=-2$.}
    \label{fig:navier-stokes}
\end{figure}

\begin{figure}[H]
    \centering
    \includegraphics[scale=0.3]{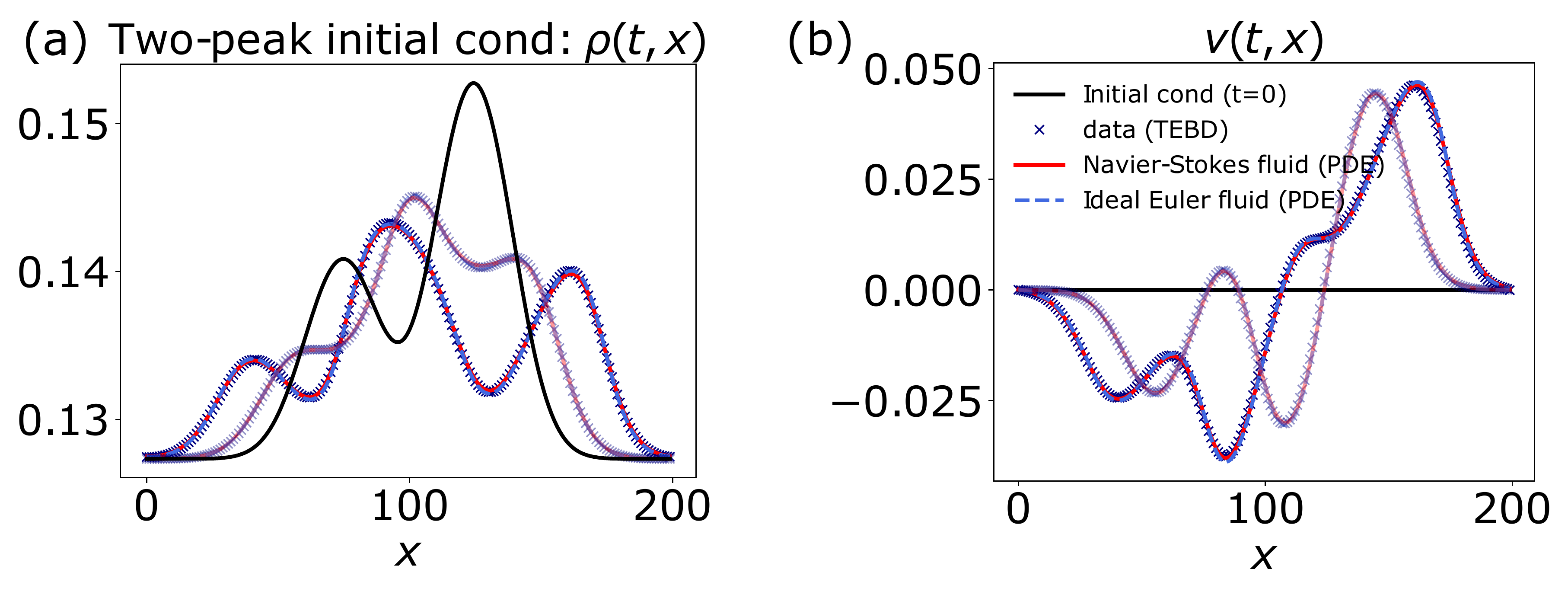}
    \caption{Hydrodynamics in the  Fermi-Hubbard model (\ref{eq:F-H_model}) in the strongly interacting regime,  $U/J=4$. (a, b) Evolution of fermion  density $\rho(t,x)$ and velocity $v(t,x)$ for an asymmetric double peak initial density profile.
    This plot demonstrates that the discovered Euler and Navier-Stokes equations (\ref{eq:kappa_term}, \ref{eq:kappa_nu}) remain valid for various initial conditions.
    }
    \label{fig:double_peak}
\end{figure}

We perform a search of hydrodynamic-type equations for the fermion density and velocity using the following form of hydrodynamic equations:
\begin{equation}
    \begin{cases}
    \rho_t+(\rho v)_x = 0,\\
    v_t + v v_x = G(\rho_x, \rho_{xx}, v_x, v_{xx}, \ldots),
    \end{cases}
\end{equation}
where the function $G(\cdot)$ is unknown. We use a truncated library of terms corresponding to Table~\ref{table:t2}, where the terms are preselected based on the $P$-parity transformation, $P=-1$.
Using the BruteForce search algorithm (with $\lambda_0=10^{-2}$), we found the following Euler equation,
\begin{eqnarray}
\label{eq:kappa_term}
v_t+v v_x + \kappa(U)\rho \rho_x = 0,
\end{eqnarray}
describing the dynamics of interacting fermions.
We now provide the summary of our analysis, consolidate some of the statements from the main text, and expand on the connection between the discovered  Euler/Navier-Stokes fluid models and the Tomonaga-Luttinger theory:
\begin{itemize}
    \item
    In the limit of vanishing interaction strength $U\to 0$, the coefficient $\kappa$ approaches the value $\kappa\approx \pi^2/m^2=4J^2\pi^2$, in agreement with the free-fermion theory for quadratic dispersion. At the same time, minor deviations from the predicted theoretical value $\kappa(U=0)\approx 4J^2\pi^2$ are primarily due to the lattice dispersion corrections, as discussed in Sec.~\ref{sec:cos_corrections}. 
    \item 
    
    In the limit of small density perturbations, $\delta\rho \ll \rho_0$, the viscosity term in Eq.~(\ref{eq:kappa_term}) can be neglected, and the density dynamics takes the form $\Bigl[\partial_t^2-v_{\rm eff}^2(U)\partial_x^2\Bigl]\rho(t,x)=0$, where $v_{\rm eff}(U)=\sqrt{\kappa(U)}$ is a renormalized quasiparticle velocity. This equation can be derived independently in the framework of the Tomonaga-Luttinger
    (T-L) theory\cite{fradkin2013field}, which predicts the renormalized velocity of the form
    \begin{equation}\label{eq:vf_TL}
    v_{\rm eff}(U)=v_{F0}\sqrt{1+\frac{U}{2\pi v_{F0}}},    
    \end{equation}
    where $v_{F0}=\pi\rho_0$ is the Fermi velocity in the non-interacting limit.
    Hence, given the relation between effective velocity and $\kappa(U)$, the value of $\kappa(U)$ in this regime must be close to $\kappa(U)=\kappa(0)\left[1+U/(2\pi^2\rho_0)\right]$.
    Indeed, we find a quantitative agreement between extracted values of $\kappa(U)$ and the T-L prediction for interaction strengths in the region $U/J \lesssim 1$. However, in the strongly interacting regime, $U/J\gg 1$, the observed values of $\kappa$ significantly deviate from the T-L theory (see Fig.~5 in the main text).
    Notably, the effective hydrodynamic model (\ref{eq:kappa_term})  works even beyond the linearized regime for relatively large density perturbations in both weakly- and strongly-interacting limits.
    \item Although the linearization of the discovered Euler equation (\ref{eq:kappa_term}) is equivalent to  the T-L theory, to the best of our knowledge, the non-linear term $\kappa(U)\rho\rho_x$  cannot be explicitly derived from the T-L theory.
    \item Larger perturbations and longer-time evolution would have notable features that cannot be captured by a renormalized $\kappa(U)$. Lowering the penalty constant $\lambda_0$ and extending the library of candidate terms, we found that the deviations are well-captured by the equation
    \begin{equation}
        v_t+v v_x + \kappa \rho \rho_x = \nu  v_{xx}.
        \label{eq:kappa_nu}
    \end{equation}
    Comparison between TEBD data and the solution of the Navier-Stokes model (\ref{eq:kappa_nu}) is shown in Fig.~\ref{fig:navier-stokes}.
    The viscosity term on the right-hand side significantly improves agreement with TEBD data at long evolution times.
    Despite the fact that the viscosity term violates time-reversal invariance of the effective Navier-Stokes model, it does not conflict with unitary dynamics, due to the presence of relaxation processes. Short-range interactions result in the entropy flow from small-scale fluid cells to large-scale structures, by analogy with classical hydrodynamics.
    More accurate analysis shows that the universal value of viscosity emerges only in late-time dynamics, see Sec.~\ref{sec:viscosity}.
    Notably, viscosity effects are beyond linear T-L theory.
    \item We verified that both Euler and Navier-Stokes PDEs~(\ref{eq:kappa_term}, \ref{eq:kappa_nu}) remain accurate for various initial states, see Fig.~\ref{fig:double_peak}.
\end{itemize}

\subsection{Analysis of an emergent Navier-Stokes equation and the viscosity coefficient}\label{sec:viscosity}

In this subsection, we discuss an emergent viscosity term in the hydrodynamics of the 1D Fermi-Hubbard model in more detail and point out limitations of our analysis.

\begin{figure}[H]
    \centering
    \includegraphics[scale=0.3]{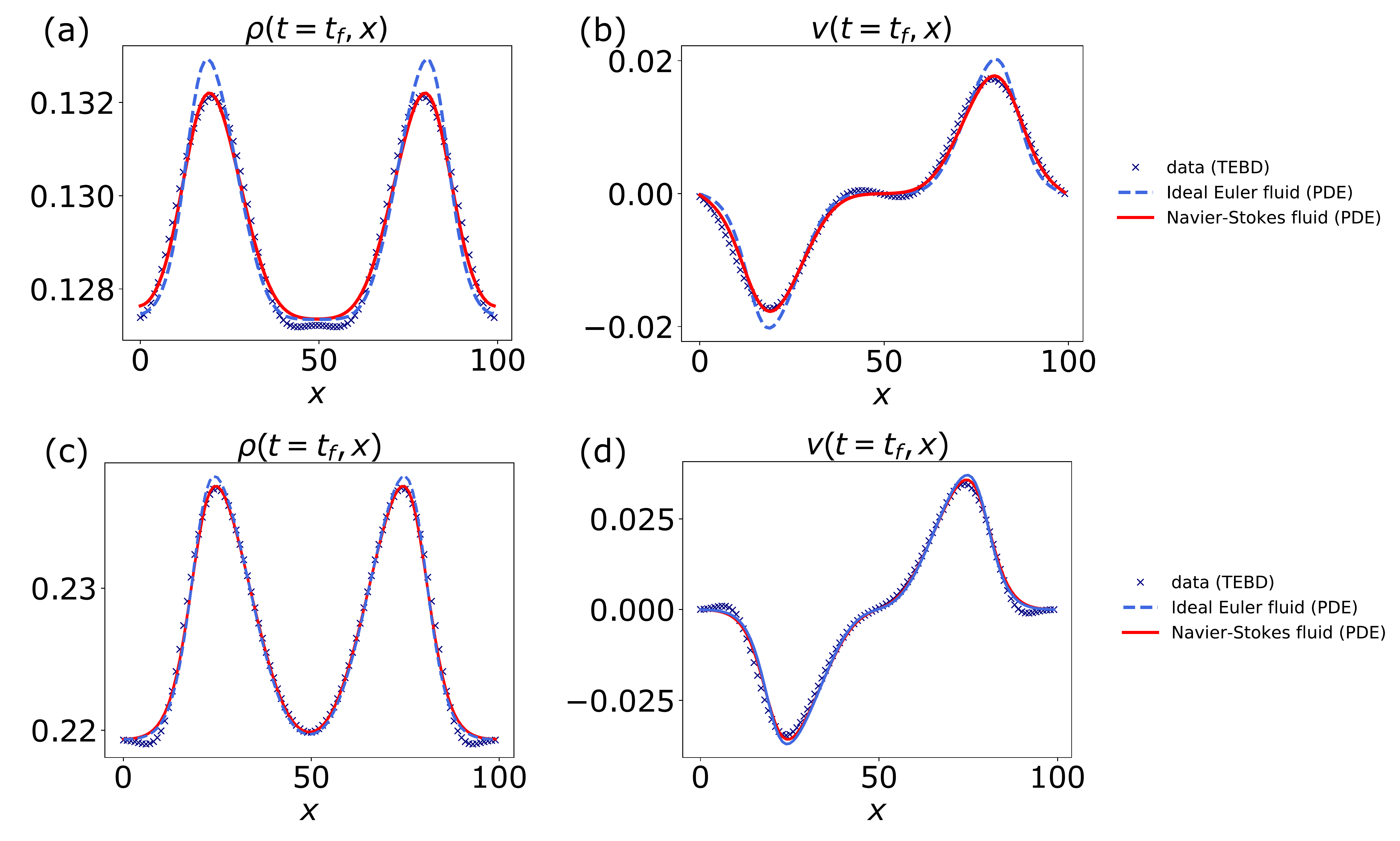}
    \caption{Fermion hydrodynamics in the  Fermi-Hubbard model with nearest-neighbor interaction strength $U$ and next-nearest-neighbor interaction strength $U_2$.
    Panels (a, c) show data  for fermion density, and panels (b, d) show data for  the velocity at the final evolution time, $t_f=60$. 
    Parameters of the Fermi-Hubbard Hamiltonian are  (a,~b) $U/J=4$, $U_2/J=0$, (c,~d)  $U/J=2$, $U_2/J=-1$. The blue and red solid lines correspond to the discovered Euler-like equation (\ref{eq:kappa_term}) and Navier-Stokes-like equation (\ref{eq:kappa_nu}), respectively.
    }
    \label{fig:fermi_hubbard_nnn}
\end{figure}

In order to break integrability of the spinless Fermi-Hubbard model and consider a more generic case, we introduce next-nearest-neighbor  fermion-fermion interactions: $H=H_{\rm f}+V_{\rm int}$. Here $H_{\rm f}$ is a tight-binding Hamiltonian and fermion-fermion interaction has the form $V_{\rm int} = U\sum_i n_i n_{i+1}+U_2\sum_i n_in_{i+2}$. Parameters $U$ and $U_2$ are the nearest-neighbor and the next-nearest-neighbor couplings, respectively. 
Remarkably, the hydrodynamic equation (\ref{eq:kappa_term})  remains valid in the presence of next-nearest-neighbor interactions: see comparison between TEBD data and the solution of the PDE in Fig.~\ref{fig:fermi_hubbard_nnn}. In this case, the pressure renormalization coefficient becomes a function of both coupling parameters: $\kappa(U, U_2)$.
Hence the hydrodynamic model (\ref{eq:kappa_term}) is applicable to a wide range of  models of interacting fermions (e.g.~the Fermi-Hubbard model with nearest-neighbor interactions, next-nearest-neighbor interactions, etc..). 
Discovered Eq.~(\ref{eq:kappa_term}) for an ideal Euler fluid works  well for short evolution times. However, we found that for longer evolution times, the viscous Navier-Stokes model (\ref{eq:kappa_nu}) becomes more accurate.

\begin{figure}[H]
    \centering
    \includegraphics[scale=0.45]{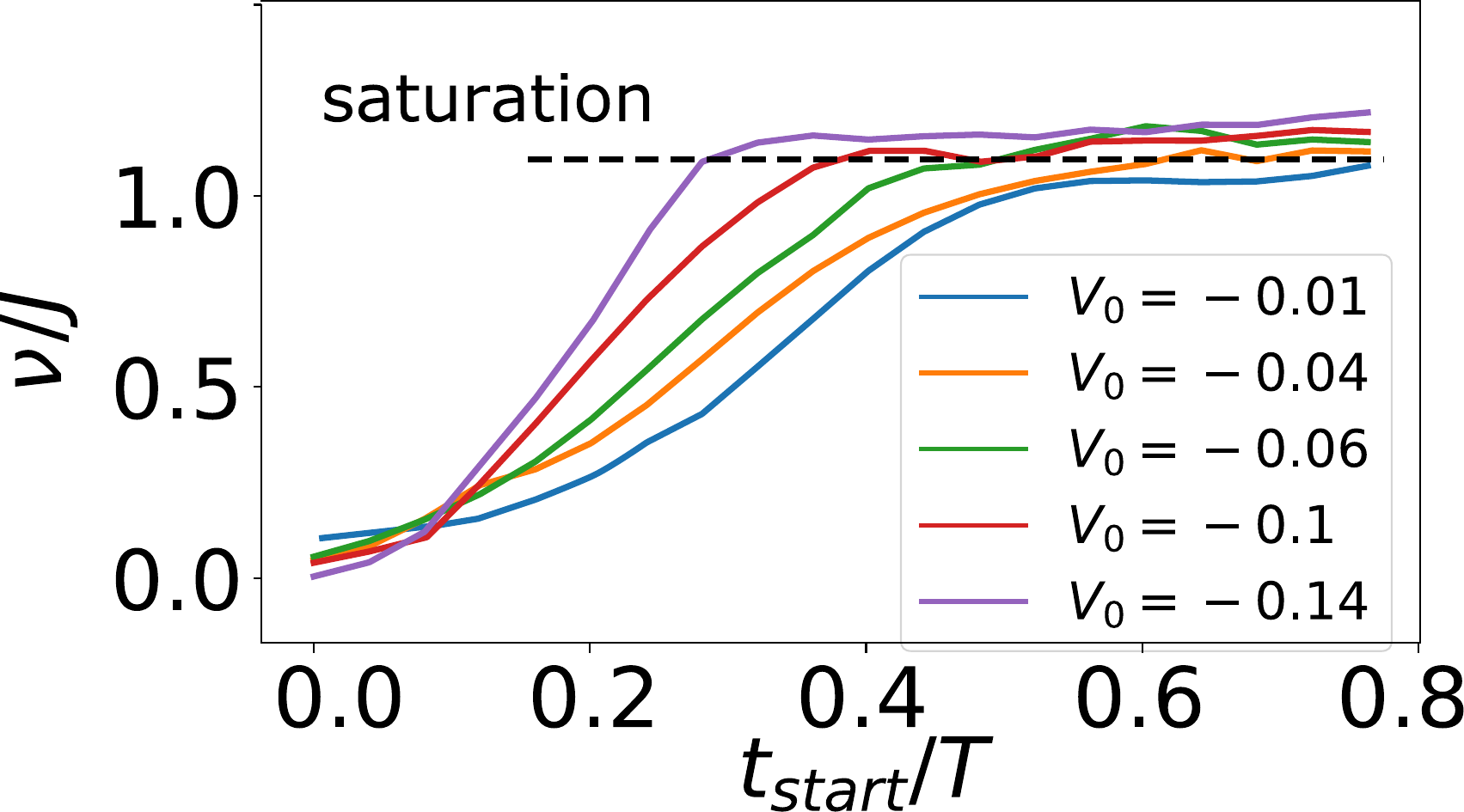}
    \caption{Time dependence of the extracted viscosity coefficient $\nu(U, U_2, t)$ in the extended Fermi-Hubbard model with nearest-neighbor interactions ($U/J=U_2/J=2$). Viscosity is extracted within the sliding time window $[t_{start}, t_{start}+0.2 T]$, where $T$ is the total evolution time.
    We present data for five initial conditions corresponding to diferent values of the amplitude of the Gaussian potential $V_0$.
    At large evolution times, the viscosity coefficient saturates to a constant value $\nu(t\to \infty, U, U_2) \sim J $.
    }
    \label{fig:viscosity_evolution}
\end{figure}

The viscosity term in Eq.~(\ref{eq:kappa_nu}) discovered by our algorithm has an important qualitative role: it prevents the formation of the gradient catastrophe.
We found that the viscosity coefficient obtained from TEBD data $\nu$ could  drift with time, see Fig.~\ref{fig:viscosity_evolution}.
We extract the time dependence of the viscosity coefficient by performing our PDE-reconstruction within a sliding temporal window $[t,\, t+0.2 T]$, where $T$ is the total evolution time.
At the beginning of the evolution, the effect of the viscosity term is negligible, and the effective dynamics is well-described by the inviscid Euler equation.
The viscosity coefficient extracted from TEBD data is close to zero at the start of the evolution and grows with evolution time
approaching a fixed value $\nu\sim J$ at long evolution times, $t\to \infty$.
We checked that the asymptotic value for the viscosity coefficient remains stable to variations of the initial state (changing the   amplitude of the external potential), see Fig.~\ref{fig:viscosity_evolution}.
Generally, the value of the viscosity coefficient will depend on the values of interaction constants $U$ and $U_2$. 
In the non-interacting limit $U, U_2\to 0$, the viscosity coefficient must vanish, $\nu\to 0$.
In our simulations, the magnitude of the effective viscosity term $\nu v_{xx}$ is much smaller compared to the dominant pressure term $\kappa \rho\rho_x$, which affects the precision of the extracted value of $\nu$.
The saturation of the viscosity coefficient $\nu(t\to\infty, U, U_2)\to const$ can be interpreted as an onset of local equilibration in the interacting fermion gas.

\section{Details of numerical simulations}
In the present Section, we provide additional details of numerical simulations for data generation. 
In Section~\ref{sec:errors}, we analyze the errors in the coefficients of  recovered PDEs and the dependence of the errors on the spatiotemporal resolution of the input data.  

Simulations of dynamics in the 1D XX spin chain and the non-interacting fermion chain were performed by exact diagonalization of the single-particle density matrix, $\mathcal G_{ij}(t)=\langle c_i^\dag (t) c_j(t)\rangle$.
In the cases where single-magnon dynamics in the XXZ spin chain was considered, we exactly solved the Schr\"odinger equation in the single-magnon sector of the full Hamiltonian and computed the observables.

TEBD simulations for the dynamics of the domain-wall initial state in the XXZ model and for the dynamics in the interacting Fermi-Hubbard model were performed with the TenPy package~\cite{tenpy}.
The matrix-product-state (MPS) bond dimension was set to $\chi=200$. We checked that an increase of the MPS bond dimension to $\chi=300$  resulted in a small change in the values of the recovered coefficients of the hydrodynamic PDEs ($<0.1 \%$).

\subsection{Error analysis}\label{sec:errors}
In this subsection, we would like to comment on the sources of error in our PDE-reconstruction procedure.

The primary sources of error encountered in PDE-learning from numerical simulations are (i) errors in numerical schemes for the evaluation of high-order derivatives from data, (ii) numerical errors in the dataset (e.g.~truncation errors in TEBD simulations), (iii) physical-model errors originating from higher-order corrections to the approximate PDE: corrections beyond the hydrodynamic approximation, high-order terms in gradient expansion, etc...
A substantial amount of noise in the data can confuse the sparse regression algorithm, thereby introducing spurious terms and/or shifting the values of the extracted coefficients.
Of course, experimental data is usually more noisy as compared to numerical simulations, thus affecting the reliability of the recovered equations.

\begin{figure}[hbtp]
    \centering
    \includegraphics[scale=0.3]{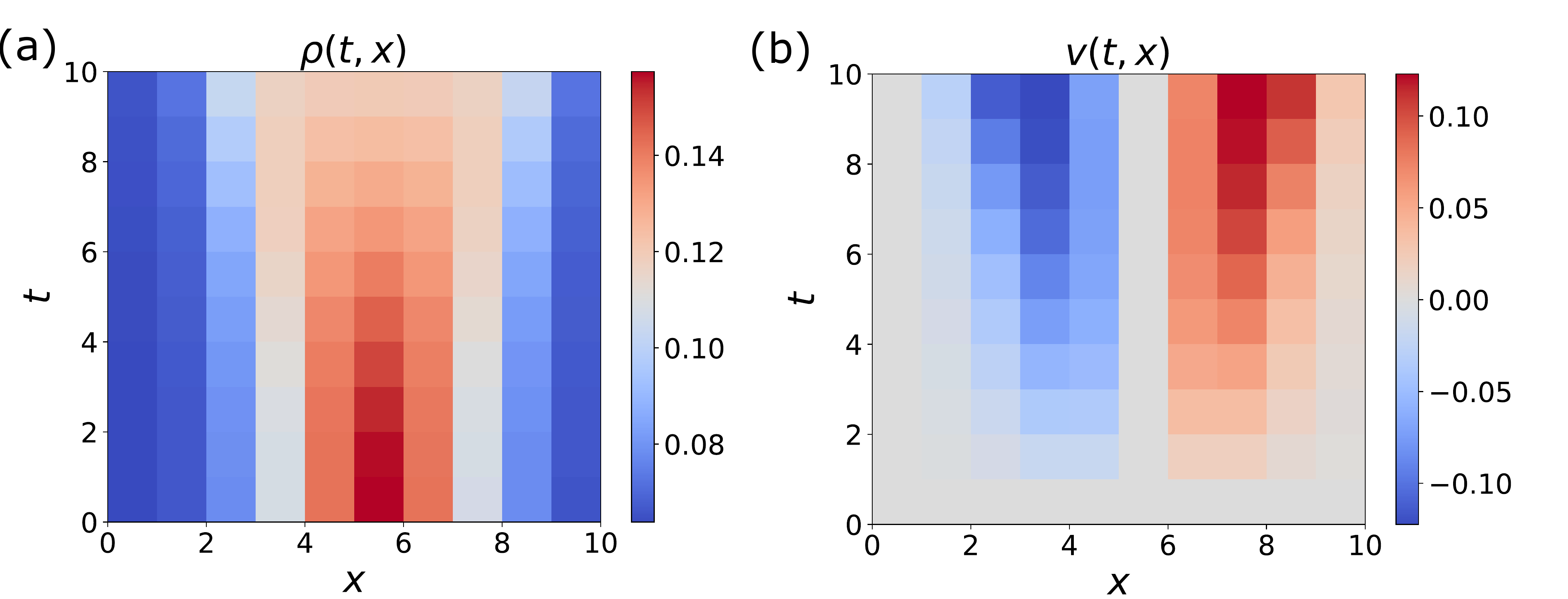}
    \caption{Fermion hydrodynamics (non-interacting fermions): PDE-learning from low resolution data for (a) density  and (b) velocity. When significantly decreasing data resolution to $(N_t,N_x)=(10,10)$, our algorithm was still able to identify the correct form of the hydrodynamic PDE, although  the error in the  values of the coefficients became significant:
    $\rho_t + 1.39\, \rho v_x + 1.14\,v \rho_x = 0$ ($\lambda_0=10^{-5}$),  and 
    $v_t+1.94\, v v_x + 10.3 \rho \rho_x = 0$ ($\lambda_0=10^{-4}$).
    }
    \label{fig:pixelized}
\end{figure}

Below we discuss the role of the spatiotemporal resolution on the quality of PDE reconstruction, see Table~\ref{tab:table_terms}.
We found that leading semiclassical terms are robustly identified with our method, even for very ``pixelated'' data, see Fig.~\ref{fig:pixelized}. 

\begin{table} [hbtp]
    \caption{Dependence of PDE reconstruction performance on the spatiotemporal resolution $(N_t, N_x)$ of the dataset for the quench problem in the non-interacting fermion gas [see Sections \ref{sec:fermion_hydro_parabol} and \ref{sec:cos_corrections}].
    While changing the resolution of the dataset, we keep the spatiotemporal extent $(T,L)$  fixed.
    The candidate terms in the Euler equation $v_t=G(\cdot)$ are preselected based on $(P, T)$ symmetry, resulting in $M=20$ terms, see Table~\ref{table:t2}.
    The leading WKB terms $v_t+v v_x + \pi^2\rho\rho_x + \ldots = 0$  were correctly identified by the BruteForce algorithm for the entire range of $(N_t,\, N_x)$ presented.
    When decreasing the number of spatiotemporal points some of the correction terms originating from the tigh-binding dispersion were  misidentified, the number of misidentified terms $(b_1 \rho^3\rho_x, b_2 v^2\rho \rho_x, b_3 \rho^2 v v_x)$ is shown in the entries of the table.}
    \label{tab:table_terms}
    \includegraphics[scale=0.2]{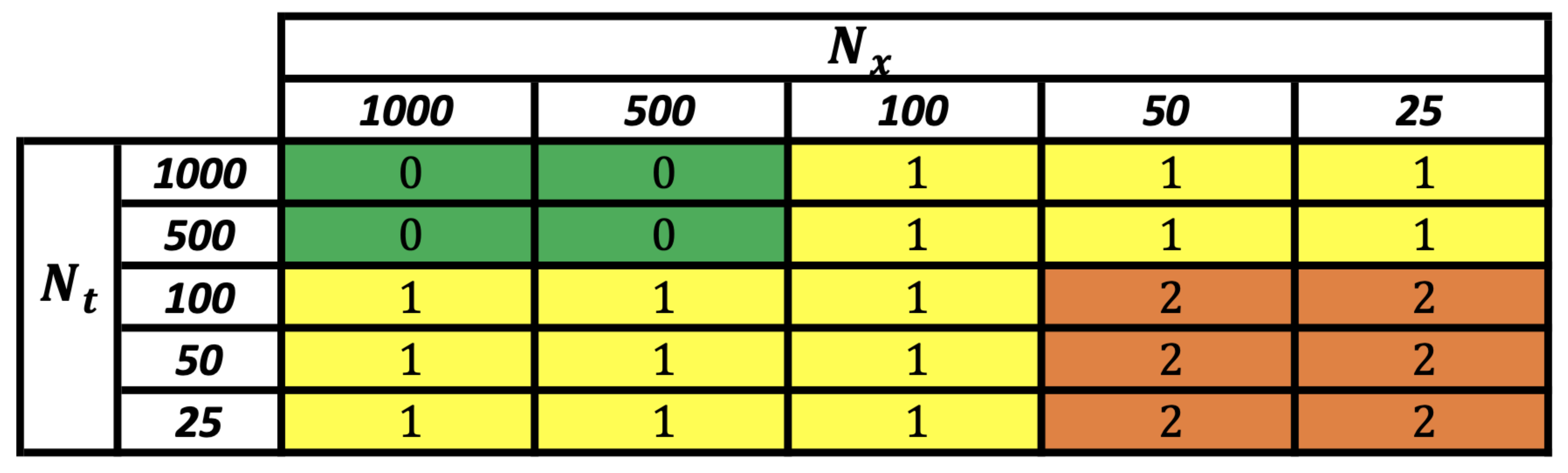}
\end{table}

\begin{figure}
    \centering
    \includegraphics[scale=0.3]{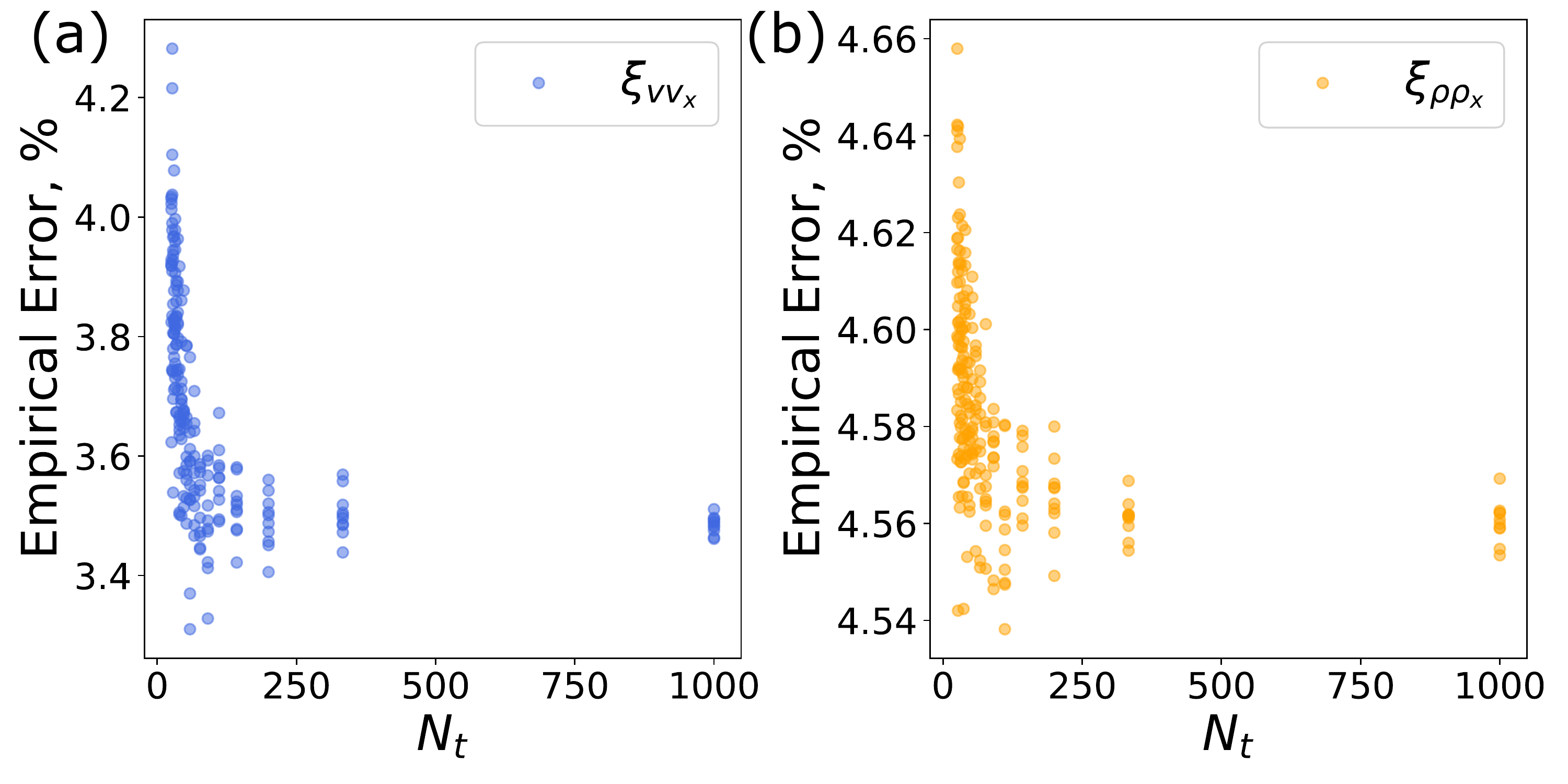}
    \caption{Dependence of the empirical error $|\xi-\xi_{true}|$ on the dataset temporal resolution $N_t$ in recovered coefficients for the  $v v_x$ and $\rho \rho_x$ terms in the PDE $v_t=\xi_{v v_x} v v_x + \xi_{\rho \rho_x}\rho\rho_x$. Here $\xi_{true}=(\xi_{v v_x},\xi_{\rho\rho_x})=-(1,\pi^2)$ corresponds to the theoretical values of the coefficients for the leading WKB terms.
    Scattered points show empirical error in the coefficients (a) $\xi_{vv_x}$ and (b) $\xi_{\rho\rho_x}$  for individual random batches of subsampled data.
    This plot illustrates that the uncertainty in the reconstruction of coefficients in nonlinear PDEs is primarily systematic, since the spread of the error distribution in batches is much smaller than the mean error for a fixed value of $N_t \in [25, 1000]$. The number of spatial points was fixed at $N_x=1000$. The input dataset corresponds to Fig.~\ref{fig:hydro_vs_exact}.
    }
    \label{fig:empir_stat_err}
\end{figure}

Statistical uncertainty in the values of regression coefficients could be estimated if the covariance matrix of the error term 
$\epsilon = y - A \xi$ is known,
$\Sigma = \mathbb{E}[\epsilon\, \epsilon^T]$.
However, for a given dataset, the residual term $\epsilon$ is a fixed vector, and therefore its covariance is not known.
As an alternative approach to estimating the statistical error in $\xi$, we randomly select a subset of rows of the regression vector and the regression matrix: we split data in 10 batches of equal size each containing $10\%$ of the data points. Next we find the regression vector $\xi$ for each batch and estimate 
the uncertainty in the regression coefficients  as an element-wise standard deviation of the values of $\xi$ across  batches.
The resulting statistical error, as well as the empirical error $|\xi-\xi_{true}|$ (deviation of recovered coefficients from the exact theoretical values), is shown in Fig.~\ref{fig:empir_stat_err}, where $\xi_{true}$ are  the theoretically expected coefficients.
Comparing left and right panels of Fig.~\ref{fig:empir_stat_err}, we see that the statistical error remains significantly lower compared to the empirical error.
Therefore, the uncertainties in the values of the  regression coefficients are primarily systematic and generated by noise in the numerically calculated derivatives, as well as by the mutual bias from  higher-order nonlinear terms.

\begin{figure}[H]
    \centering
    \includegraphics[scale=0.3]{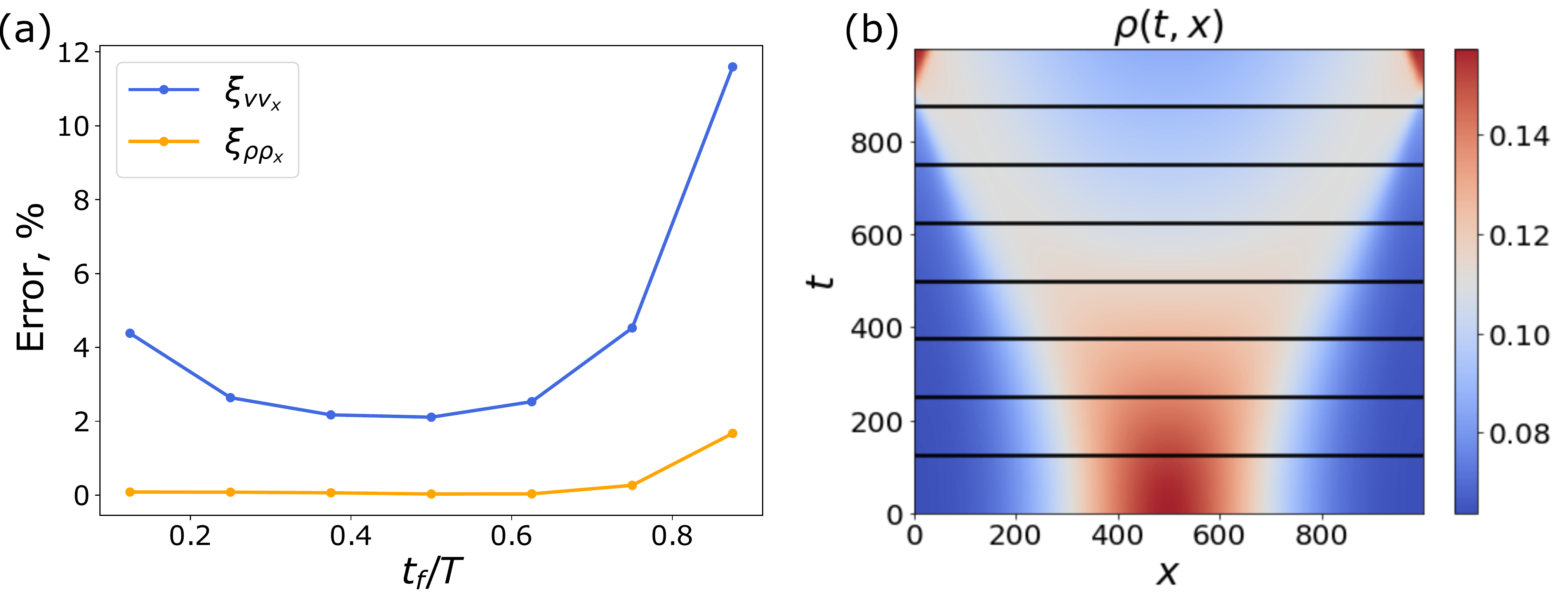}
    \caption{(a) Dependence of the reconstruction error  of the coefficients of the leading WKB terms, $v_t=\xi_{vv_x}vv_x + \xi_{\rho\rho_x}\rho\rho_x$, on the choice  of the training time window. 
    The reconstruction error is defined as $|\xi-\xi_{true}|$.
    (b) Input data for the evolution of fermion density, $\rho(t,x)$.
    The solid horizontal black lines in (b) show the upper cutoff $t_f$ of the training window $t\in [0, t_f]$. At large values $t_f\sim T$, the reconstruction error $\xi_{v v_x}$ and $\xi_{\rho \rho_x}$ grows due to high spatial gradients of the density and the velocity (the ``gradient catastrophe'').  The gradient catastrophe is a well-known feature of the semiclassical description of fermionic dynamics~\cite{mirlin2013}.}
    \label{fig:tmax_err}
\end{figure}

Additionally, we analyze the dependence of the reconstruction error on the choice of the upper cutoff of the evolution time $t_f$ in the dataset, see Fig.~\ref{fig:tmax_err}.

\section{Details of PDE-learning from experimental data: interacting bosons on an atom chip}
In this Section, we provide details of PDE-learning from experimental data,  including details of data post-processing/interpolation. 

\begin{figure}
    \centering
    \includegraphics[scale=0.4]{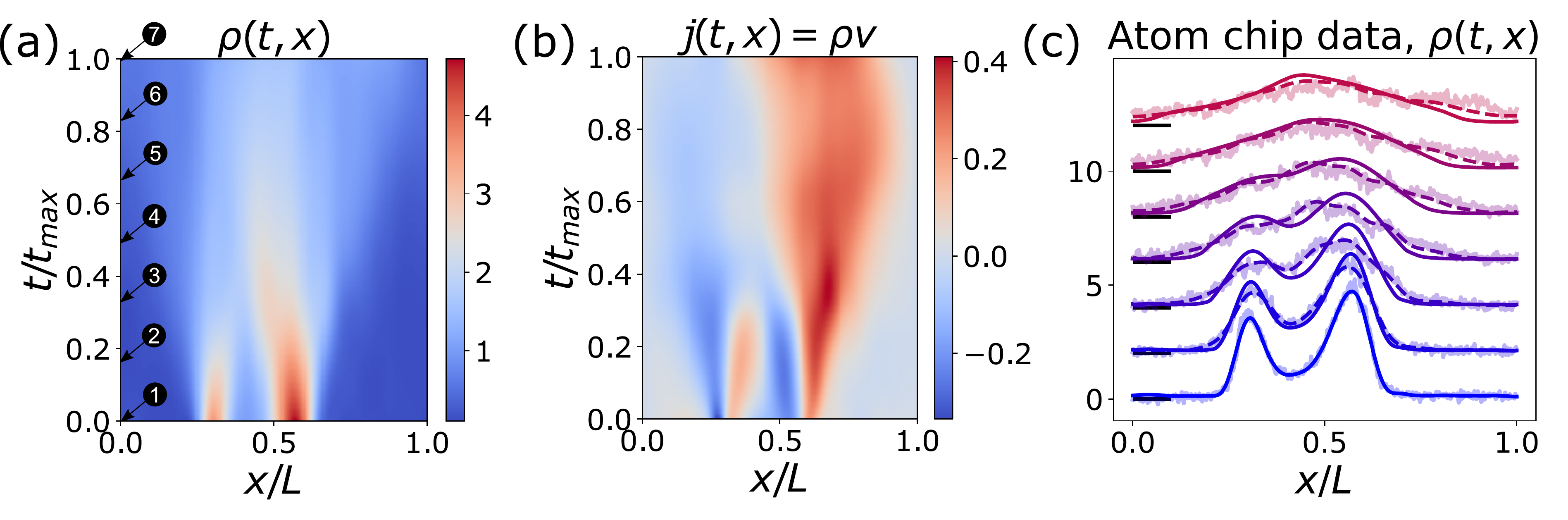}
    \caption{(a, b) Post-processed experimental data [from Ref.~\onlinecite{schemmer2019generalized}] for (a) density $\rho(t,x)$ and (b) particle current $j(t,x)=\rho v$  corresponding to boson cloud expansion  on an atom chip from a double well potential. Here $t_{max}=85$ms is the maximum evolution time in the experimental dataset.
    The particle-current data was reconstructed from the density data $\rho(t,x)$ by utilizing the continuity equation.
    (c) Atom cloud density  (in arbitrary units): original unprocessed data (thick fading line), post-processed data (dashed line), and the solution of the inferred PDE (solid line, Eq.~(20) in the main text). The individual density profiles at different times were shifted vertically relative to each for visualization purposes (black horizontal lines correspond to the origin of the vertical axis). The seven density profiles in (c) correspond to evolution times labeled in (a).}
    \label{fig:atom_chip_interp}
\end{figure}

We process the experimental data as follows. The data from Ref.~\onlinecite{schemmer2019generalized} contains density profiles of the atomic cloud. The spatiotemporal resolution of the original data is $[565 \times 7]$. 
The seven experimental snapshots at different evolution times correspond to the time range  $t\in[0, 85]$ ms. The total length of the 1D atomic cloud is $L\sim 10^{3}\mu m$.
The original post-processed experimental data contains high-frequency spatial noise  resulting in negative values of the  measured density $\rho(t,x)$. 
In order to reconstruct continuous equations, we remove the spatial noise and increase the time resolution. We first remove the high-frequency component of the spatial noise  by applying a Gaussian filter with the variance parameter $\sigma_x/L=2.5\times 10^{-2}$ for each of the seven snapshots.
Next, to suppress the remaining low-frequency noise, we apply the Savitsky-Gollay~\cite{savitzky1964smoothing} filter with a sliding window of length 41 and a polynomial of order 2. 
Finally, we perform cubic 2D interpolation of the resulting data in order to increase resolution along the temporal dimension, which results in the final dataset with spatiotemporal dimensions $[200\times 200]$. The data for the particle velocity was reconstructed by leveraging the continuity equation, Eq.~(\ref{eq:v_hidden})---the result is shown in Fig.~\ref{fig:atom_chip_interp}(b).

The interpolation process may impact the precision of the learning process. For example, the interpolated particle velocities at the start of the quench (i.e.~$t=0$) are non-zero, see Fig.~\ref{fig:atom_chip_interp} (middle panel).
In contrast,
based on the experimental quench protocol, we expect zero particle velocity immediately after the quench of the confining double-well potential.
This mismatch 
is a byproduct of insufficient temporal resolution of the original data. This issue, however, has a limited adverse impact since the inferred velocities at the initial times are not too large.
Also, while solving the inferred PDE forward, Fig.~\ref{fig:atom_chip_interp} (c), we plugged $\rho(t=0,x)=\rho_{data}(0,x)$ and $v(t=0,x) = 0$ as the initial conditions.

\end{document}